\documentclass[pre,showpacs,preprintnumbers ]{revtex4}

\usepackage{amsmath}
\usepackage{amssymb}
\usepackage{mathrsfs}
\usepackage[mathscr]{eucal}
\usepackage{bm}
\usepackage{verbatim}   

\def\be{\begin{equation}}
\def\ee{\end{equation}}
\def\beu{\begin{equation*}}
\def\eeu{\end{equation*}}

\def\bsub{\begin{subequations}}
\def\esub{\end{subequations}}

\def\ie{i.e.}  
\def\eg{e.g.}
\providecommand{\stext}[1]{\text{\tiny{#1}}}
\DeclareMathOperator{\realpart}{Re} 
\DeclareMathOperator{\impart}{Im} 
\def\realsymbol{\mathbb{R}}
\def\complexsymbol{\mathbb{C}}

\def\smallhalf{\tfrac{1}{2}}

\def\roottwo \sqrt{2}

\providecommand{\abs}[1]{\left\lvert#1\right\rvert}
\DeclareMathOperator*{\linspan}{span}

\DeclareMathOperator*{\argmin}{arg\,min}
\DeclareMathOperator*{\argmax}{arg\,max}

\providecommand{\bv}[1]{\boldsymbol{#1}}

\providecommand{\vv}[1]{\bv{#1}}  
\providecommand{\unitvec}[1]{\hat{\boldsymbol{#1}}} 
\providecommand{\vnorm}[1]{\lVert #1 \rVert}  

\def\transsymbol{\text{\tiny T}}

\providecommand{\ket}[1]{\left|#1\right\rangle}
\providecommand{\bra}[1]{\left\langle#1\right|}

\providecommand{\innerp}[2]{\left\langle#1\vert#2\right\rangle}
\providecommand{\innerpa}[2]{\left(#1\, , \, #2\right)}

\providecommand{\outerp}[2]{\left\lvert#1\left\rangle\right\langle#2\right\rvert}
\providecommand{\qamp}[3]{\left\langle#1\left\lvert#2\right\rvert#3\right\rangle}
\providecommand{\comm}[2]{\left[ #1 , #2 \right]}

\providecommand{\cc}{^{\ast}}
\providecommand{\hc}{^{\dag}}
\providecommand{\trans}{^{\transsymbol}}

\providecommand{\grad}{\vv{\nabla}}
\providecommand{\divergence}{\grad\cdot}
\providecommand{\curl}{\grad\times}
\providecommand{\laplacian}{\nabla^{2}}

\providecommand{\del}{\partial}

\DeclareMathOperator{\sinc}{sinc}

\begin{document}


\preprint{UCB/LBNL-CBP preprint}

\title{A Hilbert-space formulation of and variational principle for\\ spontaneous wiggler radiation}

\author{A.E. Charman}
\email{acharman@physics.berkeley.edu}
\author{J.S. Wurtele}
\affiliation{Department of Physics, University of California, Berkeley, Berkeley, CA 94720, USA\\
Center for Beam Physics, Lawrence Berkeley National Laboratory, USA}
\date{22 December 2004}

\begin{abstract}
Within the framework of Hilbert space theory, we derive a maximum-power variational principle applicable to classical spontaneous radiation from prescribed harmonic current sources.  Results are first derived in the paraxial limit, then appropriately generalized to non-paraxial situations.  The techniques were developed within the context of undulator radiation from relativistic electron beams, but are more broadly applicable.
\end{abstract}
\maketitle

\setlength{\parindent}{0 em}
\setlength{\parskip}{1.75ex plus 0.5ex minus 0.2ex}


\section{Introduction}\label{section:introduction}

Variational principles are ubiquitous in electromagnetism \cite{collin:60, harrington:61, mikhlin:64, cairo_kahan:65, jackson:75, kong:86, davies:90, wang:91, zhang:91, vago_gyimesi:98, hanson_yakovlev:02}.  Thomson's and Dirichlet's Principles allow the determination of potentials or capacitances in electrostatics, or the determination of steady-state currents and voltages in  electrical networks.  General reciprocity relations lead to variational estimates for eigenmodes of cavities or waveguides, impedances of cavities, apertures, or other structures, and even certain scattering coefficients.  In magnetohydrodynamics, plasma kinetic theory, and certain equilibrium or non-equilibrium thermodynamic situations, stability or dynamics accessibility of charged particle distributions has been derived from minimum energy/free energy, maximum entropy, or minimum entropy production or minimum heat production principles \cite{bernstein_et_al:58, kruskal_oberman:58, frieman_rotenberg:60, newcomb:62, gardner:63, arnold:65, holm_et_al:85, finn_sun:87, morrison_pfirsch:89}.  Density functional theory, an elaboration of the famous Thomas-Fermi and Hartree-Fock methods in quantum mechanics, has also been applied to classical and quantum Coulomb gases \cite{evans:92, ichimaru:04}.  Many numerical methods and approximation techniques \cite{harrington:68, glowinski:83, mitchell_wait:84, wang:91, blanchard_bruning:92, brenner_scott:94, steele:97} for mechanical or manual computation, associated with the names Raleigh, Ritz, Galerkin, Finite Elements, Minimum Residuals, Method of Moments, etc., may all be derived from or interpreted as variational principles.  As is well known, all of ray optics may be derived from Fermat's Principle of Least Time, and ultimately, all of  classical electrodynamics may be derived via Hamilton's Principle, a variational  formulation demanding stationarity of the action functional.

Variational techniques or approaches enjoy many advantages.  They provide unified theoretical treatments and compact mathematical descriptions of many physical phenomena, which allow for straightforward change of coordinates and incorporation of additional constraints.  Conventional equations of motion and conservation laws may be derived in an concise manner.  Connections between classical and quantum limits are often more readily apparent.  Variational principles often suggest appealing physical interpretations of the behaviors governed by them.  They provide a unified, compact  framework for performing eikonal, ponderomotive, or other averaging techniques.  Perhaps most importantly from a practical standpoint, they offer starting points for efficient approximation or numerical computation, whereby complicated partial differential equations or integro-differential equations may be replaced with more tractable mathematical beasts, namely quadratures, ordinary differential equations, systems of algebraic or even linear equations, or ordinary function minimization.  Although approximate, these simplified, projected, or parameterized variational solutions may be more easily interpreted or more easily computed or offer more insight than the exact forms.

Here, we derive another principle, which we will call the \textit{Maximum-Power Variational Principle} (MPVP), together with some surrounding mathematical formalism.  Although relatively simple in its statement and scope, somewhere between intuitively plausible and obvious, depending upon one's point of view, the MPVP may be of some use in the classical theory of radiation, in particular in the analysis of light sources relying on spontaneous radiation from relativistic electron beams in undulators or other magnetic insertion devices, or more generally, for the approximation of features of other forms of synchrotron or ``magnetic Bremsstrahlung'' radiation.  After some suitable generalization, these ideas may also be applicable to the cases of \v{C}erenkov, transition, wave-guide/cavity, Smith-Purcell, or other types of radiation emitted by classical particle beams traveling through  certain media or structures.

This work arose out of a recent treatment of coherent X-ray generation via a harmonic cascade in an electron beam which radiates in the low-gain regime while traveling through a series of undulators.  Penn, {\it et al.} \cite{penn_et_al:04a, penn_et_al:04b, penn_et_al:04c} approximate the modal structure of the radiation in terms of a paraxial mode described by certain adjustable parameters.  Some of these parameters are subsequently determined directly by dynamical considerations, but some remain free, and are determined at the end of the calculation under the assumption that the ``proper'' or ``natural'' criterion is to choose the values of these parameters which maximize the resulting radiated power assuming that mode shape.  This procedure seems so physically reasonable, appealing, and plausible that intuition suggests that, in some sense, it must be the correct approach under the circumstances.  However, at least the present authors' intuitions fail to immediately suggest a precise derivation or rigorous justification of this power-maximization idea.  Then again, it does lead us to expect that the governing principles or concepts should be quite fundamental, and applicable to more general radiation problems.  

In this paper we pursue these ideas, developing various Hilbert-space approaches to the spontaneous radiation from prescribed current sources, culminating in the Maximum-power variational principle (MPVP) applicable to such problems under very general assumptions.   Although the final results are quite simple and intuitive, the mathematics are developed in some detail in a deliberate and pedagogical fashion;  much can probably be skipped with impunity by all but the most interested readers.

We first consider the approximate but commonly-encountered case of paraxial fields, where the structure and dynamics of the fields and the development of the mathematical results are greatly simplified.  Motivated by the parallels between the Schrodinger equation in non-relativistic quantum mechanics and the paraxial wave equation of classical physical optics, we present an elementary derivation of a convenient Hilbert-space formalism for the spontaneous radiation produced by prescribed classical sources in the paraxial limit, including a derivation of the MPVP.
Guided by these results, we then employ Green function and spherical wave treatments of the general three-dimensional radiation fields to generalize these results to the case of non-paraxial fields propagating in free space.  Specifically, we have in mind applications to the spontaneous synchrotron emission from a reasonably well-collimated relativistic electron bunch in a wiggler or similar insertion device, although the results remain valid more generally.

By \textit{spontaneous} emission, we mean that the trajectories of the charged particles constituting the source for the radiation are considered prescribed functions of time, \textit{independent} of the actual radiation fields generated.  That is, the particle trajectories are assumed to be determined by initial conditions, externally applied wiggler or other guiding fields, and possibly even space-charge effects (quasi-static self-fields, either exact including collisions, or in a mean-field/Vlasov approximation), while any back-action of the radiation itself on the particles -- either recoil, absorption, or multiple scattering effects --  is neglected.  Thus the MPVP provides an approximate alternative to calculation of the fields via the usual Li\'{e}nard-Weichart potentials or related expressions \cite{jackson:75}.

For the case of a wiggler or undulator, this implies that any gain due to ponderomotive feedback and dynamic bunching remains small, so the resulting self-consistent stimulated emission component of the radiation can be neglected compared to the spontaneous component.  The spontaneously-emitted radiation and its measurable properties  (power, angular distribution, coherence, etc.) can then in principle be expressed as deterministic functions of the incoming beam phase space profile (including emittance effects and any pre-bunching) and the prescribed external fields, without having to solve the equations of motion including radiation to obtain completely self-consistent trajectories for the particles.

Although the radiation is treated classically, this terminology involving spontaneous/stimulated emission is standard in the Free Electron Laser (FEL) literature, which, of course, intentionally borrowed it from the quantum theory of lasers.  The \textit{stimulated} emission is that which requires (for ponderomotive bunching) the presence of both the static wiggler field and additional radiation, either from an external seed (FEL amplifier), from emission earlier in the same wiggler (SASE), or from a previous pass through the wiggler in the presence of a resonant cavity or mirror system (conventional FEL oscillator).   The \textit{spontaneous} emission is, by definition, that which occurs in the absence of additional (applied) radiation fields, but of course it still requires the static wiggler field.  Truly ``free'' electrons do not radiate because they are not accelerated, and in fact energy-momentum considerations would forbid even a solitary accelerated electron from radiating if the interaction providing the acceleration did not also allow for a momentum exchange with some external matter (ultimately the wiggler magnets or other electrons, in the present case).

Therefore, the description of the wiggler radiation as spontaneous is similar to the atomic case in that it occurs in the absence of applied radiation (\ie, real photons), but it differs in that electrons bound in atoms, if somehow excited, can then radiate in the absence of any external fields.  The static (\ie, virtual photon) wiggler field is therefore playing a role analogous to the nuclear Coulomb field.  In the average rest frame of a sufficiently relativistic beam, the Weizsacker-Williams method of virtual quanta may be used, since the Lorentz-transformed static wiggler field begins to resemble an oncoming beam of actual photons.  Spontaneous radiation occurs when a backwardly-moving, virtual, wiggler photon is scattered by an electron into a forward-moving, real photon.  Stimulated emission occurs in a ``three-body collision'' where an electron simultaneously scatters a virtual wiggler photon and real radiation photon into two real photons, both in the same mode as the incident photon.

Throughout this analysis, we also assume that the charge and current densities are not only prescribed, but remain \textit{localized} in space (so that the far-field, or wave zone, is defined) and time (so that certain Fourier transforms exist), in a manner more precisely specified in the course of our derivations.  We could generalize our treatment to additionally include the effects of uniform background conductive or dielectric media, but for simplicity we here assume that, apart from its generation by the prescribed microscopic sources in a bounded spacetime region, the emitted radiation otherwise propagates in vacuum.  Generalizations to allow for non-uniform dielectric tensors, representing wave-guides, lenses, windows, or other optical devices, might also be possible, but is not pursued here.  As we are interested in the causal (outgoing) radiation produced by the given sources, we neglect any incident source-free fields or fields from other, remote sources.  Because we are working in the fully linear spontaneous limit, any such fields may be added in superposition at the very end of the calculation.

\section{Fundamental Equations}\label{sec:fundamentals}

Throughout this paper, we denote in boldface type any real or complex vectors, \eg, $\bv{a}$ or $\bv{b},$ and denote with carets any real or complex unit vectors such as $\unitvec{a}$ which satisfy
\be
\vnorm{\unitvec{a}} \equiv \left(\unitvec{a} \cc \cdot \unitvec{a}\right)^{1/2} = 1,
\ee
where for any $N$-dimensional  complex-valued vectors $\bv{a}, \, \bv{b} \in \complexsymbol^N,$
\be
\bv{a} \cdot \bv{b} = \sum_{n = 1}^{N} a_{n} b_{n}
\ee
is the usual Euclidean dot product (without any implicit complex conjugation) of the components of $\bv{a}$ and $\bv{b}$ in some orthogonal basis.  Additional notation will be introduced as needed.

Conforming to widespread, if not universal, convention in beam physics, we work in the (non-covariant) Coulomb, transverse, or radiation gauge, where the vector potential $\bv{A} = \bv{A}_{\perp} = \bv{A}(\bv{x}, t)$ is chosen to be purely solenoidal (\ie, divergenceless), or everywhere transverse in the sense of Helmholtz's Theorem:
\be
\divergence \bv{A}(\bv{x}, t) = 0.
\ee
Here, $\bv{x} \in \realsymbol^3$ denotes the three-dimensional position and $t \in \realsymbol$ the time, in any one convenient Lorentz frame (typically the lab frame, or possibly the average electron beam frame, for wiggler radiation problems).  In Gaussian units, the scalar potential $\Phi = \Phi(\bv{x}, t) $ then satisfies the instantaneous (\ie, unretarded) Poisson's equation 
\be\label{poisson}
\laplacian \Phi(\bv{x}, t) = -4\pi \rho(\bv{x}, t),
\ee
given the charge density $\rho,$ while the vector potential the vector potential $\bv{A}$ satisfies the inhomogeneous wave equation
\be
\laplacian \bv{A}(\bv{x}, t) - \tfrac{1}{c^2}\tfrac{\partial^2 }{\partial t^2}\bv{A}(\bv{x}, t) =  - \tfrac{4\pi}{c}\bv{J}_{\perp}(\bv{x}, t),
\ee
where $c$ is the speed of light in vacuo, and $\bv{J}_{\perp}$ is the transverse (solenoidal) component of the full current density $\bv{J},$ and which, using (\ref{poisson}) and local charge conservation, can be shown to be given by
\be
\label{eqn:j_decomp}
\bv{J}_{\perp}(\bv{x}, t) =  \bv{J}(\bv{x}, t) -  \bv{J}_{\|}(\bv{x}, t) =   \bv{J}(\bv{x}, t) - \tfrac{1}{4\pi}\grad\tfrac{\partial }{\partial t} \Phi(\bv{x}, t).
\ee
The physical fields are, as usual, related to the potentials by
\be
\bv{E} = \bv{E}_{\perp} + \bv{E}_{\|} = -\tfrac{1}{c}\tfrac{\del }{\del t} \bv{A} - \grad \Phi,
\ee
and
\be
\bv{B} = \bv{B}_{\perp} = \curl \bv{A}.
\ee

Note again that we are describing only the fields generated by the prescribed sources $\rho$  and $\bv{J}.$  By assumption, any externally applied fields (due ultimately to some additional, remote sources) such as undulator or bend magnets are not included in the potentials $\bv{A}$ or $\phi,$ but their effects on the trajectories of the charge particles constituting the actual radiation sources are assumed to be either already included in $\bv{J}(\bv{x},t)$ and $\rho(\bv{x},t),$ or else are neglected.

Without some prior expectation for the typical magnitudes of the dominant radiation frequency $\omega_{\text{\tiny{rad}}}$ or characteristic transverse spot size $w_{0}$ of a radiation beam,  it is convenient to re-scale the space-time coordinates using the remaining spatio-temporal scales, namely those determined by the speed of light $c$ and the plasma frequency $\omega_p$ associated with the charge density of the known source.  That is, choosing some reference number density $\bar{n}$, ideally characteristic of that of the actual particle bunch, and defining the corresponding linear plasma frequency 
\be
\omega_p = \left(\frac{4\pi\bar{n} e^2}{m_{e}}\right)^{1/2},
\ee
were $m_{e}$ is the electron mass and $e$ the magnitude of the electron charge, 
we define normalized (dimensionless) variables: a normalized time $\tau = \omega_p t,$ normalized  longitudinal position $\xi = \tfrac{\omega_p}{c} z,$ normalized transverse coordinates $\bv{r} = r_x\, \unitvec{x} + r_y \,\unitvec{y} = \tfrac{\omega_p}{c}\left( x\, \unitvec{x} + y \,\unitvec{y}\right),$ normalized three-dimensional coordinates $\bv{\zeta} = \xi \unitvec{z} + \bv{r},$
the normalized vector potential $\bv{a}(\bv{\zeta}; \tau) = \bv{a}(\bv{r}, \xi; \tau)  = \tfrac{e}{mc^2} \bv{A}(\bv{x}, t),$ normalized scalar potential $\phi(\bv{\zeta}; \tau) = \phi(\bv{r}, \xi; \tau) = \tfrac{e}{mc^2}\Phi(\bv{x}, t),$ normalized charge density $\mu(\bv{\zeta}; \tau) = \mu(\bv{r}, \xi; \tau) = \tfrac{1}{e \bar{n}}\rho(\bv{x}, t),$ and a normalized current density $\bv{j}(\bv{\zeta}; \tau) = \bv{j}(\bv{r}, \xi, \tau) =  \tfrac{1}{e\bar{n} c }\bv{J}(\bv{x}, t),$ with transverse (solenoidal) part
$\bv{j}_{\perp} = \tfrac{1}{e\bar{n} c}\bv{J}_{\perp} = \bv{j} - \left(\unitvec{z}\frac{\partial }{\partial \xi} + \grad_{\perp}\right)\frac{\partial }{\partial \tau}\phi,$ where 
$\grad_{\perp} \equiv \frac{\del }{\del \bv{r}} = \unitvec{x} \frac{\del }{\del r_x} +  \unitvec{y} \frac{\del }{\del r_y}$ is the scaled, (geometrically) transverse gradient operator, and we also define $\bv{\del} = \frac{\del }{\del \bv{\zeta}} = \unitvec{z}\frac{\del }{\del \xi} + \grad_{\perp}$ as the scaled gradient operator in the full three-dimensional space.

In these variables, Poisson's equation becomes
\be
\del^2 \phi(\bv{\zeta}; \tau) = \left( \tfrac{\partial^2}{\partial \xi^2} + \laplacian_{\perp} \right) \phi(\bv{r}, \xi; \tau) = \mu(\bv{r}, \xi; \tau),
\ee
where $\laplacian_{\perp} = \grad_{\perp} \cdot \grad_{\perp}$ is the scaled (geometrically) transverse Laplacian, and $\del^2 = \bv{\del}\cdot \bv{\del}$ is the scaled three-dimensional Laplacian.  The solutions to this equation just consist of the well-known unretarded (\ie, instantaneous) Coulomb fields,
\be
\label{eqn:coulomb_sol}
\phi(\bv{\zeta}; \tau) = -\tfrac{1}{4\pi}\int d^3 \bv{\zeta}'\, \frac{\mu(\bv{\zeta}', \tau)}{\vnorm{\bv{\zeta} - \bv{\zeta}'}}
\ee
which fall off like the inverse-square distance from the instantaneous positions of the charges, so they do not contribute to the radiation fields, and can be neglected for our purposes.  (These unretarded Coulomb fields are ``attached'' to the charges and move with them even under acceleration, so do not include or in any way constitute radiation.)
The scaled wave equation for the vector potential may be written as
\be
\left(\del^2 - \frac{\del^2 }{\del \tau^2} \right) \bv{a}(\bv{\zeta}; \tau) = \left( \tfrac{\del^2 }{\del \xi^2 } + \laplacian_{\perp} - \frac{\del^2 }{\del \tau^2} \right) \bv{a}(\bv{r}, \xi; \tau) = -\bv{j}_{\perp}(\bv{r}, \xi; \tau),
\ee
where $\bv{a}$ also satisfies the gauge condition
\be
\bv{\del} \cdot \bv{a}(\bv{\zeta}; \tau) = \left( \unitvec{z}\tfrac{\partial }{\partial \xi } + \grad_{\perp}\right) \cdot \bv{a}(\bv{r}, \xi; \tau) = 0.
\ee
All of what we conventionally consider radiation is associated with the fields derived from this vector potential, although these fields also include (in the near-field region) non-radiative components needed to cancel the non-retarded features of the Coulomb fields.

To separately analyze each spectral component of the radiation, we perform Fourier transforms in (scaled) time, obtaining, for each (scaled) frequency $\omega,$  the frequency-domain wave equation
\be
\left( \del^2 + \omega^2 \right) \bv{a}(\bv{r}, \xi; \omega) = \left( \tfrac{\del^2 }{\del \xi^2 } + \laplacian_{\perp} + \omega^2 \right) \bv{a}(\bv{r}, \xi; \omega) = -\bv{j}_{\perp}(\bv{r}, \xi; \omega),
\ee
and frequency-domain gauge condition
\be
\bv{\del} \cdot \bv{a}(\bv{\zeta}; \omega) = \left( \unitvec{z}\tfrac{\partial }{\partial \xi } + \grad_{\perp} \right) \cdot \bv{a}(\bv{r}, \xi; \omega) = 0,
\ee
where 
\be
\bv{a}(\bv{\zeta}; \omega) = \tfrac{1}{\sqrt{2\pi}} \int d \tau\, e^{i \omega \tau}\bv{a}((\bv{\zeta}; \tau)
\ee
and
\be
\bv{j}_{\perp}(\bv{\zeta}; \omega) = \tfrac{1}{\sqrt{2\pi}} \int d \tau\, e^{i \omega \tau}\bv{j}_{\perp}(\bv{\zeta}; \tau)
\ee
denote the usual Fourier transforms in scaled time, and are in general complex-valued.   Because the physical, time-domain fields are real-valued, we necessarily have  $\bv{a}(\bv{\zeta}; -\omega) = \bv{a}(\bv{\zeta}; \omega)\cc,$ so we can restrict attention to the analytic signal, or positive frequency ($\omega > 0$), components of the radiation fields.  We assume that the source charge density $\rho$ and current density $\bv{j}$ are at least weakly localized in time, to the extent that all needed Fourier transforms of the sources and the resulting fields actually exist (at least as generalized functions).

Corresponding to the scaled, frequency-domain vector potential $\bv{a}(\bv{\zeta}; \omega),$ we define the scaled solenoidal electric field:
\be
\bv{\varepsilon}_{\bv{a}} = \bv{\varepsilon}_{\bv{a}}(\bv{\zeta}; \omega) =i\omega \, \bv{a}(\bv{\zeta}; \omega);
\ee
and the scaled frequency-domain magnetic field
\be
\bv{b}_{\bv{a}} = \bv{b}_{\bv{a}}(\bv{\zeta}; \omega) = \del \times \bv{a}(\bv{\zeta}; \omega).
\ee
The subscript may be dropped when it is clear from which vector potential these fields are derived.
For later convenience, we also define the scaled, frequency-domain Poynting vector
\be
\bv{s}_{\bv{a}}(\bv{\zeta}; \omega) = \bv{\varepsilon}_{\bv{a}}(\bv{\zeta}; \omega)  \times \bv{b}_{\bv{a}}(\bv{\zeta}; \omega)\cc. 
\ee
For any orientable  surface $\Sigma$ with unit outward normal $\unitvec{n},$ and positive frequency interval $0 < \omega_0 \le \omega \le \omega_1,$ the quantity 
\be
\mathcal{P}[\bv{a}](\Sigma; \omega_0, \omega_1) \equiv  \int\limits_{\omega_0}^{\omega_1}d\omega\,  \realpart\biggl[ \int\limits_{\Sigma} d^2 \sigma\,  \unitvec{n}\!\cdot\! \bv{s}_{\bv{a}}\biggr]
\ee
may be taken as the scaled, time-averaged (over an optical period) net electromagnetic power, in the bandwidth $[\omega_0, \omega_1],$ which passes through the surface $\Sigma.$

In order that the far-field, or radiation zone, even  be well-defined, we assume that the physical source $\bv{j}(\bv{\zeta}; \tau)$ remains localized in space throughout the relevant emission process, at least in the direction(s) of observed radiation propagation.  In the general, three-dimensional case, this means that  $\bv{j}(\bv{\zeta}; \omega)$ is non-negligible only in some neighborhood of some fixed point $\bv{\zeta}_0.$  However, from (\ref{eqn:j_decomp}) and (\ref{eqn:coulomb_sol}), it then follows that the solenoidal component $\bv{j}_{\perp}(\bv{\zeta}; \tau)$ will not be strictly localized, but must possess decaying tails which fall off like $O(1/\vnorm{\bv{\zeta} - \bv{\zeta}_s}^2)$ at sufficiently large distances from $\bv{\zeta}_0.$   That is, $\bv{j}$ and $\bv{j}_{\perp}$ will not both be of compact support simultaneously.   But the fields arising from any particular part of $\bv{j}_{\perp}(\bv{\zeta}'; \tau)$ will consist of the near-zone and intermediate zone fields which will fall off like  like $O(1/\vnorm{\bv{\zeta} - \bv{\zeta}'}^2)$ or faster, as well as the true radiation fields which fall off more slowly, i.e. as $O(1/\vnorm{\bv{\zeta} - \bv{\zeta}'}).$  Because our interest resides in the latter, it follows that as long as $\bv{j}(\bv{\zeta}; \tau)$  is appropriately localized,  we can always choose some suitably large but finite region beyond which the contributions  of the decaying tails of  $\bv{j}_{\perp}$ to the radiation fields may be neglected to any desired level of accuracy, and  we can then act as if  both $\bv{j}(\bv{\zeta}; \tau)$ and $\bv{j}_{\perp}(\bv{\zeta}; \tau)$ have compact support.

\section{Paraxial Case}\label{sec:paraxial}

By paraxial, we mean that the radiation from the electron bunch in the insertion device can be assumed to be highly mono-directional, in the sense that the relevant angular scales -- namely any possible overall angular offset between the electron beam and optical axis of the insertion device, the angular spread of the particle beam, and the characteristic angle of diffraction -- are assumed to be small, in a sense made precise below.  If we restrict attention to radiation observed within a sufficiently small detection solid-angle, then radiation from relativistic particles in certain non-straight devices, \ie, synchrotron radiation in circular rings, may also be treated within the paraxial approximation.  The needed local directionality is provided by the assumed finite acceptance entendue of the detector and the characteristic $O(1/\gamma)$ angular spread in the radiation resulting from relativistic $O(1/\gamma^2)$ compression effects between emitter and observer times, where $\gamma$ is the usual relativistic kinematic factor associated with the mean velocity of the source charges, which together limit the contributing portion of the electron trajectory to some small arc.

\subsection{Paraxial Approximation to the Wave Equation}\label{subsec:paraxial_approx}

We have already seen how, in the Coulomb gauge, the scalar potential $\phi$ does not contribute to the radiation fields.  The vector potential $\bv{a}$ contributes to near-field (quasi-static), intermediate (induction-zone), and far-field (radiation-zone) effects (and in fact part of it must cancel, in the resulting fields, the non-retarded effects of the scalar potential), but considerable simplification is possible if we effectively restrict our attention to those portions of the fields which, if allowed to evolve freely after propagating beyond the finite support of the sources, actually constitute radiation, and further assume that the radiation of interest propagates nearly along the $+\unitvec{z}$ axis, with a small characteristic angle of diffraction $\theta_{\text{\tiny{D}}} \ll 1$ and any overall angular offset $\delta \theta_0 \lesssim O(\theta_d).$  Then we can make the so-called paraxial approximation to the wave-equation for the vector potential $\bv{a},$ which in effect consists of an asymptotic expansion in $\delta \theta_0.$

First, for any $\omega > 0$ of interest, we express the vector potential in enveloped form:
\be
\bv{a}(\bv{r}, \xi; \omega) = \bv{\psi}(\bv{r}, \xi; \omega) e^{i k \xi},
\ee
where the (scaled), axial carrier wavenumber $k = k(\omega) = +\omega$ satisfies the vacuum dispersion relation for a local plane wave traveling along $+\unitvec{z}.$  Similarly, we define an enveloped source term such that:
\be
\bv{j}_{\perp}(\bv{r}, \xi; \omega) =  2k \,\bv{f}(\bv{r}, \xi; \omega)e^{i k \xi}.
\ee
Using the chain rule, the wave equation then becomes
\be\label{eqn:wave2}
\left(\tfrac{\del^2 }{\del \xi^2 }+  2 i k \frac{\del }{\del \xi} + \laplacian_{\perp} \right) \bv{\psi}(\bv{r}, \xi; \omega) + 2 k \bv{f}(\bv{r}, \xi; \omega) = 0,
\ee
while the radiation envelope $\bv{\psi}$ and source envelope $\bv{f}$ both satisfy envelope-transversality conditions:
\begin{subequations}\label{eqn:transversality_3}
\begin{align}
\left(\unitvec{z}\tfrac{\partial }{\partial \xi } +  i k \,\unitvec{z} + \grad_{\perp} \right) \cdot \bv{\psi}(\bv{r}, \xi; \omega) &= 0,\label{eqn:trans_a3}\\
\left( \unitvec{z}\tfrac{\partial }{\partial \xi } +  i k\, \unitvec{z} + \grad_{\perp} \right) \cdot \bv{f}(\bv{r}, \xi; \omega) &= 0.\label{eqn:trans_b3}
\end{align}
\end{subequations}

We assume that the charge density $\rho$ and full current density $\bv{j}$ remain weakly localized transversely, in the sense that they are square-integrable in any transverse plane.  Actually we will impose the slightly stronger constraint that  $\abs{\bv{f}(\bv{r}; \xi; \omega)} \to 0$ more rapidly than $1/\abs{\bv{r}},$ as $\abs{\bv{r}} \to \infty.$   We also assume that the physical currents are strongly localized longitudinally, in the sense that they are negligible outside some finite interval $\xi_0' \le \xi \le  \xi_1'$ along the propagation axis, implying in general that the solenoidal part of the current density $\bv{j}_{\perp}$ will actually have a component with non-compact support,  corresponding to the $-\left(\unitvec{z}\frac{\partial }{\partial \xi} + \grad_{\perp}\right)\frac{\partial }{\partial \tau}\phi$ term.  However, this contribution to the sources falls off no slower than the inverse-square of the distance from the source charges, so with arbitrarily small relative error in the radiation fields, which are linear in the sources and fall off with the inverse distance from them, we can safely truncate the source envelope $\bv{f}(\bv{r}; \xi; \omega)$ outside of some sufficiently large but finite range $\xi_0 < \xi_0' \le \xi \le \xi_1'  < \xi_1.$   (Again, the neglected contributions are strictly necessary to cancel the unretarded part of the Coulomb fields to ensure that the full fields remain compatible with relativistic causality, but we are confining attention only to actual radiation.)
Most of the final results will not actually depend on the size of this interval, so if necessary we may safely take $\abs{\xi_1 - \xi_0} \to \infty$ at the end of the calculation, provided the charge and current densities continue to fall off sufficiently rapidly to ensure convergence of the relevant integrals, and we can somehow distinguish the far-fields from the local quasi-static fields when the latter may then extend to infinity.

The (scaled) waist of the radiation beam, or characteristic transverse spot size $w_0,$ may be defined in terms of the minimal, power-weighted, root mean-square radius:
\be
\smallhalf w_{0}^2 = \inf_{\xi} \frac{\int d^2 \bv{r} \, \abs{\bv{r}}^2 \abs{\bv{\psi}(\bv{r}, \xi; \omega)}^2}{\int d^2 \bv{r} \, \abs{\bv{\psi}(\bv{r}, \xi; \omega)}^2}.
\ee
The corresponding characteristic diffraction angle $\theta_{\text{\tiny{D}}}$ may then be taken as
\be
 \theta_{\text{\tiny{D}}} =  \frac{1}{2\pi} \frac{\lambda }{w_0} = \frac{1}{k w_0},
\ee
and finally the (scaled) Rayleigh range $\xi_{\text{\tiny{R}}},$  or longitudinal length-scale for appreciable change in the transverse spot size due to diffraction, becomes:
\be
\xi_{\text{\tiny{R}}} = \smallhalf k w_0^2.
\ee

We now treat $\theta_{\text{\tiny{D}}}$ as a small parameter.  Specifically, if $\psi_j$ is any (non-zero) component of the radiation envelope $\bv{\psi},$ then we may consistently assume that: $\abs{\tfrac{\del}{\del \xi} \psi_j}/\abs{\psi_j} \sim O(\xi_{\text{\tiny{R}}}^{-1}),$ while $\vnorm{\grad_{\perp} \psi_j}/\abs{\psi_j} \sim O(w_0^{-1}).$  If the spot size is sufficiently large compared to the carrier wavelength $\lambda = 2\pi/k,$ such that $\theta_{\text{\tiny{D}}} \ll 1,$ then with a relative error of $O(\theta_{\text{\tiny{D}}}^2),$ we may drop the higher-order longitudinal derivative in equation (\ref{eqn:wave2}), obtaining the forced paraxial wave equation
\be\label{eqn:paraxial_wave}
\left(i \tfrac{\del }{\del \xi} + \tfrac{1}{2k} \laplacian_{\perp} \right) \bv{\psi}(\bv{r}, \xi; \omega) + \bv{f}(\bv{r},\xi;\omega) = 0.
\ee

Note that we have not yet directly imposed the same ordering assumptions on the spatial variations in the envelope $\bv{f}(\bv{r}; \xi; \omega),$ but as the latter is the actual source for the radiation, it is clear that the transverse scale-lengths of $\bv{\psi}$ must be consistent with those of $\bv{f}(\bv{r}; \xi; \omega)$ under diffractive propagation (and therefore typically comparable in the vicinity of the sources.)  However, as long as we consider sufficiently small wavelengths, the scale-length for longitudinal variations in $\bv{f}(\bv{r}; \xi; \omega)$ need not be comparable to $\xi_{\text{\tiny{R}}}$ in order for the paraxial approximation to remain accurate for the fields of interest, \ie, for radiation observed in the far-field region beyond the sources.  Specifically, we may imagine the paraxial fields as the superposition of those fields from each infinitesimal transverse slice of current, evolved forward according to the free-space paraxial propagation, so that each component will satisfy the paraxial conditions everywhere except for a discontinuous jump at its source slice, which can be replaced with an equivalent boundary condition applied to the homogeneous (source-free) equation.  The paraxial solution is expected to remain accurate for any $\xi > \xi_1$ if the paraxial conditions both hold for the calculated fields in the region $\xi > \xi_1$ and also hold for $\xi < \xi_1,$ when the fields are extrapolated backwards via the free-space (homogeneous) paraxial propagation.

This paraxial wave equation is valid when the propagation direction is sufficiently close to the fiducial ($+\unitvec{z}$) axis, and the scale-length for transverse variations in the radiation beam are large compared to the carrier wavelength and small compared to the length-scale for longitudinal envelope variation (\ie, residual variation after the faster carrier oscillations have been removed.)  Actually, although we have formally assumed that  $\theta_{\text{\tiny{D}}} \ll 1,$ numerical solutions to the full wave equation typically reveal that the paraxial approximation actually remains surprisingly accurate for focused radiation beams provided only $\theta_{\text{\tiny{D}}} \lesssim O(1).$  For still smaller spot sizes (tighter focus), the paraxial equation predicts a coherence volume $V_{\text{\tiny{coh}}} < O(\lambda^3)$ for a single oscillatory mode,  which would be in conflict with the diffraction limit dictated by the ``Heisenberg'' inequality constraining conjugate Fourier transform pairs.

From our above ordering assumptions, and the transversality condition (\ref{eqn:trans_a3}), it follows that $\abs{\psi_z}/(\abs{\psi_x}^2 + \abs{\psi_y}^2)^{1/2} \sim O(\theta_{\text{\tiny{D}}}),$ so the axial component $\psi_z$ can in certain circumstances be neglected outright, but we will retain it here in order to explicitly satisfy the gauge condition.  However, whenever convenient, the gauge constraint for fully paraxial fields may be simplified, within the same $O(\theta_{\text{\tiny{D}}}^2)$ relative accuracy as the paraxial wave equation itself, by dropping the longitudinal derivative:
\be\label{eqn:paraxial_gauge}
\left( i k \,\unitvec{z} + \grad_{\perp} \right) \cdot \bv{\psi}(\bv{r}, \xi; \omega) \approx 0.
\ee
So if the (geometrically) transverse components $\bv{\psi}_{\perp} = \bv{\psi} - \unitvec{z} \unitvec{z}\cdot \bv{\psi}$ are specified, then in this approximation $\psi_z(\bv{r}, \xi; \omega)$ may be taken as
\be\label{eqn:components}
\hat{z}\cdot \bv{\psi}(\bv{r}, \xi; \omega) = \tfrac{i}{k} \grad_{\perp}\cdot \bv{\psi}_{\perp}(\bv{r}, \xi; \omega),
\ee
or conversely, if $\psi_z(\bv{r}, \xi; \omega)$ is given, then a $\bv{\psi}_{\perp}$ can always be chosen to satisfy (\ref{eqn:components}) and ensure paraxial Helmholtz-transversality.  The solenoidal source envelope $\bv{f}$ must in general satisfy the complete envelope constraint (\ref{eqn:trans_b3}) including the longitudinal derivative, because its $\xi$-dependence may be arbitrary.

\subsection{Hilbert Space Formalism}\label{sec:hilbert_space}

Mathematically, the homogeneous (\ie, free-space, or source-free) part of (\ref{eqn:paraxial_wave}) is exactly analogous to the Schrodinger equation, in the position representation, for a quantum mechanical, spin-one particle of mass $M,$ moving non-relativistically in a potential-free, two-dimensional, Euclidean space, where the longitudinal position $\xi$ plays the role of the ``temporal'' evolution variable, $\bv{r}$ serves as the spatial coordinates, and polarization is analogous to spin angular momentum, all in units where $\hbar = 1$ and $M = k = \omega.$
 
In order to leverage the analogies to ordinary quantum mechanics, we introduce a Hilbert space, inner product, state vectors, operators, etc., using a Dirac-like notation.  The full Hilbert space may be considered a tensor product of the Hilbert spaces for the (geometrically) transverse-spatial and the spin (polarization) degrees of freedom.  (Our notation is similar, but not identical, to that in \cite{vanenk_nienhuis:03, nienhuis_allen:03}).  We associate the ket (really a spinor) $\ket{\bv{\psi }}$ with the complex vector field $\bv{\psi}(\bv{r})$ defined on the (geometrically) transverse spatial plane, \ie, as a mapping $\bv{\psi}: \realsymbol ^2 \to \complexsymbol ^3$ (possibly also parameterized by additional continuous variables such as $\xi$ and $\omega$ and by discrete mode indices, all suppressed for the moment).  The corresponding bra $\bra{\bv{\psi}}$ is associated with the dual, or conjugate transpose, vector field, $\bv{\psi}(\bv{r})\hc,$ and a combined $\mathscr{L}_2$/Euclidean inner product is defined by
\be
\innerp{\bv{\psi}_1}{\bv{\psi}_2} = \int d^2 \bv{r} \,\, \bv{\psi}_1(\bv{r})\cc \cdot \bv{\psi}_2(\bv{r}).
\ee

We define a number of Hermitian operators acting on the spatial degrees of freedom by their action on these position-representation wave vector fields $\bv{\psi}.$  Given any fixed unit vectors $\unitvec{e}, \, \unitvec{e}' \in \complexsymbol^2$ in the complex linear span of $\unitvec{x}$ and $\unitvec{y},$ the transverse position operators just act via multiplication by the corresponding transverse spatial coordinates
\be
\unitvec{e} \cdot \bv{Q}\bv{\psi}(\bv{r})  \equiv (\unitvec{e} \cdot \bv{r}) \bv{\psi}(\bv{r}),
\ee
while the conjugate ``momentum'' operators act by differentiation: 
\be
\unitvec{e}' \cdot \bv{P}\bv{\psi}(\bv{r}) \equiv \tfrac{1}{i}\left( \unitvec{e}'\cdot \grad_{\perp}\right)\bv{\psi}(\bv{r}),
\ee
so these operators then satisfy the usual canonical commutation relations
\be
\comm{\unitvec{e}\cdot \bv{Q}\,}{\,\unitvec{e}'\cdot \bv{P} } = i \left(\unitvec{e}\cdot \unitvec{e}' \right) I_{r},
\ee
where $I_{\bv{r}}$ is the identity operator on transverse position-state space.
We stress that these operators act only on the transverse spatial coordinates, so that $\hat{z} \cdot \bv{Q} = \unitvec{z} \cdot \bv{P} = 0.$  The  ``kinetic energy'' or ``free-particle Hamiltonian'' operator is then defined as:
\be
H \equiv \tfrac{1}{2M} \bv{P} \cdot \bv{P} = -\tfrac{1}{2 k} \laplacian_{\perp},
\ee
which is just the infinitesimal generator of free-space, diffractive propagation along $\xi.$  With somewhat more mathematical care about operator domains than is justified for this presentation, these Hermitian operators may be extended to fully self-adjoint operators without difficulty. 

In order to include the polarization, we also define the complex helicity basis in $\complexsymbol^{3}:$
\be
\unitvec{\epsilon}_{s} = -s \tfrac{1}{\sqrt{2}}\left(\unitvec{x} - i s \unitvec{y} \right) + (1 - s^2) \unitvec{z}
 = -\smallhalf s^2(1+ s)\unitvec{e}_{-}  +  \smallhalf s^2(1 - s)\unitvec{e}_{+} + (1-s^2)\unitvec{z},
\ee
for $s = -1, 0, +1,$ satisfying
\be
\unitvec{\epsilon}_{s}\cc \cdot \unitvec{\epsilon}_{s'} = \delta_{s\,s'},
\ee
where $\delta_{s \, s'}$ is the usual Kronecker delta symbol and $\unitvec{e}_{\pm} \equiv \tfrac{1}{\sqrt{2}}(\unitvec{x} \pm  i \unitvec{y}) = \unitvec{e}_{\mp}\cc,$
and then can define various Hermitian ``spin'' operators by their action on the polarization components of the fields $\bv{\psi}.$  In particular, we define a vector of mutually orthogonal polarization projection operators 
\be
\bv{\Pi} = \sum\limits_{s = -1, 0, +1}\!\!\!\! \unitvec{\epsilon}_{s} \Pi_{s},
\ee
where
\be
\Pi_{s} \bv{\Psi} = \unitvec{\epsilon}_{s} \left(\unitvec{\epsilon}_{s}\cc \cdot \bv{\Psi}\right),
\ee
such that 
\be
\Pi_{s}\Pi_{s'} = \Pi_{s'}\Pi_{s}  = \delta_{s\, s'} \Pi_{s}, 
\ee
and 
\be
\sum\limits_{s = -1, 0, +1}\!\!\!\! \Pi_{s} = I_3,
\ee
where 
\be
I_3 = \sum\limits_{s = -1, 0, +1}\!\!\!\! \unitvec{\epsilon}_{s} \unitvec{\epsilon}_{s}\hc
\ee
 is the identity operator on the spin degrees of freedom.
We also define the helicity, or circular polarization operator by
\be
S_z = \Pi_{1} - \Pi_{-1},
\ee
so that
\be
S_z \bv{\psi} =  \unitvec{e}_{-}\unitvec{e}_{-}\cc \cdot \bv{\psi} - \unitvec{e}_{+}\unitvec{e}_{+}\cc \cdot \bv{\psi}.
\ee
and 
\be
S_z^2 = I_3 - \Pi_0.
\ee
For an enlightening discussion of the orbital and spin components of angular momentum in electromagnetic fields, see \cite{allen_et_al:03}.

Clearly, all of these polarization operators commute with the spatial operators, so we
may define generalized (\ie, non-normalizable) simultaneous eigenkets of position and spin $\ket{\bv{r}; s}$ for any $\bv{r} \in \mathbb{R}^2$ and $s \in \left\{ -1, 0, 1 \right\}:$
\be
\bv{Q} \ket{\bv{r} ; s} = \bv{r} \ket{\bv{r} ; s},
\ee
\be
e^{i \bv{r}' \cdot \bv{P}} \ket{\bv{r} ; s} =  \ket{\bv{r} - \bv{r}' ; s},
\ee
and
\be
S_z \ket{\bv{r} ; s} = s \ket{\bv{r} ; s}.
\ee
We similarly define generalized simultaneous eigenkets of  spin and ``momentum'' (transverse wavevector)  by taking the Fourier transforms of these position kets:
\be
\ket{\bv{p} ; s} = \tfrac{1}{2\pi}\int d^2 \bv{r} \,  e^{i \bv{p} \cdot \bv{r}}\ket{\bv{r} ; s},
\ee
which satisfy
\be
\bv{P} \ket{\bv{p} ; s} = \bv{p} \ket{\bv{p} ; s},
\ee
and
\be
S_z \ket{\bv{p} ; s} = s \ket{\bv{p} ; s}.
\ee
These generalized eigenkets satisfy the orthogonality relations
\be
\innerp{ \bv{r}; s}{\bv{r}'; s'} = \delta_{s \, s'}\, \delta^{(2)}(\bv{r} - \bv{r}'),
\ee
and
\be
\innerp{ \bv{p}; s}{\bv{p}'; s'} = \delta_{s \, s'}\, \delta^{(2)}(\bv{p} - \bv{p}'),
\ee
and the completeness relation
\be
\sum_{s} \int d^2 \bv{r} \, \outerp{\bv{r}; s}{\bv{r}; s} = \sum_{s} \int d^2 \bv{p} \, \outerp{\bv{p}; s}{\bv{p}; s} = I,
\ee 
where $\delta^{(2)}(\bv{r})$ is the two-dimensional Dirac delta function, and $I =  I_3 \otimes I_{\bv{r}}$ is the identity operator on the full Hilbert space $\mathcal{H}$ consisting of all $\ket{\bv{\psi}}$ which are normalizable, \ie, for which $\innerp{ \bv{\psi} }{ \bv{\psi} } < \infty,$ implying that the corresponding paraxial radiation fields $\bv{\psi}(\bv{r})$ are of finite power (but not of finite energy).  The allowed radiation envelopes actually reside in the Hilbert sub-space $\mathcal{H}_{\perp} \subset \mathcal{H}$ consisting of vector fields which also satisfy the transverse envelope gauge constraint when evolved in $\xi$ according to free-space paraxial evolution.

Defining the coupled spin-momentum operator
\be
T(\delta) = \left( k - i \delta\frac{\del}{\del \xi}\right)\Pi_0 + \bv{P} \cdot \bv{\Pi} 
\ee
for a real parameter $\delta,$ and the constant reference ket $\ket{ \bv{\psi}_0} \in \mathcal{H}$ such that 
\be
\bv{\psi}_{0}(\bv{r})  = \tfrac{1}{\sqrt{3}}\left( \unitvec{\epsilon}_{-1} +  \unitvec{\epsilon}_{0} + \unitvec{\epsilon}_{+1} \right),
\ee
the full envelope-transversality constraints (\ref{eqn:transversality_3}) on the radiation and source may be written in our quantum notation as:
\begin{subequations}\label{eqn:qm_gauge}
\begin{align}
\qamp{\bv{\psi}_0}{T(\delta = 1)}{\bv{\psi}(\xi; \omega) }   &=  0,\label{eqn:qm_transverse_a}\\
\qamp{\bv{\psi}_0}{T(\delta = 1)}{\bv{f}(\xi; \omega) }   &=  0,\label{eqn:qm_transverse_b}
\end{align}
\end{subequations}
while the approximate paraxial gauge constraint (\ref{eqn:paraxial_gauge}) becomes:
\be\label{eqn:qm_paraxial_gauge}
\qamp{\bv{\psi}_0}{T(\delta = 0)}{\bv{\psi}(\xi; \omega) }   =  0,
\ee
Again, within the paraxial approximation, either of these constraints may be used interchangeably for the radiation, but the source must satisfy the full transversality constraint because its $\xi$ dependence is arbitrary.

It is straightforward to see how the full envelope gauge constraint (\ref{eqn:qm_transverse_a}) (including the longitudinal derivative) will be automatically maintained by the evolution, and how the approximate paraxial constraint (\ref{eqn:qm_paraxial_gauge}) (neglecting the longitudinal derivative) will be preserved by free-space propagation or by propagation in the region of a fully paraxial source.  Assuming that $\ket{\bv{\psi}(\xi; \omega)}$ satisfies the paraxial wave equation, with an initial condition satisfying
\be
\lim_{\xi \to -\infty} \qamp{\bv{\psi}_0}{T(\delta)}{\bv{\psi}(\xi; \omega) }  =  0,
\ee
(our specific choice of $\ket{\bv{\psi}(\xi; \omega) } = 0$ for all $\xi < \xi_0$ clearly works),
and that the source $\ket{\bv{f}(\xi; \omega)}$ satisfies
\be
\qamp{\bv{\psi}_0}{T(\delta)}{\bv{f}(\xi; \omega) }  = 0
\ee
for all $\xi,$ and noting that
\be
\comm{T(\delta)}{H(\omega)} = \comm{T(\delta)}{\tfrac{\del }{\del \xi}} = 0,
\ee
\be
\tfrac{\del }{\del \xi} H(\omega) = 0,
\ee
and
\be
H(\omega) \ket{\bv{\psi}_0} = 0,
\ee
then:
\be
\begin{split}
\tfrac{\del }{\del \xi} \qamp{\bv{\psi}_0}{T(\delta)}{\bv{\psi}(\xi; \omega) }  &=
\qamp{\bv{\psi}_0}{T(\delta)\tfrac{\del }{\del \xi}}{\bv{\psi}(\xi; \omega) } \\
& = -i\qamp{\bv{\psi}_0}{T(\delta)H(\omega)}{\bv{\psi}(\xi; \omega) }  + i\qamp{\bv{\psi}_0}{T(\delta)}{\bv{f}(\xi; \omega) }\\
& = -i\qamp{\bv{\psi}_0}{H(\omega) T(\delta)}{\bv{\psi}(\xi; \omega) } + 0 = 0.
\end{split}
\ee
Using the initial condition, we then have $\qamp{\bv{\psi}_0}{T(\delta)}{\bv{\psi}(\xi; \omega) } = 0 $ for all $\xi.$
This shows that the full ($\delta = 1$) envelope transversality constraint (\ref{eqn:qm_transverse_a} (including the longitudinal derivative) on the fields is preserved everywhere, if it holds initially for the fields and holds everywhere for the sources. The approximate (\ie, $\delta = 0$) paraxial gauge constraint (\ref{eqn:qm_paraxial_gauge}) is preserved by the evolution in regions free of sources and wherever the solenoidal source also satisfies this same, stronger condition.

\subsection{Green Function Solution}\label{sec:green}

In this quantum-like notation, the paraxial wave equation may be written as
\be\label{eqn:helmoltz_inh_qm}
i\tfrac{\del }{\del \xi} \ket{\bv{\psi}(\xi; \omega)} = H\ket{\bv{\psi}(\xi; \omega)}  - \ket{\bv{f}(\xi; \omega)}
\ee
Given some ``initial'' state $\ket{\bv{\psi}(\xi'; \omega)},$ the solution to the homogeneous part of the equation (\ie, for $\ket{\bv{f}(\xi; \omega)} = 0)$ representing free space propagation of the radiation fields, may be written in terms of the unitary evolution operator as
\be
\ket{\bv{\psi}(\xi; \omega)} = U(\xi, \xi'; \omega)\ket{\bv{\psi}(\xi'; \omega)} ,
\ee
where $U(\xi, \xi'; k)$ satisfies the operator-valued Schrodinger equation
\be\label{eqn:op_schrodinger}
i\tfrac{\del }{\del \xi} U(\xi, \xi'; \omega) = H U(\xi, \xi'; \omega)
\ee
with initial condition
\be
U(\xi', \xi'; \omega) = I,
\ee
and possesses the group composition properties
\be
U(\xi, \xi'; \omega)^{-1} =  U(\xi, \xi'; \omega)\hc = U(\xi', \xi; \omega),
\ee
and
\be
U(\xi, \xi''; \omega)U(\xi'', \xi'; \omega) = U(\xi, \xi'; \omega).
\ee
Here $H = H(\omega)$ (with $\omega =  k$) is the Hamiltonian, or diffraction operator, defined above. 
Because $H$ is $\xi$-independent, we can immediately integrate (\ref{eqn:op_schrodinger}) to find
\be
U(\xi, \xi'; \omega) = e^{-i (\xi - \xi') H(\omega)}.
\ee

The solution to the full inhomogeneous wave equation (\ref{eqn:helmoltz_inh_qm}) may be written in terms of the propagator, or causal Green function operator $G(\xi, \xi'; \omega)$ satisfying
\be
\left( i\tfrac{\del }{\del \xi} - H\right)  G(\xi, \xi'; \omega) = \delta( \xi - \xi')
\ee
and
\be
G(\xi, \xi'; \omega) = 0 \,\,\mbox{ for } \,\, \xi < \xi'.
\ee
It can easily be verified that this operator is given by 
\be\label{eqn:paraxial_green_1}
G(\xi, \xi'; \omega) = -i \Theta(\xi - \xi') U(\xi, \xi'; \omega).
\ee
Here $\Theta(\xi)$ is the usual Heaviside step function, and $\delta(\xi) = \tfrac{d}{d\xi}\Theta(\xi)$ is the one-dimensional Dirac delta function.
Assuming no initial (incoming or outgoing) radiation, the formal solution to the inhomogeneous equation (\ref{eqn:helmoltz_inh_qm}) (\ie, with the source term) is then
\be\begin{split}\label{eqn:green_solution}
 \ket{\bv{\psi}(\xi; \omega)} &= -\!\!\int\limits_{-\infty}^{\infty} d \xi' \, G(\xi, \xi'; \omega )\ket{\bv{f}(\xi'; \omega)} \\
& =  i  \!\!\!\!\!\! \int\limits_{\xi_0}^{\min(\xi, \xi_1)} \! \! \! \! d \xi' \, U(\xi, \xi'; \omega)\ket{\bv{f}(\xi'; \omega)},
\end{split}
\ee
where we have used the assumption that the support of $\bv{f}(\xi; \bv{r}; \omega)$ is confined to $\xi_0 < \xi < \xi_1.$  For any $\xi < \xi_0,$ we then have $\ket{\bv{\psi}(\xi; \omega)} = 0$ consistent with our assumed initial condition,  while for all $\xi \ge \xi_1,$ this just reduces to free-space propagation:  $\ket{\bv{\psi}(\xi; \omega)} = U(\xi, \xi_1; \omega)\ket{\bv{\psi}(\xi_1; \omega)}.$  In between we must actually solve the full expression (\ref{eqn:green_solution}).

\subsection{Energetics}\label{sec:energetics}

In our dimensionless variables, recall that the normalized, solenoidal electric field is given by
\be
\bv{\varepsilon}_{\bv{a}\, \perp}(\bv{r}; \xi; \tau) = -\tfrac{\del}{\del \tau} \bv{a}(\bv{r}; \xi; \tau) =  -\tfrac{\del}{\del \tau} \bv{\psi}(\bv{r}; \xi; \tau) e^{i k \xi}
\ee
in the (scaled) time domain, or equivalently
\be
\bv{\varepsilon}_{\bv{a}\, \perp}(\bv{r}; \xi; \omega) = i \omega\, \bv{a}(\bv{r}; \xi; \omega) =  i \omega\, \bv{\psi}(\bv{r}; \xi; \omega)e^{i k \xi}
\ee
in the (scaled) frequency domain.
Within the paraxial approximation, the inner product $\innerp{\bv{\psi}(\xi; \omega)}{\bv{\psi}(\xi; \omega)}$ is then proportional to the (time-averaged) power spectral density for radiation of frequency $\omega$ passing through the transverse plane at the longitudinal position $\xi.$  In fact, in scaled units we may define the (normalized) power spectral density as
\be
\begin{split}
\tfrac{\del}{\del \omega}\mathscr{P}_{\stext{EM}}[\bv{a}](\xi; \omega) & = 
\omega^2 \innerp{\bv{\psi}(\xi; \omega)}{\bv{\psi}(\xi; \omega)} = \omega^2\int d^{2}\bv{r}\,\, \abs{\bv{\psi}(\bv{r}; \xi; \omega)}^2\\
&= \omega^2 \int d^{2}\bv{r}\,\,\abs{\bv{a}(\bv{r}; \xi; \omega)}^2 = \int d^{2}\bv{r}\,\, \abs{\bv{\varepsilon}_{\perp}(\bv{r}; \xi; \omega)}^2.
\end{split}
\ee
Strictly speaking, this yields the time-averaged Poynting flux only to $O(\theta_{\text{\tiny{D}}})$ at any finite distance from the sources, but it is exact as $\xi \to \infty$ and we consider only the truly radiative transport of energy to spatial infinity.  The unitary (inner-product preserving) nature of the paraxial evolution in vacuum then corresponds to energy conservation for paraxial fields, where the (time-averaged) power crossing any transverse plane in any frequency bandwidth, remains constant (with respect to propagation distance $\xi$) during free-space evolution.
Note, however, that while it is particularly convenient to formulate paraxial propagation in terms of the transformation of the fields in successive transverse planes, the actual wavefronts (level surfaces of phase) of a general paraxial beam are not planar except at the focus.

From the actual solution $\ket{\bv{\psi}(\xi_2; \omega)}$ in any conveniently-chosen  transverse plane $\xi_2 \ge  \xi_1$ beyond (\ie, to the right of) the sources, we can construct the extrapolated free-space field
\be\label{eqn:extrapolated_qm}
\ket{\bv{\chi}(\xi; \omega)} = U(\xi, \xi_2; \omega) \ket{ \bv{\psi}(\xi_2; \omega)} = 
i \int\limits_{\xi_0}^{\xi_1}  d \xi' \, U(\xi, \xi'; \omega)\ket{\bv{f}(\xi'; \omega)},
\ee
which represents the post-source paraxial radiation envelope extrapolated throughout all space via free-space propagation, including in the backward or ``upwind'' direction into the region where sources are actually present.  Thus $\ket{\bv{\chi}(\xi; \omega)}$ and $\ket{\bv{\psi}(\xi; \omega)}$ coincide for all $\xi > \xi_1,$ but not in general for $\xi < \xi_1,$ where $\ket{\bv{\chi}(\xi; \omega)}$ satisfies the homogeneous (source-free) Schrodinger equation with no incoming radiation as $\xi \to -\infty,$ whereas $\ket{\bv{\psi}(\xi; \omega)}$ satisfies the inhomogeneous equation everywhere, so that all outgoing radiation at $\xi \to \infty$ is matched by incoming radiation at $\xi \to -\infty.$  For any particular $\xi,$ these extrapolated fields are those fields which would have been present if the fields actually observed beyond the sources at some $\xi_{2} > \xi_1$ were instead produced by some effective source $\bv{g}(\bv{r}; \xi; \bar{\xi}; \omega)$ located in some sufficiently remote region  $\xi_0 - \bar{\xi}<  \xi  < \xi_1 - \bar{\xi},$ instead of by the actual sources within $\xi_0 < \xi < \xi_1:$
\be\label{eqn:eff_source_1}
\ket{\bv{\chi}(\xi; \omega)} = i \int\limits_{\xi_{0} - \bar{\xi}}^{\xi_{1} - \bar{\xi}} d\xi' \,\, U(\xi, \xi'; \omega) \ket{\bv{g}(\xi'; \bar{\xi}; \omega)},
\ee
where the effective source is determined from the actual source simply by longitudinal translation and compensation for diffraction:
\be
\ket{\bv{g}(\xi'; \bar{\xi}; \omega)} = U(\xi', \xi + \bar{\xi})\ket{\bv{f}(\xi' +\bar{\xi}; \omega)}, 
\ee
and
$\bar{\xi} = \bar{\xi}(\xi; \xi_1) \ge 0$ may  any positive constant satisfying 
\be
\bar{\xi} > \xi_1 - \xi.
\ee
Choosing any $\bar{\xi} > \xi_1 - \xi_0,$ the expression (\ref{eqn:eff_source_1}) is then valid for any $\xi \ge \xi_0,$ or we may actually take $\bar{\xi} \to +\infty$ if convenient, moving the effective sources to an abitrarily remote ``upwind'' location.

Using the Green function solution (\ref{eqn:green_solution}), and taking advantage of the assumed finite support of $\bv{f},$ we find that for any $\xi > \xi_1,$
\be\label{eqn:overlap_1}
\begin{split}
\innerp{ \bv{\psi}(\xi; \omega) }{ \bv{\psi}(\xi; \omega) } &=
i \!\!\!\!\!\!\! \int\limits_{\xi_0}^{\min(\xi, \xi_1)} \! \! \! \! d \xi' \,   \qamp{  \bv{\psi}(\xi; \omega)  } {U(\xi, \xi'; \omega)}{\bv{f}(\xi'; \omega)}\\
&=i \!\!\int\limits_{-\infty}^{\infty} d \xi' \, \innerp{ \bv{\chi}(\xi'; \omega) }{ \bv{f}(\xi'; \omega)} \\
&= i  \int d\xi' \!\! \int d^{2} \bv{r} \,\,  \bv{\chi}(\bv{r}; \xi'; \omega)\cc \cdot \bv{f}(\bv{r}; \xi'; \omega),
\end{split}
\ee
which is just a three-dimensional inner product, or overlap integral, between the source envelope $\bv{f}(\bv{r}; \xi'; \omega)$ and the free-space radiation envelope $\bv{\chi}(\bv{r}; \xi'; \omega)$ which has been extrapolated via free-space propagation backwards into the region where sources are present.  

While solenoidal and irrotational vector fields are not in general locally orthogonal in the geometric sense, they are Hilbert-space orthogonal when their Euclidean inner product is integrated over all space.  Because $\ket{\bv{\psi}(\xi; \omega)}$ and hence $\ket{\bv{\chi}(\xi; \omega)}$ satisfy the transverse gauge constraint, this overlap integral may be simplified to
\be\label{eqn:overlap_2}
\begin{split}
\innerp{ \bv{\psi}(\xi; \omega) }{ \bv{\psi}(\xi; \omega) } &=
i \int d\xi' \!\! \int d^{2} \bv{r} \,\,  \bv{\chi}(\bv{r}; \xi'; \omega)\cc \cdot \bv{f}(\bv{r}; \xi'; \omega)\\
&= -\tfrac{1}{2\omega^2}\int d\xi' \!\! \int d^{2} \bv{r} \,\,  \left[i\omega \,\bv{\chi} e^{i k \xi}\right]\cc \cdot \left[2k\, \bv{f}e^{i k\xi}\right]\\
&=  -\tfrac{1}{2\omega^2}\int d\xi' \!\! \int d^{2} \bv{r} \,\,  \bv{\varepsilon}_{\bv{\chi}\, \perp}(\bv{r}; \xi'; \omega)\cc \cdot \bv{j}_{\perp}(\bv{r}; \xi'; \omega)\\
&=  -\tfrac{1}{2\omega^2}\int d\xi' \!\! \int d^{2} \bv{r} \,\,  \bv{\varepsilon}_{\bv{\chi}\,\perp}(\bv{r}; \xi'; \omega)\cc \cdot \bv{j}(\bv{r}; \xi'; \omega)\\
&= -\tfrac{1}{2\omega^2}\int d\xi' \, \innerp{\bv{\varepsilon}_{\bv{\chi}\, \perp}(\xi'; \omega)}{ \bv{j}(\xi'; \omega) },
\end{split}
\ee
where $\bv{\varepsilon}_{\bv{\chi}\, \perp}(\bv{r}; \xi'; \omega)$ is the scaled, solenoidal electric field associated with the extrapolated radiation envelope $\bv{\chi}.$  In order to evaluate the power, we therefore need never explicitly decompose the current density into its solenoidal and irrotational components, nor worry about the precise cutoff beyond which the solenoidal component may be neglected.

As one would expect from energy conservation considerations, this  overlap integral may be interpreted in terms of  the rate of energy transfer between the charges constituting the sources and the fields.  We may define the scaled power spectral density associated with the rate of (time-averaged) mechanical work done by scaled electric fields $\bv{\varepsilon}(\bv{r}; \xi; \omega)$ on the charges associated with the scaled current density $\bv{j}(\bv{r}; \xi'; \omega)$ as
\be
\begin{split}
\tfrac{\del}{\del \omega}\mathscr{P}_{\stext{mech}}[ \bv{\varepsilon}; \bv{j}](\omega) 
&= \realpart \int d\xi' \, \innerp{\bv{\varepsilon}(\xi'; \omega)}{ \bv{j}(\xi'; \omega) }\\
&= \realpart \int d\xi' \!\! \int d^{2} \bv{r} \,\,  \bv{\varepsilon}(\bv{r}; \xi'; \omega)\cc \cdot \bv{j}(\bv{r}; \xi'; \omega).
\end{split}
\ee
From (\ref{eqn:overlap_2}), we see that $\int d\xi' \, \innerp{\bv{\varepsilon}_{\bv{\chi}\, \perp}(\xi'; \omega)}{ \bv{j}(\xi'; \omega) }$ must be purely real and non-positive, because $\innerp{ \bv{\psi} }{\bv{\psi} } = \vnorm{\bv{\psi}}^2$ is necessarily real and non-negative.  Physically, this reflects the fact that the back-extrapolated fields, if actually present, would lead to no reactive energy transfer to local fields associated with re-arrangement of the source charges, and mechanical energy would be transferred monodirectionally from the charges to the radiation fields.  Therefore (\ref{eqn:overlap_2}) is equivalent to:
\be\label{eqn:energy_2}
\tfrac{\del}{\del \omega}\mathscr{P}_{\stext{EM}}[\bv{a}](\xi; \omega) = -\smallhalf \tfrac{\del}{\del \omega}\mathscr{P}_{\stext{mech}}[ \bv{\varepsilon}_{\bv{\chi}\,\perp}; \bv{j}](\omega) 
\ee
That is, the average power seen in the far field is exactly one-half of that which would be delivered by the sources to the extrapolated homogeneous fields if these fields were those actually present in the region of the sources.

The presence of the extra factor of $\smallhalf$ is intuitively plausible.  The actual paraxial fields from each transverse current slice consist only of the forward half of the corresponding extrapolated fields from this slice, and hence the former fields could interact only with ``downstream'' current slices while propagating in the forward ($+\unitvec{z}$) direction, while the extrapolated fields from any current slice impinge on all other current slices, both upstream and downstream.   If we adopt the symmetric convention  $\Theta(0) = \smallhalf$ for the Heaviside step function (which is anyway the natural choice when Fourier transforms are involved), each current slice also appears to interact only with one-half of its own field.  Assuming a reciprocity arising from Newton's Third Law (magnetic forces can violate this law, but do no mechanical work, and radiation reaction forces can violate the third law as well, but are ignored here), we then anticipate an over-counting of the work by exactly a factor of $2$ when we use the extrapolated fields and integrate over all current slices.  This is a dynamical, electromagnetic analog of the well known electrostatic result that the potential energy associated with a given charge distribution $\rho$ in a given external potential $\Phi_{\text{\tiny{ext}}}$ is given by $U = \int d^3 \bv{x}\, \rho \Phi_{\text{\tiny{ext}}},$ while the self-energy (potential energy of formation) of a charge distribution resulting in a potential $\Phi$ is  $U = \smallhalf \int d^3 \bv{x}\, \rho \Phi.$ 

Below, we will provide an elementary proof of this result in the general case, but it is informative to verify it with an explicit calculation for paraxial fields.  By using the Green function solution and inserting a resolution of the identity, we have, for any $\xi > \xi_1,$
\be\label{eqn:overlap_3}
\begin{split}
\tfrac{\del}{\del \omega}\mathscr{P}_{\stext{EM}}[\bv{a}](\xi; \omega) &=  \omega^2\innerp{ \bv{\psi}(\xi; \omega) }{ \bv{\psi}(\xi; \omega) } = i \omega^2\int\limits_{\xi_0}^{\xi_1} d\xi' \,\,\innerp{ \bv{\chi}(\xi'; \omega)}{\bv{f}( \xi'; \omega)}\\
&= \omega^2 \int\limits_{\xi_0}^{\xi_1} d\xi' \!\! \int\limits_{\xi_0}^{\xi_1} d\xi'' \,\, \qamp{ \bv{f}(\xi''; \omega)}{U(\xi'', \xi'; \omega)}{\bv{f}( \xi'; \omega)}\\
&= \omega^2\sum_{s}\int d^{2}\bv{r}  \Biggl\lvert \, \int\limits_{\xi_0}^{\xi_1} d\xi' \,\,\qamp{\bv{r}; s}{U(\xi_{0}, \xi'; \omega)}{\bv{f}( \xi'; \omega)}\, \Biggr\rvert^2\\
&= \omega^2\sum_{s}\int d^{2}\bv{r}  \Biggl\lvert \, \int\limits_{\xi_0}^{\xi_1} d\xi' \, \bv{g}(\bv{r}; \xi_0; \xi' - \xi_0;\omega)    \Biggr\rvert^2\\
\end{split}
\ee
which explicitly demonstrates its reality and non-negativity.  Using the same sort of manipulations, we then have
\be\label{eqn:overlap_actual}
\begin{split}
&\tfrac{\del}{\del \omega}\mathscr{P}_{\stext{mech}}[ \bv{\varepsilon}_{\bv{\psi}\,\perp}; \bv{j}](\omega) = \realpart\Biggl[  \int\limits_{\xi_0}^{\xi_1} d\xi' \,\,\innerp{ \bv{\varepsilon}_{\psi\, \perp}(\xi'; \omega)}{\bv{j}( \xi'; \omega)}\Biggr]\\
 &\quad = \realpart\Biggl[  -2i \omega k \int\limits_{\xi_0}^{\xi_1} d\xi' \,\,\innerp{ \bv{\psi}(\xi'; \omega)}{\bv{f}( \xi'; \omega)}\Biggr]\\
 &\quad = -2\omega^2 \realpart\Biggl[\int\limits_{\xi_0}^{\xi_1} d\xi' \!\! \int\limits_{\xi_0}^{\xi'} d\xi'' \, \qamp{ \bv{f}(\xi''; \omega)}{U(\xi'', \xi'; \omega)}{\bv{f}( \xi'; \omega)}\Biggr]\\
 &\quad = -2\omega^2 \sum_{s}\!\!\int \! d^{2}\bv{r} \, \realpart\Biggl[ \int\limits_{\xi_0}^{\xi_1} d\xi' \,  \bv{g}(\bv{r}; \xi_0; \xi' - \xi_0;\omega) \int\limits_{\xi_0}^{\xi'} d\xi'' \,  \bv{g}(\bv{r}; \xi_0; \xi'' - \xi_0;\omega)\cc \Biggr]
\end{split}
\ee
But using integration by parts, this simplifies to 
\be\label{eqn:overlap_actual_2}
\begin{split}
&\tfrac{\del}{\del \omega}\mathscr{P}_{\stext{mech}}[ \bv{\varepsilon}_{\bv{\psi}\,\perp}; \bv{j}](\omega)\\ 
&\quad = (-2\omega^2 )\Biggl[\smallhalf  \sum_{s}\!\!\int \! d^{2}\bv{r} \,  \Biggl\lvert \int\limits_{\xi_0}^{\xi_1} d\xi' \,  \bv{g}(\bv{r}; \xi_0; \xi' - \xi_0;\omega) \Biggr\rvert^2\Biggr]\\
 &\quad  = -\omega^2 \innerp{ \bv{\psi}(\xi; \omega) }{ \bv{\psi}(\xi; \omega) } = -\tfrac{\del}{\del \omega}\mathscr{P}_{\bv{a}}(\xi; \omega)
 \end{split}.
\ee
So, indeed it is the case that:
\be\label{eqn:paraxial_energy_balance_2}
\tfrac{\del}{\del \omega}\mathscr{P}_{\stext{EM}}[\bv{a}](\xi; \omega) = -\smallhalf \tfrac{\del}{\del \omega}\mathscr{P}_{\stext{mech}}[ \bv{\varepsilon}_{\bv{\chi}\,\perp}; \bv{j}](\omega) =
-\tfrac{\del}{\del \omega}\mathscr{P}_{\stext{mech}}[ \bv{\varepsilon}_{\bv{a}\,\perp}; \bv{j}](\omega),
\ee
consistent with expectations of energy conservation.  Note that while the  extrapolated source-free solution $\bv{\chi}$ coincides ,by construction, with $\bv{a}$ in the downstream region $\xi > \xi_1,$ and so exhibits identical out-flowing power, in order to actually satisfy the source-free equation, it must have an equal magnitude of power flowing into the interaction region through any upstream plane at $\xi < \xi_0,$ which accounts for the extra factor of two.

\subsection{Basis-Set Approximation}\label{sec:basis_approx}

As we have seen, equation (\ref{eqn:green_solution}) in principle gives the exact solution to the paraxial wave equation with prescribed sources, without any further approximations, and is compatible with energy conservation for paraxial electromagnetic fields.  For given sources, the field at an arbitrary longitudinal plane $\xi$ can be determined via a convolution integral over the longitudinal source position $\xi'$ and a two-dimensional integral over either the transverse position-dependence (real space) or transverse wavevector-dependence (reciprocal space) of the solenoidal component of the current.  In position space, we have
\be
\ket{\bv{\psi}(\xi; \omega)} =
\sum\limits_{s} \int d^{2} \bv{r} \int d^{2} \bv{r}' \!\!\!\!\int\limits_{\xi_0}^{\min(\xi, \xi_1)} \!\! d \xi' \, 
\frac{k \, e^{i\frac{k\abs{\bv{r} - \bv{r}'}^2}{2(\xi - \xi')}}}{2\pi i (\xi - \xi')} \innerp{\bv{r}'; s}{\bv{f}(\xi';\omega)}\,\ket{\bv{r};s},
\ee
where we have used the well-known result for the two-dimensional, position-space, free-particle propagator in quantum mechanics.  In momentum space representation, the propagator is diagonal, and we have
\be
\ket{\bv{\psi}(\xi; \omega)} = i \sum\limits_{s} \int d^{2} \bv{p} \!\!\!\! \int\limits_{\xi_0}^{\min(\xi, \xi_1)} \!\!\!\! d \xi' \, 
e^{-i(\xi - \xi')\frac{\abs{\bv{p}}^2}{2k}} \innerp{\bv{p}; s}{\bv{f}(\xi'; \omega)}\,\ket{\bv{p}; s},
\ee
where $\innerp{\bv{p}; s}{\bv{f}(\xi'; \omega)}$ is just  the Fourier transform (with respect to the scaled transverse wavevector) of the solenoidal current density slice at longitudinal position $\xi'.$  Now, using either either of these approaches, we require something like $10$ integrations to determine a given frequency component of the  field at a given spatial location, starting from the full current density as a function of space and time coordinates.  In practice, we often only know the source probabilistically, so we will typically  have to perform additional averages over the particle distribution function, involving up to $6$ more integrations over the full particle phase space.

A complementary approach is to decompose the paraxial radiation fields into a complete, orthogonal set of modes.   Because we are actually only interested in the radiation observed beyond the region of interaction with the localized sources, these modes are naturally chosen to satisfy the homogeneous (source-free) paraxial wave equation for all longitudinal  positions $\xi > \xi_1,$ but we may then also choose these modes to be extrapolated free-space solutions everywhere, including within the support of the sources.  That is, we consider a countable set of explicitly $\xi$-dependent, solenoidal, envelope modes $\ket{\bv{u}_{n}(\xi; \omega)} \in \mathcal{H}_{\perp},$ where $n$ denotes some set of integral transverse-spatial and polarization modal indices, such that each mode envelope satisfies
\be
i \tfrac{\del }{\del \xi} \ket{\bv{u}_{n}(\xi; \omega)} = H(\omega)  \ket{\bv{u}_{n}(\xi; \omega)},
\ee
or equivalently
\be
\ket{\bv{u}_{n}(\xi; \omega)} = U(\xi, \xi'; \omega) \ket{\bv{u}_{n}(\xi'; \omega)}
\ee
together with the gauge constraint (\ref{eqn:qm_transverse_a}), for all longitudinal positions $-\infty < \xi < \infty.$  Because free-space propagation is unitary, these modes may be chosen to be orthonormal in each transverse plane, \ie,
\be
\innerp{\bv{u}_{n}(\xi; \omega) }{\bv{u}_{n'}(\xi; \omega) } = \delta_{n \, n'}
\ee
and complete in each transverse plane, in the sense that
\be
\linspan\big[ \ket{\bv{u}_{n}(\xi; \omega)}  \big] \cong \mathcal{H}_{\perp}.
\ee
or equivalently
\be
 \sum_{n} \outerp{\bv{u}_{n}(\xi; \omega)}{\bv{u}_{n}(\xi; \omega)} = I_{\perp},
\ee
where $I_{\perp}$ is the identity operator restricted to the solenoidal sub-space $\mathcal{H}_{\perp},$ or equivalently the Hermitian projection from $\mathcal{H}$ into $\mathcal{H}_{\perp}.$  Familiar and convenient choices are the usual free-space Gauss-Hermite modes or Gauss-Laguerre modes in paraxial optics.  The spatial profiles of these modes may be characterized by specifying the longitudinal location of the focal plane and the eigenvalues of certain Hermitian operators of which they are eigenfunctions in that plane.  For example, in the focal plane, the components of the Gauss-Hermite modes are simultaneous  number states of two harmonic oscillator Hamiltonians (one for each transverse Cartesian coordinate), while the Gauss-Laguerre modes are simultaneous eigenstates of a radial harmonic oscillator Hamiltonian and the longitudinal component of orbital angular momentum \cite{nienhuis_allen:03, vanenk_nienhuis:03}.  Modes with slightly off-axis propagation directions can be defined in terms of the coherent states associated with these quadratic Hamiltonians, and many other generalizations are possible to suit particular problems.

Any linear combinations of the $\bv{u}_n$ modes satisfy the homogeneous paraxial wave equation, so cannot possibly represent the fields in the region actually containing the sources, but they can exactly reproduce any fields in the ``downstream'' vacuum region beyond the sources, and then be extrapolated upstream.  That is, for all $\xi > \xi_1,$ the actual radiation envelope $\ket{\bv{\psi}(\xi; \omega)}$ satisfies the homogeneous wave equation,  lies entirely in the solenoidal sub-space  $\mathcal{H}_{\perp},$ and may be decomposed as
\be
\begin{split}
\ket{\bv{\psi}(\xi; \omega)} & = \biggl[ \sum_{n} \outerp{\bv{u}_{n}(\xi; \omega)}{\bv{u}_{n}(\xi; \omega)} \biggr]\ket{\bv{\psi}(\xi; \omega)} \\
&= \phantom{[} \sum_{n} \innerp{\bv{u}_{n}(\xi; \omega)}{\bv{\psi}(\xi; \omega)} \, \ket{\bv{u}_{n}(\xi; \omega)},
\end{split}
\ee
and both the modulus and argument of the complex expansion coefficients appearing in the sum have a simple interpretation.  Using the orthonormality of the modes, the normalized power spectral density at a given transverse plane $\xi > \xi_1$ is given by
\be
\tfrac{\del}{\del \omega}\mathscr{P}_{\stext{EM}}[\bv{a}](\xi; \omega) = \omega^2 \innerp{{\bv{\psi}(\xi; \omega)}}{{\bv{\psi}(\xi; \omega)} } = \omega^2 \sum_{n} \abs{ \innerp{\bv{u}_{n}(\xi; \omega)}{\bv{\psi}(\xi; \omega)}}^2;
 \ee
and, individually,  each 
\be
\tfrac{\del}{\del \omega}\tilde{\mathscr{P}}_{\stext{EM}}[\bv{u}_{n}](\xi; \omega) \equiv \omega^2 \abs{ \innerp{\bv{u}_{n}(\xi; \omega)}{\bv{\psi}(\xi; \omega)}}^2 \ge 0
\ee
provides the normalized power-spectral-density contribution from the $n$th mode, while 
\be
\theta_n(\xi; \omega) = \arg\bigl[  \innerp{\bv{u}_{n}(\xi; \omega)}{\bv{\psi}(\xi; \omega)}  \bigr]
\ee
determines the slowly-varying envelope phase (\ie, phase apart from  the fast $k\xi - \omega\tau$ dependence) of the the $n$th mode.  From unitarity, it follows that $\tfrac{\del}{\del \omega}\mathscr{P}_{\stext{EM}}[\bv{u}_{n}](\xi; \omega),$ $\theta_n(\xi; \omega)$ and $\tfrac{\del}{\del \omega}\mathscr{P}_{\stext{EM}}[\bv{a}](\xi; \omega)= \sum\limits_{n}\tfrac{\del}{\del \omega}\tilde{\mathscr{P}}_{\stext{EM}}[\bv{u}_{n}](\xi; \omega)$ are all independent of $\xi$ for any $\xi > \xi_1.$  With perhaps a slight abuse of notation, we have used $\tfrac{\del}{\del \omega}\tilde{\mathscr{P}}_{\stext{EM}}[\bv{u}_{n}]$ with a tilde to denote the scaled power (spectral density) contribution from the $n$th normalized mode \ie, the power spectral density associated through Poynting's theorem with the ket $\innerp{\bv{u}_{n}(\xi; \omega)}{\bv{\psi}(\xi; \omega)} \, \ket{\bv{u}_{n}(\xi; \omega)},$ and not the power spectral density $\tfrac{\del}{\del \omega}\mathscr{P}_{\stext{EM}}[\bv{u}_{n}]$ literally associated with the ket $\ket{\bv{u}_{n}(\xi; \omega)},$ which is fixed at the value $\omega^2$ by our normalization conventions. 

In general, a countably infinite number of orthogonal modes are required to exactly describe the radiation fields, but if the modes are chosen appropriately, a finite basis set $\mathcal{B}_{N}$ of size $N, \,\, 1 \le N < \infty,$ such that attention is confined to $n \in \mathcal{B}_{N},$ may provide a sufficiently accurate representation.  For example, if the mode shape is approximately Gaussian in cross section, then it should be accurately represented by the fundamental and perhaps a few higher-order Gauss-Hermite modes, with proper choice of the waist size and location.  Formally, the expansion coefficients may be determined by again using the Green function solution (\ref{eqn:green_solution}), taking advantage of the finite support of $\ket{\bv{f}(\omega)}.$  In a derivation exactly analogous to that in the previous section, we find that for any $\xi > \xi_1,$
\be\label{eqn:ip_solution}
\begin{split}
\innerp{\bv{u}_n(\xi; \omega) }{ \bv{\psi}(\xi; \omega) } &=
i \!\!\!\!\!\! \int\limits_{\xi_0}^{\min(\xi, \xi_1)} \! \! \! \! d \xi' \,   \qamp{  \bv{u}_{n}(\xi; \omega)  } {U(\xi, \xi'; \omega)}{\bv{f}(\xi'; \omega)}\\
&= i \int\limits_{-\infty}^{\infty} d \xi' \,  \innerp{  \bv{u}_{n}(\xi'; \omega)}{\bv{f}(\xi'; \omega)} \\
&= i  \int d\xi' \!\! \int d^{2} \bv{r} \,\,  \bv{u}_{n}(\bv{r}; \xi'; \omega)\cc \cdot \bv{f}(\bv{r}; \xi'; \omega),
\end{split}
\ee
which apart from an overall constant is just a three-dimensional inner product, or overlap integral, between the solenoidal source envelope $\bv{f}(\bv{r}; \xi'; \omega)$ and the free-space mode envelope $\bv{u}_{n}(\bv{r}; \xi'; \omega).$  Equivalently, we can write this as the overlap integral between the full current density $\bv{j}(\bv{r}; \xi'; \omega),$ and the normalized electric field $\bv{\varepsilon}_{n\, \perp}(\xi'; \omega) = i\omega\, \bv{u}_{n}(\bv{r}; \xi'; \omega)$ associated with the $n$th mode:
\be\label{eqn:ip_solution2}
\begin{split}
\innerp{\bv{u}_{n}(\xi; \omega) }{ \bv{\psi}(\xi; \omega) } &=  -\tfrac{1}{2\omega^2}\int d\xi' \!\! \int d^{2} \bv{r} \,\,  \bv{\varepsilon}_{n\, \perp}(\bv{r}; \xi; \omega)\cc \cdot \bv{j}(\bv{r}; \xi; \omega)\\
&= -\tfrac{1}{2\omega^2}\int d\xi' \, \innerp{ \bv{\varepsilon}_{n\, \perp}(\xi'; \omega)}{ \bv{j}(\xi'; \omega) }.
\end{split}
\ee
So, again, we do not need to explicitly decompose the current density into its solenoidal and irrotational components in order to calculate these expansion coefficents.  The factor of $\smallhalf$ is analogous to that appearing in (\ref{eqn:energy_2}), and in fact (\ref{eqn:ip_solution2}) reduces to the full energy conservation result if we are lucky enough to choose a basis which contains the extrapolated solution $\bv{\chi}(\bv{r}; \xi; \omega)$ in its span.
More generally, it is clear that the finite-basis set approximation is just the orthogonal projection of the actual extrapolated free-space solution into the subspace $\linspan\limits_{n \in \mathcal{B}_{N}}\big[ \ket{\bv{u}_n(\xi; \omega)}\big]$ spanned by the modes in $\mathcal{B}_{N}:$
 \be
\ket{ \bv{\nu}(\xi; \omega; \mathcal{B}_{N}) } = \sum_{n \in \mathcal{B}_{N}} \outerp{\bv{u}_{n}(\xi; \omega) }{\bv{u}_{n}(\xi; \omega) } \ket{\bv{\chi}(\xi; \omega)}.
\ee

From linearity, it immediately follows that if $\ket{ \bv{\nu}_{m}(\xi; \omega) }$ is the basis-set approximation to  $\ket{ \bv{\chi}_{m}(\xi; \omega) }$ resulting form a source $\ket{ \bv{f}_{m}(\xi; \omega)}$ then
$\sum\limits_{m} c_m(\omega) \ket{ \bv{\nu}_{m}(\xi; \omega) }$ is the approximation 
to $\sum\limits_{m} c_m(\omega) \ket{ \bv{\chi}_{m}(\xi; \omega) }$ resulting from the source envelope 
$\sum\limits_{m} c_m(\omega) \ket{ \bv{f}_{m}(\xi; \omega) },$ where the $c_m(\omega) \in \complexsymbol$ are arbitrary coefficients.  So we may also determine the basis set approximation for any source by first finding the basis set approximation for the impulse source $\ket{\bv{r}; s}$ and then performing a convolution over the actual source.  That is, we can use as an approximate propagator, the restriction of the full Green operator to the subspace $\linspan\limits_{n \in \mathcal{B}_{N}}\big[ \ket{\bv{u}_n(\xi; \omega)}\big]:$
\be
\ket{ \bv{\nu}(\xi; \omega; \mathcal{B}_{N}) } = i \int\limits_{\xi_0}^{\xi_1}d \xi' \,   
\sum_{n \in \mathcal{B}_{N}} \outerp{\bv{u}_{n}(\xi; \omega) }{\bv{u}_{n}(\xi'; \omega) } \ket{\bv{f}(\xi'; \omega)},
\ee
which clearly agrees with previous results after exchanging the order of the integral and sum.

It follows from the properties of orthogonal projections in Hilbert space that the projected basis-set approximation is also uniquely determined by the following constrained minimum-distance criteria:
\begin{subequations}\label{eqn:min_1a}
\begin{align}
\ket{ \bv{\nu}(\xi; \omega; \mathcal{B}_{N}) } &= \argmin\limits_{\bv{\nu}} \Bigl[ \vnorm{\, \ket{\bv{\nu}(\xi; \omega)} -  \ket{\bv{\chi}(\xi; \omega)}  \,}   \Bigr] \\
 \mbox{s.t. } \ket{\bv{\nu}(\xi; \omega)} &\in  \linspan\limits_{n \in \mathcal{B}_{N}}\Bigl[ \ket{\bv{u}_n(\xi; \omega)}\Bigr].
\end{align}
\end{subequations}
The estimated radiated power (spectral density) is then always a lower bound on the actual power (spectral density) in any plane $\xi > \xi_1:$
\be\label{eqn:paraxial_basis_2}
\tfrac{\del}{\del \omega}\mathscr{P}_{\stext{EM}}[\bv{\nu}](\xi; \omega; \mathcal{B}_{N} ) = \sum_{n \in \mathcal{B}_{N} }\tfrac{\del}{\del \omega}\tilde{\mathscr{P}}_{\stext{EM}}[\bv{u}_{n}](\xi; \omega) \le \tfrac{\del}{\del \omega}\mathscr{P}_{\stext{EM}}[\bv{\chi}](\xi; \omega) = \tfrac{\del}{\del \omega}\mathscr{P}_{\stext{EM}}[\bv{a}](\xi; \omega),
\ee
with equality if and only if $\bv{\nu}(\bv{r}; \xi; \omega; \mathcal{B}_{N}) =  \bv{\chi}(\bv{r}; \xi; \omega).$  In fact, it is straightforward to establish that, at the constrained optimum (\ref{eqn:min_1a}), the squared-distance between the basis-set approximation and the actual extrapolated fields is just proportional to their difference in power spectral density:
\be
\omega^2\vnorm{\, \ket{\bv{\nu}(\xi; \omega; \mathcal{B}_{N} )} -  \ket{\bv{\chi}(\xi; \omega)}}^2 = \tfrac{\del}{\del \omega}\mathscr{P}_{\stext{EM}}[\bv{\chi}](\xi; \omega) - \tfrac{\del}{\del \omega}\mathscr{P}_{\stext{EM}}[\bv{\nu}](\xi; \omega; \mathcal{B}_{N} ) \ge 0.
\ee

\subsection{Variational Approximation}\label{sec:variational_approx}

Such connections revealed in the basis-set approach between the orthogonal projection, minimum distance, maximum radiated power, minimal negative mechanical work, and maximal source/field overlap suggest that an approximate field profile for the radiation and a lower bound on the radiated power may be obtained directly through a general variational principle.  Consider a parameterized family of trial envelope modes $\bv{v}(\bv{r}; \xi; \omega; \bv{\alpha})$ depending on some set of adjustable parameters denoted by the vector $\bv{\alpha},$ where, for any fixed choice of allowed parameter values, we assume the the trial mode satisfies the free-space wave equation, is solenoidal, and is normalized in any transverse plane.  In principle, the trail mode can be written formally as a linear combination of the above basis modes:
\be
\ket{ \bv{v}(\xi; \omega; \bv{\alpha})} = \sum_{n} c_{n}(\omega; \bv{\alpha})\ket{ \bv{u}_{n}(\xi; \omega) }
\ee
for some complex expansion coefficients 
$c_n = c_{n}(\omega; \bv{\alpha}) \in \mathbb{C}$ which satisfy the constraint
\be
\sum_{n} \abs{c_{n}(\omega; \bv{\alpha})}^2 = 1,
\ee
so that 
\be\label{eqn:paraxial_var_norm_1}
\innerp{\bv{v}(\xi; \omega; \bv{\alpha})}{\bv{v}(\xi; \omega; \bv{\alpha})}  = 1,
\ee
but are otherwise arbitrary.  Note that the expansion coefficients are here assumed independent of $\xi,$ but may in general depend on the adjustable parameters in some complicated,  and perhaps nonlinear fashion.  (More generally, we could allow dependence on $\xi$ in the $c_n,$ leading to a variational problem involving the solution of a set of coupled ODEs rather than miniization of a function.  This extra generalization is not really needed for the case of spontaneous radiation.)  Using the Cauchy-Schwarz inequality, we have, for any $\xi,$
\be
\abs{  \innerp{ \bv{v}(\xi; \omega; \bv{\alpha}) }{ \bv{\psi}(\xi; \omega) }  }^2  \le 
 \innerp{ \bv{v}(\xi; \omega; \bv{\alpha}) }{ \bv{v}(\xi; \omega; \bv{\alpha}) } 
\innerp{ \bv{\psi}(\xi; \omega)  }{ \bv{\psi}(\xi; \omega) },
\ee
which using (\ref{eqn:paraxial_var_norm_1}) simplifies to 
\be
\omega^2 \abs{  \innerp{ \bv{v}(\xi; \omega; \bv{\alpha}) }{ \bv{\psi}(\xi; \omega) }  }^2  \le  \omega^2 \big(\innerp{ \bv{\psi}(\xi; \omega)  }{ \bv{\psi}(\xi; \omega) }\big),
\ee
or equivalently
\be\label{eqn:inequality_2}
\tfrac{\del}{\del \omega}\tilde{\mathscr{P}}_{\stext{EM}}[\bv{v}](\xi; \omega; \bv{\alpha}) \le  \tfrac{\del}{\del \omega}\mathscr{P}_{\stext{EM}}[\bv{a}](\xi; \omega),
\ee
with strict equality if and only if  $\bv{v}(\bv{r}; \xi; \omega; \bv{\alpha})$ exactly mirrors the shape and polarization of the actual  paraxial solution $\bv{\psi}(\bv{r}; \xi; \omega)$ in the plane $\xi,$ up to some overall phase; \ie,
\be
\ket{ \bv{v}(\xi; \omega; \bv{\alpha})} = e^{i\theta_{\bv{v}}( \omega; \bv{\alpha})} \bigl( \innerp{ \bv{v}(\xi; \omega)  }{ \bv{\psi}(\xi; \omega) } \bigr)^{-1/2} \ket{ \bv{\psi}(\xi; \omega)}
\ee
for some real angle $\theta_{\bv{v}}( \omega; \bv{\alpha}).$

We have indeed arrived at a variational principle for the paraxial radiation, where we may approximate both the relative spatial/polarization profile and overall amplitude of the radiation fields by
maximizing the power-spectral density in a normalized trial mode $\bv{v}(\xi; \omega; \bv{\alpha})$ measured  in some post-source plane $\xi > \xi_1,$ as a function of the variational parameters $\bv{\alpha}$ determining the trial mode's shape and polarization.  That is, we take as the approximate radiation envelope
\be
\ket{ \bv{\nu}(\xi; \omega; \tilde{\bv{\alpha}}) } = \bigl(  \innerp{ \bv{v}(\xi; \omega; \tilde{\bv{\alpha}})  }{  \bv{\psi}(\xi; \omega) } \bigr) \ket{ \bv{v}(\xi; \omega; \tilde{\bv{\alpha}})},
\ee
where the optimal parameter vector $\tilde{\bv{\alpha}} = \tilde{\bv{\alpha}}[\bv{j}](\omega)$ is chosen so as to maximize 
\be
\tfrac{\del}{\del \omega}\mathscr{P}_{\stext{EM}}[\bv{\nu}](\xi; \omega; \bv{\alpha})  =
\tfrac{\del}{\del \omega}\tilde{\mathscr{P}}_{\stext{EM}}[\bv{v}](\xi; \omega; \bv{\alpha}) 
= \omega^2\abs{  \innerp{ \bv{v}(\xi; \omega; \bv{\alpha}) }{ \bv{\psi}(\xi; \omega) }  }^2.
\ee
The optimized mode shape is then the best guess, within the manifold of possibilities allowed by the shapes parameterized by $\bv{\alpha},$ of the actual paraxial field profile beyond the sources, and the squared-norm yields a lower bound on the actual paraxial power spectral density of the radiation at the frequency under consideration.

However, as written, this variational principle appears vacuous at best, since if we knew the actual $\bv{\psi}(\bv{r}; \xi; \omega)$ so as to be able to calculate its overlap with the trial field $\bv{v}(\bv{r}; \xi; \omega; \bv{\alpha}),$ there would of course be no need for a variational approximation in the first place.  But from (\ref{eqn:ip_solution2}), we may also write
\be\label{eqn:variational_2}
\begin{split}
\innerp{ \bv{v}(\xi; \omega; \bv{\alpha}) }{ \bv{\psi}(\xi; \omega) } & =
-\tfrac{1}{2\omega^2}\int d\xi' \!\! \int d^{2} \bv{r} \,\,  \bv{\varepsilon}_{\bv{v}\,\perp}(\bv{r}; \xi; \omega; \bv{\alpha})\cc \cdot \bv{j}(\bv{r}; \xi; \omega)\\
&= -\tfrac{1}{2\omega^2}\int\limits_{\xi_0}^{\xi_1} d\xi '\, \innerp{ \bv{\varepsilon}_{\bv{v}\,\perp}(\xi'; \omega; \bv{\alpha}) }{ \bv{j}(\xi'; \omega) }
\end{split}
\ee
where
$\bv{\varepsilon}_{\bv{v}\,\perp}(\bv{r}; \xi; \omega; \bv{\alpha}) = i\omega\,\bv{v}(\bv{r}; \xi; \omega; \bv{\alpha})e^{i k \xi}$ is the normalized solenoidal frequency-domain electric field associated with the unit-norm trial vector potential envelope $\bv{v}(\bv{r}; \xi; \omega; \bv{\alpha}).$  In terms of the trial  envelope $\bv{\nu}(\bv{r}; \xi; \omega; \tilde{\bv{\alpha}}),$ we see that
\be\label{eqn:variational_2b}
\begin{split}
\tfrac{\del}{\del \omega}\mathscr{P}_{\bv{\nu}}(\xi; \omega; \bv{\alpha}) &= \omega^2 \abs{ \innerp{ \bv{v}(\xi; \omega; \bv{\alpha})  }{  \bv{\psi}(\xi; \omega; \bv{\alpha}) } }^2
= \omega^2 \innerp{ \bv{\nu}(\xi; \omega; \bv{\alpha})  }{  \bv{\nu}(\xi; \omega;  \bv{\alpha}) }\\
&=  -\tfrac{1}{2}\int d\xi '\, \innerp{ \bv{\varepsilon}_{\bv{\nu}\,\perp}(\xi'; \omega; \bv{\alpha}) }{ \bv{j}(\xi'; \omega) },
\end{split}
\ee
where $\bv{\varepsilon}_{\bv{\nu}\,\perp}(\bv{r}; \omega; \bv{\alpha}) = i\omega\,\bv{\nu}(\xi; \omega; \bv{\alpha})$ is the solenoidal electric field associated with the unnormalized trial envelope .
Because the left-hand side is purely real and non-negative, the right hand side must already be so as well, so in fact:
\be\label{eqn:variational_3}
\begin{split}
\tfrac{\del}{\del \omega}\mathscr{P}_{\bv{\nu}}(\xi; \omega; \bv{\alpha}) &= -\tfrac{1}{2}\realpart\left[ \int d\xi '\, \innerp{ \bv{\varepsilon}_{\bv{\nu}\,\perp}(\xi'; \omega; \bv{\alpha}) }{ \bv{j}(\xi'; \omega) }\right]\\
&= -\smallhalf \tfrac{\del}{\del \omega}\mathscr{P}_{\text{\tiny{mech}}}[\bv{\varepsilon}_{\bv{\nu}\,\perp}; \bv{j}](\omega; \bv{\alpha}).
\end{split}
\ee
From the  inequality (\ref{eqn:inequality_2}) and the energy-conservation result (\ref{eqn:energy_2}) we arrive at an alternative  formulation of the variational principle:
\be
\begin{split}\label{eqn:variational_4}
\tfrac{\del}{\del \omega}\mathscr{P}_{\bv{\nu}}(\xi; \omega; \bv{\alpha}) 
& \le \tfrac{\del}{\del \omega}\mathscr{P}_{\bv{\nu}}(\xi; \omega; \tilde{\bv{\alpha}})
= -\smallhalf \tfrac{\del}{\del \omega}\mathscr{P}_{\text{\tiny{mech}}}[\bv{\varepsilon}_{\bv{\nu}\,\perp}; \bv{j}](\omega; \bv{\alpha})\\
& \le \tfrac{\del}{\del \omega}\mathscr{P}_{\bv{a}}(\xi; \omega)
=-\tfrac{\del}{\del \omega}\mathscr{P}_{\text{\tiny{mech}}}[\bv{\varepsilon}_{\bv{a}\,\perp}; \bv{j}](\omega)\\
&= -\smallhalf \tfrac{\del}{\del \omega}\mathscr{P}_{\text{\tiny{mech}}}[\bv{\varepsilon}_{\bv{\chi}\,\perp}; \bv{j}](\omega).
\end{split}
\ee

In addition, since 
\be
\innerp{ \bv{\nu}(\xi; \omega; \bv{\alpha})}{ \bv{\nu}(\xi; \omega; \bv{\alpha}) } 
= \abs{ \innerp{ \bv{v}(\xi; \omega; \bv{\alpha})}{\bv{\chi}(\xi; \omega)}  }^2
=  \innerp{ \bv{\nu}(\xi; \omega; \bv{\alpha})}{ \bv{\chi}(\xi; \omega)},
\ee
for all $\xi,$ it follows that the unnormalized trial envelope $\bv{\nu}(\bv{r}; \xi; \omega; \tilde{\bv{\alpha}})$ (\ie, with the absolute amplitude included as an adjustable parameter) also minimizes the Hilbert-space distance between the trial solution and  the actual extrapolated paraxial solution in any transverse plane $\xi:$
\be\label{eqn:paraxial_var_dist_1}
\begin{split}
\vnorm{\, \ket{ \bv{\nu}(\xi; \omega; \bv{\alpha})\, } - \ket{ \bv{\chi}(\xi; \omega)}  } &\ge
\vnorm{\, \ket{ \bv{\nu}(\xi; \omega; \tilde{\bv{\alpha}})} - \ket{ \bv{\chi}(\xi; \omega)}  \,}\\
&= \omega^{-2}\left[ \tfrac{\del}{\del \omega}\mathscr{P}_{\bv{\chi}}(\xi; \omega) - \tfrac{\del}{\del \omega}\mathscr{P}_{\bv{\nu}}(\xi; \omega; \tilde{\bv{\alpha}})\right] \ge 0.
\end{split}
\ee
The results (\ref{eqn:variational_4}) and (\ref{eqn:paraxial_var_dist_1}) are obviously closely related to the results (\ref{eqn:paraxial_basis_2}) and (\ref{eqn:min_1a}) derived in the basis-set expansion, which first led us to the variational formulation.  Equivalently, we can express our variational principle most succinctly in terms of the unnormalized (amplitude-included) trial wavefield which solves a constrained power-maximization problem for all parameters $\bv{\alpha}$ defining the trial-solution $\bv{\nu}(\bv{r}; \xi; \omega; \tilde{\bv{\alpha}})$:
\begin{subequations}\label{eqn:max_3}
\begin{align}
\tilde{\bv{\alpha}} &= \argmax\limits_{\bv{\alpha}} \left[  \tfrac{\del}{\del \omega}\mathscr{P}_{\bv{\nu}}(\xi; \omega; \bv{\alpha})   \right]\\
\mbox{s.t. }\tfrac{\del}{\del \omega}\mathscr{P}_{\bv{\nu}}(\xi; \omega; \bv{\alpha}) &=  -\smallhalf \tfrac{\del}{\del \omega}\mathscr{P}_{\text{\tiny{mech}}}[\bv{\varepsilon}_{\bv{\nu}\,\perp}; \bv{j}](\omega; \bv{\alpha}).
\end{align}
\end{subequations}

These several equivalent formulations of the maximum-power variational principle (MPVP) are intuitively appealing, perhaps so much so that they might be guessed immediately, without the entire build-up of mathematical machinery.  The variational approximation, which maximizes the resemblance (in the Hilbert-space metric), within the considered family of possibilities, to the actual paraxial field in the post-source region, or equivalently, to the extrapolated source-free paraxial field everywhere in space, can be obtained by maximizing the source/field spatial overlap $\abs{\int d\xi '\, \innerp{ \bv{\varepsilon}_{\bv{v}\,\perp}(\xi'; \omega; \bv{\alpha}) }{ \bv{j}(\xi'; \omega)}},$ \ie, the magnitude of the three-dimensional inner product between the normalized source-free trial field and the actual current density.  That is to say, our best approximation to the free-space radiation fields beyond the sources is found by maximizing the physical resemblance of these fields, when extrapolated backward into the region of the sources assuming free-space propagation, to the profile of the sources.

Equivalently, we can say the optimal profile is that which, if it actually were to be incident on the sources, would experience the maximal small-signal gain (\ie, neglecting saturation effects or indeed any back-action on the sources), due to energy absorbed from those sources, and in this case the virtual ``gross'' power gain (\ie, due to stimulated emission, but neglecting the stimulated absorption) is numerically proportional  to the estimated spectral density of spontaneously-radiated power.  This approximation also optimizes the total radiated power spectral density in the full variational radiation envelope $\bv{\nu}(\bv{r}; \xi; \omega; \bv{\alpha}),$ consistent with the constraint, arising from energy-conservation, that this power could have arisen from work done by the sources, with appropriate compensation (\ie, scaling by $\tfrac{1}{2}$) for the fact that we have replaced the inhomogeneous fields with their homogeneous extrapolation in the region of these sources.

Note that this variational principle is reminiscent of the Rayleigh-Ritz variational principle familiar from ordinary quantum mechanics, but despite the analogies between paraxial radiation propagation and non-relativistic quantum mechanics, these variational principles are actually distinct.  In quantum mechanics, ``the'' Rayleigh-Ritz method is applicable to approximating the low-lying energy eigenvalues and corresponding stationary states of the homogeneous, time-independent Schrodinger equation, whereas the present variational principle is associated with solutions to the inhomogeneous, time-dependent Schrodinger equation.  

However, the MPVP does share with the quantum-mechanical Raleigh-Ritz technique many of the same features and limitations common to variational approximations and optimization problems.  Because we are determining a stationary point (in fact, a maximum) of a power (spectral density) functional, the actual value obtained for the power spectrum at the extremum is relatively insensitive to the precise shape of the trial field -- generally, if the characteristic relative error in the latter is $O(\epsilon)$ in some sense, then the relative error in the former is $O(\epsilon^2),$ because the first-order variations vanish.  In order to estimate the power to say $1\%,$ we need only match the variational parameters describing the field shape to about $10\%.$   This is, of course, great news if we are actually most interested in estimating the power, but, conversely, much less satisfying if we are interested in determining, say, the spot size $w_0,$ or some characterization of the the angular spectrum.  This is analogous to the situation in quantum mechanics, where it is well known that the Raleigh-Ritz approximation estimates the ground-state energy eigenvalue more accurately than its corresponding wavefunction.

The variational approximation for the field profile is expected to be accurate to the extent that the trial mode can overlap, or mimic, the actual mode in any transverse plane $\xi > \xi_1$ in the free-space region beyond  the sources, and obviously the approximation is expected to improve as the number of functionally-independent shape parameters is increased.  Since $\bv{\nu}(\xi; \omega; \tilde{\bv{\alpha}})$ satisfies the free-space equation, it is actually an approximation throughout all $\xi$ to the extrapolated envelope $\bv{\chi}(\bv{r}; \xi; \omega).$  The corresponding power yields a lower bound for the actual power radiated, but in the absence of an upper bound, one does not have any foolproof means to  estimate the closeness of this approximated to actual power.  We can consider a so-called ``minimizing'' sequence of variational problems where at each stage we expand the parametric family of trial envelopes to include additional possible details of the field profiles which we anticipate might be present in the actual radiation but which could not be captured in the previous trial fields.  If the extended family continues to include as a possibility the previous variational solution, then the estimated power monotonically approaches the actual (unknown) power from below, and the Hilbert-space distance between the trial envelope and the actual (unknown) envelope in any post-source plane monotonically approaches zero from above.  In practice, we can stop when the difference (in power or Hilbert-space distance) upon adding finer details to the trial envelope becomes sufficiently small, or else when we have expended as much computational effort as we can spare.  From above, we know that an additional  parameter leading to an $O(\epsilon)$ relative increase in the power is expected to produce only an $O(\sqrt{\epsilon})$ relative improvement in the local field profile.

Although we formally expressed the normalized ket $\ket{\bv{v}(\xi; \omega; \bv{\alpha})}$ as a sum over the orthonormal modes $\ket{\bv{u}_{n}(\xi; \omega; \bv{\alpha})},$ we stress that the variational envelope can be explicitly written in any arbitrary form, with adjustable parameters which appear linearly or non-linearly,  as long as it satisfies the homogeneous wave equation, gauge constraint, and normalization constraint.  If the parameters $\bv{\alpha}$ represent the expansion coefficients in some finite linear sum of orthonormal modes, which is the assumption of the classic Raleigh-Ritz and Ritz-Galerkin techniques in the calculus of variations,  then the variational approach formally reduces to the basis-set approach, but it is often more convenient or efficient in practice to use an explicitly nonlinear family of trial solutions where the variational parameters do not appear linearly.  For example, one might use a Gaussian beam with indeterminate waist size and location, as discussed in the introduction, or perhaps even additional parameters to include skew or eccentricity or higher-order structure in the spot profile. 

\section{Non-Paraxial Generalization}\label{sec:non_paraxial}

When translated from the quantum mechanical into conventional notation, none of the energy-balance and variational results derived above in the paraxial limit depend in any essential way on the assumed paraxial nature of the fields.  This suggests that analogous results should hold more generally, beyond the paraxial approximation, for any radiation emitted spontaneously by charges moving along prescribed trajectories.  The essential features which emerged in the paraxial case, which we now abstract to more general problems, are: approximation of the actual radiation fields arising from the sources by homogeneous (\ie, free-space, or source-free) fields; a variational principle, derived from the Cauchy-Schwarz inequality in a Hilbert space picture,  mandating the maximization of (only) outgoing power spectral density; energy exchange between fields and charges which can be characterized as the three-dimensional inner product, or overlap integral,  between the extrapolated, solenoidal electric field and either the full or solenoidal current density, as convenient; and, finally,  a functional relationship between the mechanical work performed on/by the charges by/on the extrapolated fields within a region and the radiation flux through a boundary surface, arising, physically, from energy conservation, and, mathematically, either explicitly from the formal solution or else from some form of the the Fundamental Theorem of Calculus (\eg, Gauss's theorem or Green's Identities in multiple dimensions.)

\subsection{Vector Spherical Harmonics and the Spherical Wave Basis}

In order to generalize our results to the non-paraxial case, we first introduce a bit more mathematical notation and machinery.  We use the scaled spatial position $\bv{\zeta},$ and introduce spherical coordinates: the scaled radius  $\zeta = \vnorm{\bv{\zeta}} \ge 0,$ the polar angle $\theta,$ and the azimuthal angle $\varphi,$  so that $\bv{\zeta} = \zeta \unitvec{\zeta} = \zeta \unitvec{x}.$  We let $R$ denote any (closed) spatial region in $\realsymbol^{3},$ and $\del R$ will denote its boundary.  For any position $\bv{\zeta}_0$ and any nonnegative $\zeta'  \ge  0,$ we define the closed ball of radius $\zeta'$ centered at $\bv{\zeta}_0$ as $B(\zeta'; \bv{\zeta}_0) \equiv \left\{\bv{\zeta} \,\,\vert\,\, \vnorm{\bv{\zeta} - \bv{\zeta}_0} \le \zeta'  \right\},$ and we will take $V(\zeta'; \bv{\zeta}_0) \subset \realsymbol^3$ to denote  a closed, simply-connected region satisfying $B(\zeta'; \bv{\zeta}_0) \subseteq V(\zeta'; \bv{\zeta}_0).$  For simplicity, whenever the central position coincides with the origin ($\bv{\zeta}_0 = \bv{0}$), the dependence will be suppressed in the notation;  \ie, $B(\zeta') \equiv  B(\zeta'; \bv{0})$ and $V(\zeta') = V(\zeta';  \bv{0}).$

As in the paraxial case, we take the (scaled, frequency-domain) current density $\bv{j}(\bv{\zeta}; \omega),$ assumed known, is to be localized in space, vanishing identically outside some finite-radius ball $B(\zeta_1'),$ for some $0 < \zeta_1' < \infty.$  Again, the solenoidal component $\bv{j}_{\perp}(\bv{\zeta}; \omega)$  will therefore not have have compact support in general, but because of the rapid falloff  of its spatial dependence ($O(\zeta^{-2})$ as $\zeta \to \infty$), we may, with arbitrary small error in the calculated radiation fields, neglect it beyond some sufficiently large but finite radius $\zeta_1 \ge \zeta_1'.$   (Again, at the end of the calculation, we could attempt to take this support to be infinite, provided that the full current density continues to fall off sufficiently rapidly to ensure convergence of the integrals, and to enable somehow the unambiguous characterization of the ``far-zone'' versus non-radiative fields, when the latter extend to infinity.)  

It will prove very convenient to have some explicit basis in which to decompose the vector potential.  The plane-wave (spatial-Fourier)  basis immediately comes to mind, but certain cumbersome singularities inevitably arise in this representation.  Namely, whenever  the vacuum dispersion relation is satisfied, \ie, $\vnorm{\bv{k}} = \omega,$ solutions to the homogeneous Helmholtz equation possess delta-function singularities, while solutions to the inhomogeneous solution exhibit second-order divergences whenever the corresponding Fourier components $\bv{j}_{\perp}(\bv{k}, \omega) \neq  \bv{0}.$ 
In order to avoid such difficulties, we will instead use the spherical wave basis \cite{jackson:75, george_gamliel:90, morehead:01}.  The conventional vector spherical harmonics are defined as
\be
\bv{X}_{\ell \, m} = \bv{X}_{\ell \, m}(\unitvec{\zeta}) = \bv{X}_{\ell \, m}(\theta, \varphi) = \tfrac{1}{\sqrt{\ell(\ell+1)}}\bv{L}Y_{\ell \, m}(\theta, \varphi),
\ee
where, in analogy to quantum mechanics, we have defined the Hermitian (scaled) orbital angular momentum operator as 
\be
\bv{L} \equiv \bv{Q}_{\stext{3D}} \times \bv{P}_{\stext{3D}} \equiv \bv{\zeta} \times \left(\tfrac{1}{i}\bv{\del}\right) = -i\left(\bv{\zeta} \times \bv{\del}\right),
\ee
which acts only on the angular degrees of freedom (\ie, not on the radial or the polarization degrees-of-freedom),
\be
\bv{L} = \tfrac{1}{\sqrt{2}} \unitvec{e}_{+} e^{-i \varphi} \left(-\tfrac{\del}{\del \theta} + i \cot(\theta)\tfrac{\del}{\del \varphi} \,\, \right)
+\tfrac{1}{\sqrt{2}} \unitvec{e}_{-} e^{+i \varphi} \left(+\tfrac{\del}{\del \theta} + i \cot(\theta)\tfrac{\del}{\del \varphi}  \,\,\right)
-i \unitvec{z}\tfrac{\del}{\del \varphi};
\ee
and the functions $Y_{\ell\, m}(\theta, \varphi)$ denote the usual scalar spherical harmonics for nonnegative integers $\ell = 0, 1, 2, \dotsc,$ and integer $m = -\ell, -\ell + 1, \dotsc, 0, \dotsc, \ell-1, \ell.$  In addition, we define two other related sets of basis vector fields:
\be
\bv{Z}_{\ell \, m} = \bv{Z}_{\ell \, m}(\unitvec{\zeta}) = \bv{Z}_{\ell \, m}(\theta, \varphi) = \unitvec{\zeta} \times \bv{X}_{\ell \, m},
\ee
and
\be
\bv{N}_{\ell \, m} = \bv{N}_{\ell \, m}(\unitvec{\zeta}) = \bv{N}_{\ell \, m}(\theta, \varphi) = \unitvec{\zeta} \, Y_{\ell \, m}.
\ee
(By definition, we take $\bv{X}_{0 \, 0}  = \bv{Z}_{0 \, 0} \equiv \bv{0}$ identically, ultimately reflecting the fact that spherically-symmetric solutions to the free-space Maxwell equations can exist only in the non-radiative $\omega \to 0$ limit.)

The orbital angular momentum operator $\bv{L}$ effects infinitesimal (inverse) rotations of the spatial degrees of freedom, \ie, observation position or field point, and satisfies the useful identities:
\begin{subequations}
\begin{align}
\unitvec{\zeta} \cdot \bv{L}\phantom{]} &= 0,\\
\bv{L} \times \bv{L} &= i\bv{L},\\
\comm{L^2}{\bv{L}} &= \bv{0},\\
\comm{P^2_{\stext{3D}}   }{\bv{L}} &= \bv{0},\\
L^2 &= r \tfrac{\del^2}{\del r^2}\left(r \,\,\,\,\, \right) + r^2 P^2_{\stext{3D}}.
\end{align}
\end{subequations}
The scalar spherical harmonics themselves may be interpreted as simultaneous eigenstates of $L^2$ and $L_z.$  Each of the basis vector fields may then be interpreted as a simultaneous eigenstate of $J^2 = J_x^2 + J_y^2 + J_z^2$ and  $J_z$ for some spin-one object, where $\bv{J} = \bv{L} + \bv{S}$ is the total (spatial plus spin/polarization) angular momentum operator.  For fixed $\ell$ and fixed $\bv{W} = \bv{X},\, \bv{Z}, \, \mbox{ or } \bv{N},$ the members  $\bv{W}_{\ell \, m}(\theta, \varphi)$ of each of these three families therefore transform amongst themselves under total spatial rotations (of both spatial and polarization degrees of freedom).  The spin angular momentum operator $\bv{S}$ generates rotations of the polarization vector, and  in the Cartesian basis, $\bv{S} = \unitvec{x}S_1 + \unitvec{y}S_2 + \unitvec{z}S_3$ may be defined as a vector of $3 \times 3$ Hermitian matrices $S_j$ for  $j = 1,\, 2,\, 3,$ whose components are given by
\be
(S_j)_{k n} = -i \epsilon_{jkn},
\ee
where $\epsilon_{jkn}$ is the completely antisymmetric Levi-Civita tensor in three dimensions.  
From this definition one can prove the useful operator identity
\be
\bv{\del}\times \, \equiv \bv{P}_{3D} \cdot \bv{S}.
\ee

Using various vector identities, these spherical vector harmonic basis fields also can be shown to satisfy:
\begin{subequations}
\begin{align}
\bv{L} \cdot \!\bv{X}_{\ell \, m} &= \sqrt{\ell(\ell+1)}Y_{\ell\, m},\\
\bv{L} \cdot \bv{Z}_{\ell \, m} &=0,\\
\bv{L} \cdot \bv{N}_{\ell \, m} &= 0;
\end{align}
\end{subequations}
and
\begin{subequations}
\begin{align}
\bv{L} \times \!\bv{X}_{\ell \, m} &= i\bv{X}_{\ell \, m},\\
\bv{L} \times \bv{Z}_{\ell \, m} &= \sqrt{\ell(\ell+1)}\bv{N}_{\ell \, m},\\
\bv{L} \times \bv{N}_{\ell \, m} &= \phantom{i}\bv{0};
\end{align}
\end{subequations}
as well as
\begin{subequations}
\begin{align}
\unitvec{\zeta} \cdot \!\bv{X}_{\ell \, m} &= 0,\\
\unitvec{\zeta} \cdot \bv{Z}_{\ell \, m} &= 0,\\
\unitvec{\zeta} \cdot \bv{N}_{\ell \, m} &= Y_{\ell\, m};
\end{align}
\end{subequations}
and
\begin{subequations}
\begin{align}
\unitvec{\zeta} \times \!\bv{X}_{\ell \, m} &= \phantom{-}\bv{Z}_{\ell \, m},\\
\unitvec{\zeta} \times \bv{Z}_{\ell \, m} &=-\bv{X}_{\ell \, m},\\
\unitvec{\zeta} \times \bv{N}_{\ell \, m} &= \phantom{-}\bv{0};
\end{align}
\end{subequations}
and also
\begin{subequations}
\begin{align}
\bv{\del} \cdot \!\bv{X}_{\ell \, m} &= \phantom{+}0,\\
\bv{\del} \cdot \bv{Z}_{\ell \, m} &= -\tfrac{i}{\zeta}\sqrt{\ell(\ell+1)} Y_{\ell\, m},\\
\bv{\del} \cdot \bv{N}_{\ell \, m} &= \phantom{+}\tfrac{2}{\zeta} Y_{\ell\, m};
\end{align}
\end{subequations}
together with
\begin{subequations}
\begin{align}
\bv{\del} \times \!\bv{X}_{\ell \, m} &=  \phantom{-}\tfrac{1}{\zeta}\left(i \sqrt{\ell(\ell+1)}\bv{N}_{\ell \, m} + \bv{Z}_{\ell \, m}\right),\\
\bv{\del} \times \bv{Z}_{\ell \, m} &= -\tfrac{1}{\zeta}\bv{X}_{\ell \, m},\\
\bv{\del} \times \bv{N}_{\ell \, m} &= -\tfrac{i}{\zeta} \sqrt{\ell(\ell+1)} \bv{X}_{\ell \, m}.
\end{align}
\end{subequations}
For any allowed $\ell$ and $m,$ the vector spherical harmonics are geometrically orthogonal as vectors at every point $\unitvec{\zeta}$ on the unit sphere:
\be
\bv{X}_{\ell \, m}(\unitvec{\zeta})\cc \cdot \bv{Z}_{\ell \, m}(\unitvec{\zeta})= \bv{X}_{\ell \, m}(\unitvec{\zeta})\cc \cdot \bv{N}_{\ell \, m}(\unitvec{\zeta}) = \bv{Z}_{\ell \, m}(\unitvec{\zeta})\cc \cdot \bv{N}_{\ell \, m}(\unitvec{\zeta}) = 0;
\ee
are (Hilbert-space) orthonormal when integrated over solid angle (except in the trivial case where both vector fields identically vanish for $\ell = 0$):
\be
\int \!d\theta \,\sin\theta \!\int \!d\varphi \, \bv{W}_{\ell\,m}(\theta, \varphi)\cc \cdot  \bv{W}'_{\ell'\,m'}(\theta, \varphi) = \delta_{\bv{W}\,\bv{W}'}\delta_{\ell\, \ell'}\delta_{m \,m'}\left(1 - \delta_{\bv{W}\bv{X}}\delta_{\ell\, 0}\right)\left(1 - \delta_{\bv{W}\bv{Z}}\delta_{\ell\, 0}\right);
\ee
and also collectively constitute a complete set for vector fields defined on the unit sphere:
\be
\sum_{\bv{W} = \bv{X}, \bv{Z}, \bv{N}} \sum_{\ell = 0}^{\infty} \sum_{m = -\ell}^{\ell}  \bv{W}_{\ell\,m}(\theta, \varphi)\cdot  \bv{W}_{\ell\,m}(\theta', \varphi')\hc =   \delta(\cos \theta - \cos \theta')\delta(\varphi - \varphi')I_{3},
\ee
where $I_3$ is the identity matrix on the polarization degrees of freedom.

Now, outside the (effective) support of the solenoidal sources, \ie, for any $\zeta \ge \zeta_2 > \zeta_1$, the actual Coulomb-gauge vector potential $\bv{a}(\bv{\zeta}; \omega)$ is assumed to satisfy the homogeneous Helmholtz equation, and the method of  separation-of-variables may be used in spherical coordinates to show that there the solution may always be written in the form:
\be\label{eqn:inhom_spherical}
\bv{a}(\bv{\zeta}; \omega) = \sum_{\ell,\, m} \tfrac{1}{\omega^2} a^{\stext{E}}_{\ell\, m}(\omega)\bv{\del}\times\bigl[ h^{(1)}_{\ell}(k\zeta) \bv{X}_{\ell\, m}(\theta, \varphi)\bigr] - \tfrac{i}{\omega} a^{\stext{M}}_{\ell\, m}(\omega) h^{(1)}_{\ell}(k\zeta) \bv{X}_{\ell\, m}(\theta, \varphi), 
\ee
where the spherical Hankel functions of the first and second kind,
\begin{subequations}
\begin{align}
h^{(1)}_{\ell}(x) &= j_{\ell}(x) + i\, n_{\ell}(x),\\
h^{(2)}_{\ell}(x) &= j_{\ell}(x) - i\, n_{\ell}(x).
\end{align}
\end{subequations}
are defined in terms of the usual spherical Bessel functions,
\be
j_{\ell}(x) = (-x)^{\ell} \left(\frac{1}{x}\frac{d}{d x} \right)^{\ell} \left(\frac{\sin x}{x} \right),
\ee
(which should not be confused with any current density), and the spherical Neumann functions,
\be
n_{\ell}(x) = -(-x)^{\ell} \left(\frac{1}{x}\frac{d}{d x} \right)^{\ell} \left(\frac{\cos x}{x} \right).
\ee
Any two of these four families of special functions may be taken as the linearly-independent solutions to the radial differential equation obtained through separation of variables.  The appearance of only the $h^{(1)}_{\ell}(kr) \propto \frac{e^{i k \zeta}}{k\zeta}$ in the radial part of our formal solution reflects the facts that the solution holds in the extra-source region $\zeta > \zeta_1$ which excludes the origin, and that we have imposed the so-called Sommerfeld boundary conditions, requiring only outgoing radiation at spatial infinity.

Each term in the superposition (\ref{eqn:inhom_spherical}) is everywhere divergenceless, so the total vector field is solenoidal, as required.  The expansion coefficients $a^{\stext{E}}_{\ell\, m}(\omega) \in \complexsymbol$ and $a^{\stext{M}}_{\ell\, m}(\omega) \in \complexsymbol$ are, respectively, the (scaled) electric and magnetic multipole moments of the transverse current density $\bv{j}_{\perp}(\bv{\zeta}; \omega),$ and are uniquely determined by quadratures over $\bv{j}$, or equivalently by field values in the far-field (or, in fact,  by just the radial components of the fields on any two spherical shells outside the sources.)  For example, using the far-field asymptotic behavior, one can, with a bit of algebra, show that
\begin{subequations}
\begin{align}
a^{\stext{M}}_{\ell\, m}(\omega) &= - \phantom{i}\frac{\omega^2}{\sqrt{\ell(\ell+1)}} \sum_{m' = \min[0,m-1]}^{\max[m+1, \ell]} \bv{M}_{\ell\, m'}(\omega)\cdot
\int \!d\theta \,\sin\theta \!\!\int \!d\varphi \, Y_{\ell \, m} (\theta, \varphi)\cc \bv{L} Y_{\ell \, m'} (\theta, \varphi),\\
a^{\stext{E}}_{\ell\, m}(\omega) &= \phantom{+}i \frac{\omega}{\sqrt{\ell(\ell+1)}} \sum_{m' = \min[0,m-1]}^{\max[m+1, \ell]} \bv{R}_{\ell\, m'}(\omega)\cdot
\int \!d\theta \,\sin\theta \!\!\int\! d\varphi \, Y_{\ell \, m} (\theta, \varphi)\cc \bv{L} Y_{\ell \, m'} (\theta, \varphi);
\end{align}
\end{subequations}  
where we have defined
\begin{subequations}
\begin{align}
\bv{M}_{\ell\, m}(\omega) &= \int d^3\bv{\zeta} \, \left[j_{\ell}(kr) Y_{\ell \, m} (\unitvec{r})\cc \right]\bv{j}_{\perp}(\bv{\zeta}; \omega),\\
\bv{R}_{\ell\, m}(\omega) &= \int d^3\bv{\zeta} \, \left[j_{\ell}(kr) Y_{\ell \, m} (\unitvec{r})\cc \right]\bv{\del}\!\times\!\bv{j}_{\perp}(\bv{\zeta}; \omega).
\end{align}
\end{subequations}  
Many other equivalent forms are possible, depending on precisely what data are used and where, but explicit forms for the multipole expansion coefficients  will not actually be needed here.  (After all, if they were known or easily calculable, we would not need to resort to any sort of approximation.)  Note that time-reversed, or equivalently conjugate, or incoming-wave, solutions can obviously be written in an analogous manner, with the $h^{(1)}_{\ell}(k\zeta)$ replaced with $h^{(2)}_{\ell}(k\zeta)  \propto \frac{e^{-i k \zeta}}{k\zeta}.$ 

By assumption, any solenoidal, source-free  ``trial'' vector potential $\bv{\chi}(\bv{\zeta}; \omega; \bv{\alpha})$ satisfies the homogeneous equation everywhere in space, and may be expressed, for any $\bv{\zeta}$, in a similar form: 
\be\label{eqn:hom_spherical}
\bv{\chi}(\bv{\zeta}; \omega; \bv{\alpha}) = \sum_{\ell,\, m} \tfrac{1}{\omega^2} \chi^{\stext{E}}_{\ell\, m}(\omega; \bv{\alpha})\bv{\del}\!\times\!\bigl[ j_{\ell}(k\zeta) \bv{X}_{\ell\, m}(\theta, \varphi)\bigr] + - \tfrac{i}{\omega}\chi^{\stext{M}}_{\ell\, m}(\omega; \bv{\alpha}) j_{\ell}(k\zeta) \bv{X}_{\ell\, m}(\theta, \varphi). 
\ee
The use of $j_{\ell}(k\zeta)$ is the only choice for the radial dependence which solves the Helmholtz equation but is regular everywhere in space, including the origin.  Each term in (\ref{eqn:hom_spherical}) is also automatically divergenceless everywhere, so their sum (when convergent) is solenoidal.  The complex coefficients  $\chi^{\stext{E}}_{\ell\, m}(\omega; \bv{\alpha})$ and $\chi^{\stext{M}}_{\ell\, m}(\omega; \bv{\alpha})$ for this source-free solution may nevertheless be interpreted as some effective (scaled) multipole moments which would describe what we might term an ``effective source'' $\bv{g}(\bv{\zeta}; \omega; \bv{\alpha}),$ or may just be regarded as free expansion coefficients determined by (or even taken to be equal to) the actual variational parameters $\bv{\alpha}$ appearing in $\bv{\chi}(\bv{\zeta}; \omega; \bv{\alpha})$.   The familiar expansions
\be
e^{i\bv{k}\cdot \bv{\zeta}} = 4\pi \sum_{\ell = 0}^{\infty} i^{\ell} j_{\ell}(k\zeta) \!\!\sum_{m = -\ell}^{\ell} Y_{\ell \, m}(\unitvec{\zeta})\cc Y_{\ell \, m}(\unitvec{k})
= \sum_{\ell = 0}^{\infty} i^{\ell}\sqrt{4\pi(2\ell+1)} j_{\ell}(k\zeta)Y_{1 \, 0}(\arccos(\unitvec{\zeta}\cdot \unitvec{k}), 0),
\ee
for any fixed (scaled) wavevector $\bv{k} = k\unitvec{k},$ may be used to establish the connection between the plane wave and spherical wave representations, and to confirm that they both span the same space of source-free solutions.

\subsection{Green Functions}

Alternatively, the exact solution to the inhomogeneous problem at any position $\bv{\zeta}$ may be expressed in terms of a convolution over the retarded Green function:
\be
G^{\stext{ret}}(\bv{\zeta}; \bv{\zeta}'; \omega) = \frac{ e^{+i k \vnorm{\bv{\zeta} - \bv{\zeta}'}} }{4\pi \vnorm{\bv{\zeta} - \bv{\zeta}'}},
\ee
which is invariant under simultaneous translations, rotations, or reflections of both spatial arguments, 
and satisfies the scalar Helmholtz equation with impulsive source (and with negative unit charge, conforming to convention for electromagnetic problems),
\be
\left(\del^2 + \omega^2 \right)G^{\stext{ret}}(\bv{\zeta}; \bv{\zeta}'; \omega) = -\delta(\bv{\zeta} - \bv{\zeta}'),
\ee
together with the Sommerfeld asymptotic radiation conditions for outgoing waves:
\begin{subequations}
\begin{align}
G^{\stext{ret}}(\bv{\zeta}; \bv{\zeta}'; \omega)= O(1/\zeta) \,\mbox{ as }\,  \zeta \to \infty\\
\bigl(\tfrac{\del}{\del \zeta} - i k \bigr)G^{\stext{ret}}(\bv{\zeta}; \bv{\zeta}'; \omega) = o(1/\zeta) \,\mbox{ as }\,  \zeta  \to \infty,
\end{align}
\end{subequations}
for any fixed emitter position $\bv{\zeta}'.$  (Because we are working with solenoidal sources in otherwise free space, we can use a scalar rather than dyadic Green function \cite{Tai:93}.) 
The retarded Green function $G^{\stext{ret}}(\bv{\zeta}; \bv{\zeta}'; \omega),$  then represents an outgoing (scalar) spherical wave at the observation position $\bv{\zeta}$ emanating from a harmonic point source at $\bv{\zeta}'.$  (By symmetry, analogous conditions and interpretations obviously hold true under permutation of the two spatial positions $\bv{\zeta}$ and $\bv{\zeta}',$ from which follow various well-known electromagnetic reciprocity results such as the Raleigh-Carson and Lorentz  theorems.)  The advanced Green function
\be
G^{\stext{adv}}(\bv{\zeta}; \bv{\zeta}'; \omega) =  \bigl[G^{\stext{ret}}(\bv{\zeta}; \bv{\zeta}'; \omega)\bigr]\cc =
\frac{ e^{-i k \vnorm{\bv{\zeta} - \bv{\zeta}'}} }{4\pi \vnorm{\bv{\zeta} - \bv{\zeta}'}},
\ee
satisfies the same impulsive Helmholtz equation, and in fact may be interpreted as the time-reversal of the retarded impulse response, representing an incoming spherical wave, observed at $\bv{\zeta},$ and ultimately converging to a point absorber at $\bv{\zeta}'.$  Given any current density $\bv{j}(\bv{\zeta}'; \omega)$ with compact support in the neighborhood $V(\zeta_1)$ of the origin, and supposing no incoming radiation from distant sources by assumption, the solenoidal component $\bv{j}_{\perp}(\bv{\zeta}'; \omega)$ can result only in outgoing radiation asymptotically as $\zeta \to \infty,$ so the Coulomb-gauge vector potential can always be expressed as:
\be
\bv{a}(\bv{\zeta}; \omega) = \int d^3 \bv{\zeta}'\, G^{\stext{ret}}(\bv{\zeta}; \bv{\zeta}'; \omega) \bv{j}_{\perp}(\bv{\zeta}'; \omega)
= \int\limits_{V(\zeta_1)} \!\!\!d^3 \bv{\zeta}'\, G^{\stext{ret}}(\bv{\zeta}; \bv{\zeta}'; \omega) \bv{j}_{\perp}(\bv{\zeta}'; \omega).
\ee
Clearly, this satisfies the appropriate inhomogeneous Helmholtz equation:
\be
\begin{split}
(\del^2 + \omega^2)\bv{a}(\bv{\zeta}; \omega) &= \int d^3 \bv{\zeta}'\, (\del^2 + \omega^2)G^{\stext{ret}}(\bv{\zeta}; \bv{\zeta}'; \omega) \bv{j}_{\perp}(\bv{\zeta}'; \omega)\\
&= -\int d^3 \bv{\zeta}'\, \delta(\bv{\zeta} - \bv{\zeta}' )\,\bv{j}_{\perp}(\bv{\zeta}'; \omega),
= - \bv{j}_{\perp}(\bv{\zeta}; \omega)
\end{split}
\ee
and also shares with $G^{\stext{ret}}(\bv{\zeta}; \bv{\zeta}'; \omega)$ the appropriate outgoing-wave Sommerfeld boundary conditions.  Because $G^{\stext{ret}}(\bv{\zeta}; \bv{\zeta}'; \omega)$ is radially symmetric, depending on the spatial coordinates only through $\vnorm{\bv{\zeta} -\bv{\zeta}' },$ and because $\bv{\del}\cdot \bv{j}_{\perp}(\bv{\zeta}; \omega) = 0$ by construction, we have
\be
\begin{split}
\bv{\del} \cdot \bigl[ G^{\stext{ret}}(\bv{\zeta}; \bv{\zeta}'; \omega)\bv{j}_{\perp}(\bv{\zeta}'; \omega)\bigr] &= \bv{\del} G^{\stext{ret}}(\bv{\zeta}; \bv{\zeta}'; \omega) \cdot \bv{j}_{\perp}(\bv{\zeta}'; \omega)
= -\bv{\del}' G^{\stext{ret}}(\bv{\zeta}; \bv{\zeta}'; \omega) \cdot \bv{j}_{\perp}(\bv{\zeta}'; \omega)\\
&= -\bv{\del}' \cdot \left[ G^{\stext{ret}}(\bv{\zeta}; \bv{\zeta}'; \omega)\bv{j}_{\perp}(\bv{\zeta}'; \omega)\right].
\end{split}
\ee
Since $G^{\stext{ret}}(\bv{\zeta}; \bv{\zeta}'; \omega) = O(1/\abs{\bv{\zeta} - \bv{\zeta}'})$ and $\bv{j}_{\perp}(\bv{\zeta}; \omega) \lesssim O(1/\zeta^2)$ as $\zeta \to \infty,$ Gauss's law then implies
\be
\begin{split}
\bv{\del} \cdot \bv{a}(\bv{\zeta}; \omega) &= 
\int d^3 \bv{\zeta}'\,  \bv{\del}\cdot [G^{\stext{ret}}(\bv{\zeta}; \bv{\zeta}'; \omega) \bv{j}_{\perp}(\bv{\zeta}'; \omega)]
= -\int d^3 \bv{\zeta}'\,  \bv{\del}'\!\cdot\! [G^{\stext{ret}}(\bv{\zeta}; \bv{\zeta}'; \omega) \bv{j}_{\perp}(\bv{\zeta}'; \omega)]\\
&= -\lim_{\zeta_2 \to \infty} \int\limits_{V(\zeta_2)} \!\!\!d\sigma' \unitvec{n}\cdot [G^{\stext{ret}}(\bv{\zeta}; \bv{\zeta}'; \omega) \bv{j}_{\perp}(\bv{\zeta}'; \omega)]  = 0
\end{split}
\ee
so indeed $\bv{a}(\bv{\zeta}; \omega)$ is solenoidal, as required.  (Subsequently,
we will make use of a number of similar multidimensional ``integrations by parts,'' relying on the  the sufficiently rapid fall-off of $\bv{j}_{\perp}$ and radiative fall-off of $G^{\stext{ret}},$ but for brevity we will not always provide the explicit demonstrations.)

Outside the support of sources, say for $\zeta \ge \zeta_2 > \zeta_1,$ the solution to the inhomogeneous Helmholtz equation may also be written in terms of a surface integral involving the Green function and boundary data.  Specifically, by applying Green's second identity to the region $\lim\limits_{\zeta_3 \to \infty}\!\!V(\zeta_3) - V(\zeta_2),$ using the Helmholtz equations satisfied by the vector potential and the retarded Green function, and making use of the assumed Sommerfeld radiation conditions, we find:
\be\label{eqn:kirchoff}
\bv{a}(\bv{\zeta}; \omega) = \!\!\int\limits_{\del V(\zeta_2)} \!\!\!d\sigma' \, \bigl[ \bv{a}(\bv{\zeta}'; \omega)(\unitvec{n}'\!\cdot\!\bv{\del}') G^{\stext{ret}}(\bv{\zeta}; \bv{\zeta}'; \omega) - G^{\stext{ret}}(\bv{\zeta}; \bv{\zeta}'; \omega)(\unitvec{n}'\!\cdot\!\bv{\del}')\bv{a}(\bv{\zeta}'; \omega) \bigr],
\ee
which is the vector form of the well-known Kirchoff diffraction integral.  Here the unit vector $\unitvec{n}' = \unitvec{n}(\bv{\zeta}')$ denotes an outward normal to the boundary surface at any point $\bv{\zeta}' \in \del V(\zeta_2).$

By piecing together different spherical wave solutions to produce just the right discontinuity at $\bv{\zeta} = \bv{\zeta}',$ it is straightforward to establish the well-known Bessel decomposition of the retarded Green function:
\be\label{eqn:green_spherical_decomp}
G^{\stext{ret}}(\bv{\zeta}; \bv{\zeta}'; \omega) = i\omega \sum_{\ell = 0}^{\infty} j_{\ell}(k\zeta_{<}) h^{(1)}_{\ell}(k\zeta_{>}) \sum_{m = -\ell}^{\ell} Y_{\ell\, m}(\unitvec{\zeta})Y_{\ell\, m}(\unitvec{\zeta}')\cc
\ee
where 
\begin{subequations}
\begin{align}
\zeta_{<} &= \min[\zeta, \zeta'],\\
\zeta_{>} &= \max[\zeta, \zeta'].
\end{align}
\end{subequations}
From linearity, we can see that the difference between the retarded and advanced Green functions,
\be
D(\bv{\zeta}; \bv{\zeta}'; \omega) \equiv G^{\stext{ret}}(\bv{\zeta}; \bv{\zeta}'; \omega) - G^{\stext{adv}}(\bv{\zeta}; \bv{\zeta}'; \omega) = -\frac{ \sin\left( k \vnorm{\bv{\zeta} - \bv{\zeta}'}\right) }{2\pi i \vnorm{\bv{\zeta} - \bv{\zeta}'}} =  \frac{i}{2\pi} \sinc\left( k \vnorm{\bv{\zeta} - \bv{\zeta}'}\right),
\ee
must be a solution (with respect to either spatial coordinate) to the source-free scalar Helmholtz equation everywhere in space.
Substituting in the decomposition (\ref{eqn:green_spherical_decomp}) for the Green function, this homogeneous solution $D(\bv{\zeta}; \bv{\zeta}'; \omega)$ may be written as:
\be\label{eqn:fundamental__spherical_decomp}
D(\bv{\zeta}; \bv{\zeta}'; \omega) = 2i\omega \sum_{\ell\, , m} j_{\ell}(k\zeta) j_{\ell}(k\zeta')\cc Y_{\ell\, m}(\unitvec{\zeta})Y_{\ell\, m}(\unitvec{\zeta}')\cc,
\ee
where we have used the facts that $\sum\limits_{m = -\ell}^{\ell} \!\!Y_{\ell\, m}(\unitvec{\zeta})Y_{\ell\, m}(\unitvec{\zeta}')\cc$ and $ j_{\ell}(k\zeta)$ are both real.

Now, it turns out that, just as a general solution to the inhomogeneous equation with outgoing Sommerfeld radiation boundary conditions may be written as a convolution between the retarded Green function and the solenoidal portion of the current density, any arbitrary solenoidal solution to the homogeneous problem may be written as a convolution between what we may term the fundamental solution $D(\bv{\zeta}; \bv{\zeta}'; \omega)$ and the solenoidal component $\bv{g}_{\perp}(\bv{\zeta}'; \omega),$ of some effective source term $\bv{g}(\bv{\zeta}'; \omega)$ whose support can be confined to any finite-radius ball $B(\zeta_0)$ where $\zeta_0 > 0:$
\be\label{eqn:fundamental_sol}
\bv{\chi}(\bv{\zeta}; \omega) = \int d^3 \bv{\zeta}'\, D(\bv{\zeta}; \bv{\zeta}'; \omega) \bv{g}_{\perp}(\bv{\zeta}'; \omega)
= \int\limits_{V(\zeta_0)} \!\!\!d^3 \bv{\zeta}'\, D(\bv{\zeta}; \bv{\zeta}'; \omega) \bv{g}_{\perp}(\bv{\zeta}'; \omega)
\ee
 To see this, first note that if we choose 
 \be
\bv{g}(\bv{\zeta}; \omega) =  \bv{g}_{\perp}(\bv{\zeta}; \omega) \propto \Theta( \zeta_0 - \zeta) j_{\ell}(k\zeta) \bv{X}_{\ell\, m}(\theta, \varphi),
 \ee
(which is clearly divergenceless by the chain rule and the transverse nature of  $\bv{X}_{\ell\, m}(\unitvec{\zeta})$), then, by using the explicit decomposition (\ref{eqn:fundamental__spherical_decomp}) for $D(\bv{\zeta}; \bv{\zeta}'; \omega)$ and the orthogonality properties of the spherical harmonics as well as their behavior under the action of the orbital angular momentum operator $\bv{L},$ we find
\be
\bv{\chi}(\bv{\zeta}; \omega) \propto  \biggl[\int\limits_{0}^{\zeta_0} d\zeta'\, \abs{\zeta'  j_{\ell}(k\zeta')}^2\biggr]  j_{\ell}(k\zeta) \bv{X}_{\ell\, m}(\theta, \varphi),
\ee
and thus any term of the magnetic multipole form in (\ref{eqn:hom_spherical}) can be generated in this manner.  Using standard vector identities, the symmetry of  $D(\bv{\zeta}; \bv{\zeta}'; \omega)$ in its spatial arguments, and integration by parts relying on the rapid fall-off of $\bv{g}_{\perp},$ we also have
\be
\begin{split}
\bv{\del}\!\times\! \int d^3 \bv{\zeta}'\, D(\bv{\zeta}; \bv{\zeta}'; \omega) \,\bv{g}(\bv{\zeta}'; \omega) &= 
\int d^3 \bv{\zeta}'\, \bv{\del}\!\times\! \bigl[D(\bv{\zeta}; \bv{\zeta}'; \omega)\, \bv{g}_{\perp}(\bv{\zeta}'; \omega)\bigr]\\
&=  \int d^3 \bv{\zeta}'\, D(\bv{\zeta}; \bv{\zeta}'; \omega) \,\bv{\del}'\!\times\! \bv{g}_{\perp}(\bv{\zeta}'; \omega),
\end{split}
\ee
So any required term of the electric multipole form can also be so generated.  By linearity, an arbitrary homogeneous solution may then be constructed as a convolution over the fundamental homogeneous solution.  We will see below that the effective source $\bv{g}(\bv{\zeta}; \omega)$ appearing in the convolutional representation may be taken to be identical to that determining the multipole coefficients in the spherical wave expansion.

\subsection{Hilbert Space Results}

In the non-paraxial case, the wave-equation is no longer equivalent to a time-dependent Schrodinger equation, although the source-free Helmholtz equation is analogous to a time-independent, free-particle Schrodinger equation, albeit typically with non-normalizable, scattering-state boundary conditions.  However, when convenient, we can still recast our non-paraxial mathematical results into a Hilbert-space-like formalism.  Without a distinguished optic axis, we extend the spatial dependence of spinors and observables from two dimensions to three dimensions, each coordinate now treated on an equal footing, and we allow for more general polarization bases, using some given family (labeled here by $n,$ standing for some set of integer indices or ``quantum numbers'')  of smooth, possibly complex-valued vector fields $\unitvec{\epsilon}^{n}_{s}(\bv{\zeta}; \omega)$ (still indexed, say by $s = -1, 0, +1$)  which constitute a local orthonormal triad at every spatial position $\bv{\zeta}:$
\be
\bv{\epsilon}^{n}_{s}(\bv{\zeta}; \omega )\cc \cdot \bv{\epsilon}^{n}_{s'}(\bv{\zeta}; \omega) = \delta_{s\, s'}.
\ee
For example, we could use, as convenient, a fixed Cartesian basis $\unitvec{x},$ $\unitvec{y},$ $\unitvec{z};$ the spherical-coordinate basis $\unitvec{\zeta},$ $\unitvec{\theta},$ $\unitvec{\varphi};$ the helicity basis $\unitvec{e}_{+},$ $\unitvec{e}_{-},$ $\unitvec{z}$ (also known as the spherical basis because of its role in defining spherical tensors,  not to be confused with the above basis vectors for spherical coordinates); normalized vector spherical harmonic basis vector field for some $\ell \ge 1$ and $m,$ \ie, $\unitvec{X}_{\ell\, m}(\unitvec{\zeta}),$ $\unitvec{N}_{\ell\, m}(\unitvec{\zeta}),$ $\unitvec{Z}_{\ell\, m}(\unitvec{\zeta});$ or a natural radiation basis field derived from some reference vector potential $\bv{a}(\bv{\zeta}; \omega),$ namely $\unitvec{\varepsilon}_{\bv{a}\,\perp},$ $\unitvec{s}_{\bv{a}},$ $\unitvec{b}_{\bv{a}}.$

Much of the ``quantum''-like formalism developed in the paraxial case can be generalized in the obvious way to a three-dimensional Hilbert space $\mathcal{H}_{\stext{3D}}$ consisting of square-integrable functions $\bv{g}: \realsymbol^3 \to \complexsymbol^3,$ and will not be reproduced here, except  to note that we will now denote the (possibly generalized, \ie, non-normalizable) eigenket corresponding to the vector field $\bv{g}(\bv{\zeta}; \omega)$ by $\ket{\bv{g}(\omega)},$ and the generalized 3D position/polarization eigenkets by $\ket{\bv{\zeta}; \unitvec{\epsilon}^{n}_{s}(\bv{\zeta}; \omega)},$ such that
\be
\innerp{\bv{\zeta}; \unitvec{\epsilon}^{n}_{s}(\bv{\zeta}; \omega)}{\bv{g}(\omega)} = \unitvec{\epsilon}^{n}_{s}(\bv{\zeta}; \omega)\cc \cdot \bv{g}(\bv{\zeta}; \omega),
\ee 
where the Hilbert-space inner product now involves integration over all spatial variables:
\be
\innerp{\bv{g}}{\bv{g}'} = \int d^3\bv{\zeta}\,\, \bv{g}(\bv{\zeta})\cc\!\cdot\! \bv{g}'(\bv{\zeta}).
\ee

Now, consider the Hilbert-space operator $K = K(\omega)$ defined via its spatial/polarization kernel:
\be
K(\bv{\zeta}, s; \bv{\zeta}', s'; \omega) \equiv \qamp{\bv{\zeta}; \unitvec{\epsilon}^{n}_{s}(\bv{\zeta}; \omega)}{K(\omega)}{\bv{\zeta}'; \unitvec{\epsilon}^{n}_{s'}(\bv{\zeta}; \omega)} \equiv
-\smallhalf i\omega \,\delta_{s\, s'}\, D(\bv{\zeta}; \bv{\zeta}'; \omega),
\ee
which is given explicitly by
\be
K(\bv{\zeta}, 0; \bv{\zeta}', 0; \omega) =  \frac{\omega}{4\pi} \sinc\left( k \vnorm{\bv{\zeta} - \bv{\zeta}'}\right)
= \omega^2 \sum_{\ell\, , m} j_{\ell}(k\zeta) j_{\ell}(k\zeta')\cc Y_{\ell\, m}(\unitvec{\zeta})Y_{\ell\, m}(\unitvec{\zeta}')\cc.
\ee
The kernel is seen to be real-valued and symmetric in all spin/spatial arguments, so the corresponding operator $K = K\hc$ is Hermitian.  Because the kernel acts as the identity on polarization degrees of freedom, the operator is independent of spin observables, and may be formally written as a function only of spatial observables.  In fact, because the kernel is translationally and rotationally invariant, depending  on the spatial coordinates only through $\vnorm{\bv{\zeta} - \bv{\zeta}'},$  $K$ must be a function solely of the momentum magnitude $\vnorm{\bv{P}_{\stext{3D}}} = \sqrt{-\del^2}.$   Crucially, $K$ is also positive semidefinite, in the sense that
\be\label{eqn:pos_def1}
\qamp{\bv{g}(\omega)}{K(\omega)}{\bv{g}'(\omega)} = \int \!d^3 \bv{\zeta}\!\int \!d^3 \bv{\zeta}' \,\,  K(\bv{\zeta}, 0; \bv{\zeta}', 0; \omega)\,\bv{g}(\bv{\zeta}; \omega)\cc \!\cdot\! \bv{g}(\bv{\zeta}'; \omega) \ge 0,
\ee
for any complex vector field $\bv{g}(\bv{\zeta}; \omega)$ which is sufficiently well-behaved for the integral to exist.  This can be seen most easily from the diagonal expansion in spherical waves:
\be
\qamp{\bv{g}(\omega)}{K(\omega)}{\bv{g}'(\omega)}
= \omega^2 \sum_{\ell\, , m}\sum_{s} \abs{ \int d^3 \bv{\zeta}\,\, \unitvec{\epsilon}^{n}_{s}(\bv{\zeta})\cc \!\cdot\! \bv{g}(\bv{\zeta}; \omega)  j_{\ell}(k\zeta)\cc Y_{\ell\, m}(\unitvec{\zeta})\cc }^2 \ge 0.
\ee
It is also a direct consequence of the well-known Bochner theorem, which says that a continuous function in real-space is nonnegative definite as a spatial kernel if and only if its Fourier transform is proportional to a positive measure over reciprocal space (and positive definite if this measure is everywhere non-zero).  Transforming to the scaled momentum (\ie, wavevector, or reciprocal-space) representation where the operator $K$ is diagonal, we find:
\be
K(\omega) = \pi \delta(\bv{P}_{\stext{3D}}\!\cdot\!\bv{P}_{\stext{3D}} - \omega^2),
\ee
so that
\be
\qamp{\bv{g}(\omega)}{K(\omega)}{\bv{g}'(\omega)}  = \pi \int d^{3}\bv{k}\, \abs{\bv{g}(\bv{k}; \omega)}^2\delta(\vnorm{\bv{k}}^2 -\omega^2) \ge 0,
\ee
where $\bv{k}$ is the scaled wavevector conjugate to $\bv{\xi},$ and $\bv{g}(\bv{k}; \omega)$ is the (scaled) spatial Fourier transform of $\bv{g}(\bv{\zeta}; \omega).$  Note that $K(\omega)$ acts to restrict the spectral content of any vector field $\bv{g}(\bv{\zeta}; \omega)$ in its domain to the manifold of wavevectors satisfying the vacuum dispersion relation $\omega^2 = \vnorm{\bv{k}}^2.$  However, $K$ is not idempotent, so cannot be considered a true projection.

That $K(\bv{\zeta}; \bv{\zeta}'; \omega)$ is only positive semidefinite and not strictly positive definite is a consequence of the well-known existence of non-radiating sources for the Helmholtz problem, which invariably complicate uniqueness results and inverse-scattering calculations.  Specifically, suppose we take
\be
\bv{g}_0(\bv{\zeta}; \omega) =  \bv{g}_{0\, \perp}(\bv{\zeta}; \omega) \propto \Theta( \zeta_0 - \zeta) g_{\ell}(k\zeta) \bv{X}_{\ell\, m}(\theta, \varphi),
\ee
for any allowed $\ell$ and $m,$ some $\zeta_0 > 0,$ and any well-behaved, non-trivial scalar function $g_{\ell}(k\zeta)$ which is chosen to satisfy
\begin{subequations}
\begin{align}
\int\limits_{0}^{\zeta_0} d\zeta \, \zeta^2 g_{\ell}(k\zeta) j_{\ell}(k\zeta) &= 0,\\
\int\limits_{0}^{\zeta_0} d\zeta \, \zeta^2 g_{\ell}(k\zeta) g_{\ell}(k\zeta) &> 0,
\end{align}
\end{subequations}
Then using the spherical wave expansions, it immediately follows for this source that, for all $\bv{\zeta},$
\be
\bv{\chi}_{0}(\bv{\zeta}; \omega) = \int d^3 \bv{\zeta}' \, D(\bv{\zeta}; \bv{\zeta}'; \omega) \bv{g}_{0}(\bv{\zeta}'; \omega) = \bv{0};
\ee
and also, anywhere in the extra-source region, \ie, for $\zeta > \zeta_0,$ 
\begin{subequations}
\begin{align}
\bv{a}^{\stext{out}}_{0}(\bv{\zeta}; \omega) &= \int d^3 \bv{\zeta}' \, G^{\stext{ret}}(\bv{\zeta}; \bv{\zeta}'; \omega) \bv{g}_{0}(\bv{\zeta}'; \omega) = \bv{0},\\
\bv{a}^{\stext{in}}_{0}(\bv{\zeta}; \omega) &= \int d^3 \bv{\zeta}' \, G^{\stext{adv}}(\bv{\zeta}; \bv{\zeta}'; \omega) \bv{g}_{0}(\bv{\zeta}'; \omega) = \bv{0},
\end{align}
\end{subequations}
and indeed no radiation is produced by $\bv{g}_0$ acting either as a real, time-reversed, or effective (homogeneous) source.  Any such vector field $\bv{g}_0(\bv{\zeta}; \omega)$ corresponds to a ket $\ket{\bv{g}_0(\omega)}$ in the nullspace of the operator $K;$ \ie,  $K(\omega)\ket{\bv{g}_0(\omega)} = 0.$

The nullspace of the Hermitian operator $K$ is therefore infinite-dimensional (and in fact non-denumerable).  Modulo some (admittedly thorny) normalization issues due to the unbounded nature of $K,$ the basis of homogeneous spherical waves solutions (\ref{eqn:hom_spherical}) are the orthogonal eigenfunctions of $K$ in the position basis corresponding to non-zero eigenvalue.  Although it would involve more mathematical detail than justified in this presentation, the properties and behavior of $K$ can be analyzed rigorously within the theory of Reproducing Kernel Hilbert Spaces (RKHS), where the spherical wave expansion of the kernel emerges as a consequence of Mercer's theorem \cite{weinhart:82}.  In a similar manner, we may define Green operators $G^{\stext{ret}}(\omega)$ and $G^{\stext{adv}}(\omega)$ from their spatial representations; these operators are complex-symmetric, but neither  Hermitian nor positive semidefinite.

However, the problems encountered with normalization cannot be completely ignored.  While this Hilbert space $\mathcal{H}_{\stext{3D}},$ equipped with the Euclidean/$L_2$ inner product, is the natural setting for the source-terms such as $\bv{j}(\bv{\zeta}; \omega),$ $\bv{j}_{\perp}(\bv{\zeta}; \omega),$ $\bv{g}(\bv{\zeta}; \omega),$ and $\bv{g}_{\perp}(\bv{\zeta}; \omega),$ it is not always a comfortable home for the vector potentials $\bv{a}(\bv{\zeta}; \omega)$ or $\bv{\chi}(\bv{\zeta}; \omega)$ or the corresponding fields, since the presence of far-zone radiation fields, falling off like $O(1/\zeta),$ implies that the vector potentials and fields are not normalizable with respect to this inner product, so do not strictly belong to the Hilbert space $\mathcal{H}_{\stext{3D}}.$  However, they may be approximated arbitrarily closely (point-wise) by vector fields in the Hilbert-space, and inner products between radiation and sources, \eg, $\innerp{\bv{j}(\omega)}{\bv{a}(\omega)}$ will exist for allowed current densities, which must be localized.

The lack of nomalizability may be traced to the fact that radiation fields can contain, in principle, an infinite amount of energy when integrated over all space, while transmitting only a finite amount of power.
If necessary, we could use the familiar regularizing device of introducing a finite ``quantization'' volume $V(\zeta_{Q})$ together with periodic or conducting-wall boundary conditions, and then take the limit $\zeta_{Q} \to \infty$ at the very end of the calculation.   For the currently envisioned  application, this is not terribly convenient, as we are interested primarily in the far-field from the start.  Alternatively, we can choose a different inner product, which effectively normalizes the radiation fields based on power (or power spectral density) rather than energy.  We can define a Hilbert space $\mathcal{H}_{\stext{out}}$ spanned by the basis consisting of outgoing solenoidal vector spherical waves, with an inner product defined as a Euclidian dot product on the multipole expansion coefficients:
\be
\innerpa{\bv{a}(\omega)}{\bv{a}'(\omega)}_{\stext{out}} =  \sum_{\ell\, m} \bigl[   a^{\stext{E}\ast}_{\ell\, m}(\omega)a'^{\stext{E}}_{\ell\, m}(\omega) 
+   a^{\stext{M}\ast}_{\ell\, m}(\omega)a'^{\stext{M}}_{\ell\, m}(\omega) \bigr].
\ee
The exact relationship between this inner product and the scaled power spectral density will be seen shortly, but it should be apparent that this quantity may be expressed soley in in terms of far-field data, rather than on the field throughout all space, and physical radiation fields, with finite power, will be normalizable in the sense $\innerpa{\bv{a}(\omega)}{\bv{a}(\omega)} < \infty.$  An isomorphic Hilbert space $\mathcal{H}_{\stext{in}} \equiv  \mathcal{H}_{\stext{out}}\cc \cong \mathcal{H}_{\stext{out}}$ may be defined in an exactly analogous manner for incoming-wave solutions, such as $\bv{a}(\bv{\zeta}; \omega)\cc.$  Complex conjugation provides a natural isometry between them, or instead one may use the mapping which simply swaps the spherical Hankel functions $h^{(1)}(kr)$ and $h^{(2)}(kr)$ in the spherical wave expansions without changing the multipole expansion coefficients or the vector spherical harmonics.  Strictly outside the support of the sources (if any), any solenoidal solution $\bv{\psi}$ to the vector Helmholtz equation (homogeneous or inhomogeneous), of finite power but with otherwise arbitrary boundary conditions, may be expressed as the superposition of incoming and outgoing spherical waves, so it necessarily lies in the Hilbert space $\mathcal{H}_{\stext{out}} \oplus \mathcal{H}_{\stext{in}}.$  Thus, we can in this case uniquely decompose (for $\zeta > \zeta_2$) any such solution $\bv{\psi}(\bv{\zeta}; \omega)$ as
\be
\bv{\psi}(\bv{\zeta}; \omega) = \bv{\psi}^{\stext{out}}(\bv{\zeta}; \omega) + \bv{\psi}^{\stext{in}}(\bv{\zeta}; \omega),
\ee 
where $\bv{\psi}^{\stext{out}}$  and  $\bv{\psi}^{\stext{in}}$ are, respectively, the outgoing and incoming components of the vector field, in the sense that they satisfy the appropriate Sommerfeld boundary conditions as $\zeta \to \infty.$
The inherited inner product is given by
\be
\innerpa{\bv{\psi}(\omega)}{\bv{\psi}'(\omega)} = \innerpa{\bv{\psi}^{\stext{out}}(\omega)}{\bv{\psi}'^{\stext{out}}(\omega)}_{\stext{out}} + \innerpa{\bv{\psi}^{\stext{in}}(\omega)}{\bv{\psi}'^{\stext{in}}(\omega)}_{\stext{in}}.
\ee
Our approximating ``trial'' vector potentials $\bv{\chi}$ will lie in the proper Hilbert subspace  $\mathcal{H}_{\stext{hom}} \subsetneq  \mathcal{H}_{\stext{out}} \oplus \mathcal{H}_{\stext{in}},$ consisting of source-free solutions which are regular everywhere in space.  This space of homogeneous solutions is also isomorphic to the outgoing or incoming wave spaces: $\mathcal{H}_{\stext{hom}} \cong \mathcal{H}_{\stext{out}} \cong \mathcal{H}_{\stext{in}},$ as can be seen by imagining the mapping that replaces, in the spherical wave expansions, each real-valued spherical Bessel function $j_{\ell}(kr)$ with the corresponding complex spherical Hankel function $h^{(1)}(kr)$ or $h^{(2)}(kr),$ respectively, without altering the angular dependence or the expansion coefficients.

\subsection{Energy Balance and Poynting Flux}

Using the Helmholtz equation, gauge constraint, and various vector identities, one can derive a generalized impedance relation, or complexified power-balance equation, for the fields in the frequency domain over any arbitrary spatial region $R,$ and valid for any consistent boundary conditions imposed on the radiation at spatial infinity:
\be\label{eqn:poynting2}
\int\limits_{R} \!d^3 \bv{\zeta}\, \left[\bv{j}_{\perp}(\bv{\zeta}; \omega)\cc\!\cdot\!\bv{\varepsilon}_{\bv{a}\, \perp}(\bv{\zeta}; \omega)\right]
+i\omega \!\int\limits_{R}\! d^3\! \bv{\zeta}\, \left[ \vnorm{\bv{\varepsilon}_{\bv{a}\, \perp}(\bv{\zeta}; \omega)}^2 - \vnorm{\bv{b}_{\bv{a}}(\bv{\zeta}; \omega)}^2\right] 
+ \int\limits_{\del R}\!d^{2}\sigma\, \left[\unitvec{n} \!\cdot\! \bv{s}_{\bv{a}}(\bv{\zeta}; \omega)\right] = 0,
\ee
where $\unitvec{n} = \unitvec{n}(\bv{\zeta})$ is a outward normal unit vector on the boundary surface at $\bv{\zeta} \in \del R.$ This relation is quite  similar to the usual conservation law derived in textbooks from the time-harmonic formulation of Maxwell's equations \cite{harrington:61, jackson:75}, except that it involves only the solenoidal components of the sources and the fields -- the longitudinal component $\bv{j}_{\|}$ of the current density and the scalar potential $\phi$ and its derivatives do not appear.

In the most general, inhomogeneous case, this relation implies:
\be
-\realpart \int\limits_{R} d^3 \bv{\zeta}\,  \left[\bv{j}_{\perp}\cc\!\cdot\!\bv{\varepsilon}_{\bv{a}\, \perp}\right]
= \realpart \int\limits_{\del R}d^{2}\sigma\,\, \left[\unitvec{n} \!\cdot\! \bv{s}_{\bv{a}}\right],
\ee
which is an expression of time-averaged energy conservation, relating the (negative) rate of mechanical work done on the solenoidal sources within $R$ to the rate of energy escape in the form of radiation through the boundary $\del R.$  If we let $R = V(\zeta_2)$ for any $\zeta_2 > \zeta_1,$ we can replace $\bv{j}_{\perp}$ with the the full current density $\bv{j}$ on the right-hand-side, and this becomes
\be
-\tfrac{\del}{\del \omega}\mathscr{P}_{\stext{mech}}[ \bv{\varepsilon}_{\bv{a}\,\perp}; \bv{j}](\omega) 
\equiv -\realpart \int d^3 \bv{\zeta}\,\,  \left[\bv{j}\cc\!\cdot\!\bv{\varepsilon}_{\bv{a}\, \perp}\right] =
\realpart \!\!\int\limits_{\del V(\zeta_2)}\!\!d^{2}\sigma\,\, \left[\unitvec{n} \!\cdot\! \bv{s}_{\bv{a}}\right] \equiv \tfrac{\del}{\del \omega}\mathscr{P}_{\stext{EM}}[\bv{a}](\zeta_2; \omega) \ee
The radiated power spectral density is independent of radius $\zeta_2$ as long as $\zeta_2 > \zeta_1,$ as assumed, so we may take $\zeta_2 \to \infty$ on the left-hand-side whenever convenient.  Using Green's Second Identity and other vector identities, the spectral density of the radiative flux can also be written in the following useful forms within the Coulomb gauge:
\be
\label{eqn:radiated_power_alt}
\tfrac{\del}{\del \omega}\mathscr{P}_{\stext{EM}}[\bv{a}](\zeta_2; \omega)  = \frac{i\omega}{2}\!\! \int\limits_{V(\zeta_2)}\!\!d^{3}\bv{\zeta}\,\left[\bv{a}\!\cdot\!\laplacian \bv{a}\cc - \bv{a}\cc\!\cdot\!\laplacian\bv{a} \right] = 
\frac{i\omega}{2}\!\!\int\limits_{\del V(\zeta_2)}\!\!d^{2}\sigma\, \left[ \bv{a}\!\cdot\!(\unitvec{n}\!\cdot\!\bv{\del})\bv{a}\cc -      \bv{a}\cc\!\cdot\!(\unitvec{n}\!\cdot\!\bv{\del})\bv{a} \right]. 
\ee
Similarly, if $\bv{a}(\bv{\zeta}; \omega)$ is the exact solenoidal radiation corresponding to $\bv{j}_{\perp}(\bv{\zeta}; \omega),$ and $\bv{\chi}(\bv{\zeta}; \omega)$ is any arbitrary solenoidal solution to the source-free Helmholtz equation, then for $\zeta_2 > \zeta_1,$
\be\label{eqn:work_alt2}
\begin{split}
-\tfrac{\del}{\del \omega}\mathscr{P}_{\stext{mech}}[ \bv{\varepsilon}_{\bv{\chi}\,\perp}; \bv{j}](\omega) &=
\frac{i\omega}{2}\int\limits_{V(\zeta_2)}\!\!d^{3}\bv{\zeta}\,\left[\bv{\chi}\!\cdot\!\del^2 \bv{a}\cc - \bv{a}\cc\!\cdot\!\del^2\bv{\chi}
+ \bv{a}\!\cdot\!\del^2 \bv{\chi}\cc - \bv{\chi}\cc\!\cdot\!\del^2\bv{a}\right]\\
& =  \frac{i\omega}{2}\!\!\int\limits_{\del V(\zeta_2)}\!\!d^{2}\sigma\, \left[ \bv{\chi}\!\cdot\!(\unitvec{n}\!\cdot\!\bv{\del})\bv{a}\cc -      \bv{a}\cc\!\cdot\!(\unitvec{n}\!\cdot\!\bv{\del})\bv{\chi}
+ \bv{a}\!\cdot\!(\unitvec{n}\!\cdot\!\bv{\del})\bv{\chi}\cc -   \bv{\chi}\cc\!\cdot\!(\unitvec{n}\!\cdot\!\bv{\del})\bv{a} \right]. 
\end{split}
\ee

For any such solution $\bv{\chi}(\bv{\zeta}; \omega)$ to the homogeneous Helmholtz equation (\ie, where $\del^2\bv{\chi} = -\omega^2\bv{\chi}$), the Lorentz invariant $\bigl[\vnorm{\bv{\varepsilon}_{\bv{\chi}\,\perp}    }^2 - \vnorm{ \bv{b}_{\bv{\chi}}}^2\bigr]$ vanishes everywhere (as can be seen easily, for example, from an expansion in transverse plane waves, or from the spherical wave basis), and from (\ref{eqn:poynting2}) or  (\ref{eqn:work_alt2}) we have:
\be
\int\limits_{\del R}d^{2}\sigma\,\, \left[\unitvec{n} \!\cdot\! \bv{s}_{\bv{\chi}}\right] =
\int\limits_{R} d^3 \bv{\zeta}\,\, \left[\bv{\del}\!\cdot\!\bv{s}_{\bv{\chi}}\right] = \frac{i\omega}{2} \!\!\int\limits_{R} d^3 \bv{\zeta}\,\, \left[\bv{\chi}\!\cdot\!\del^2\bv{\chi}\cc -  \bv{\chi}\cc\!\cdot\!\del^2\bv{\chi}\right] = 0.
\ee
That is, both real and imaginary parts of the Poynting flux vanish when integrated over any closed surface, reflecting the fact that, as is intuitively evident from the convolution representation (\ref{eqn:fundamental_sol}) or from a plane-wave decomposition, any vector potential which satisfies the homogeneous wave equation everywhere must, in fact, result in an equal amount of averaged radiative flux passing into and out from any closed surface, without any dissipation of field energy or reactive storage of energy in localized, quasi-static fields.  

As we saw above, however,  we can locally decompose the homogeneous fields into what represent, or at least eventually (as $\zeta \to \infty$)  become, separately outwardly-radiating and inwardly-radiating waves:
\be
\bv{\chi}(\bv{\zeta}; \omega) = \bv{\chi}^{\stext{out}}(\bv{\zeta}; \omega) + \bv{\chi}^{\stext{in}}(\bv{\zeta}; \omega),
\ee
where for all  $\bv{\zeta} \in \del V(\zeta_2),$ $\realpart \left[\unitvec{n}(\bv{\zeta})\cdot  \bv{s}^{\stext{out}}_{\bv{\chi}}\right] \ge 0$ and  $\realpart \left[\unitvec{n}(\bv{\zeta})\cdot  \bv{s}^{\stext{in}}_{\bv{\chi}}\right] \le 0$ as $\zeta_2 \to \infty.$  This is most easily accomplished simply by separating the fundamental representation into retarded (outgoing) and advanced (incoming) parts:
\be
\bv{\chi}^{\stext{out}}(\bv{\zeta}; \omega)  = \phantom{-}\int d^3 \bv{\zeta}'\, G^{\stext{ret}}(\bv{\zeta}; \bv{\zeta}'; \omega) \bv{g}_{\perp}(\bv{\zeta}'; \omega);
\ee
and
\be
\bv{\chi}^{\stext{in}}(\bv{\zeta}; \omega)  = -\int d^3 \bv{\zeta}'\, G^{\stext{adv}}(\bv{\zeta}; \bv{\zeta}'; \omega) \bv{g}_{\perp}(\bv{\zeta}'; \omega).
\ee
Note that these vector potentials separately satisfy the homogeneous equation only for $\zeta > \zeta_0,$ but satisfy the inhomogeneous equation everywhere for the appropriate effective source terms and specified boundary conditions.  Although $\bv{\chi}^{\stext{in}}(\bv{\zeta}; \omega)$ involves the time-reversed (advanced) Green function, it is not the time-reversal of  $\bv{\chi}^{\stext{out}}(\bv{\zeta}; \omega),$ because it involves the negation, not complex conjugation, of the effective source $g_{\perp}(\bv{\zeta}; \omega).$  Indeed, the conjugate wave $\bv{\chi}^{\stext{in}}(\bv{\zeta}; \omega)\cc$ may be interpreted as  the usual outgoing radiation from the effective source $-\bv{g}_{\perp}(\bv{\zeta}'; \omega)\cc.$

Using the symmetry of the Green functions and a simple change of variables, it is easily shown that for any $\zeta_2 > \zeta_0,$
\be
\begin{split}
\tfrac{\del}{\del \omega}\mathscr{P}_{\stext{EM}}[\bv{\chi}^{\stext{out}}](\zeta_2; \omega) &=
-\tfrac{\del}{\del \omega}\mathscr{P}_{ \text{\tiny{mech}}}[ \bv{\varepsilon}^{\stext{out}}_{\bv{\chi}\,\perp}; \bv{g}](\omega)
 =\!\int \!d^3 \bv{\zeta}\!\int \!d^3 \bv{\zeta}' \,\,  K(\bv{\zeta}; \bv{\zeta}'; \omega)\,\bv{g}_{\perp}(\bv{\zeta}; \omega)\cc \!\cdot\! \bv{g}_{\perp}(\bv{\zeta}'; \omega)\\ &= \qamp{\bv{g}_{\perp}(\omega)}{K(\omega)}{\bv{g}_{\perp}(\omega)}   \ge 0,
\end{split}
\ee
which vanishes if and only if $\bv{g}_{\perp}(\bv{\zeta}; \omega)$ is a non-radiating source at frequency $\omega.$
In the same manner, we also find
\be
\begin{split}
\tfrac{\del}{\del \omega}\mathscr{P}_{\stext{EM}}[\bv{\chi}^{\stext{in}}](\zeta_2; \omega) &=
-\tfrac{\del}{\del \omega}\mathscr{P}_{ \text{\tiny{mech}}}[ \bv{\varepsilon}^{\stext{in}}_{\bv{\chi}\,\perp}; -\bv{g}](\omega) 
= -\!\int \!d^3 \bv{\zeta}\!\int \!d^3 \bv{\zeta}' \,\,  K(\bv{\zeta}; \bv{\zeta}'; \omega)\,\bv{g}_{\perp}(\bv{\zeta}; \omega)\cc \!\cdot\! \bv{g}_{\perp}(\bv{\zeta}'; \omega)\\
&= -\qamp{\bv{g}_{\perp}(\omega)}{K(\omega)}{\bv{g}_{\perp}(\omega)} \le 0,
\end{split}
\ee
also vanishing if and only if $\bv{g}_{\perp}(\bv{\zeta}; \omega)$ is a non-radiating source.  Thus, we see that, as anticipated,
\be\label{eqn:hom_power_2}
\tfrac{\del}{\del \omega}\mathscr{P}_{\stext{EM}}[\bv{\chi}^{\stext{out}}](\zeta_2; \omega) = -\tfrac{\del}{\del \omega}\mathscr{P}_{\stext{EM}}[\bv{\chi}^{\stext{in}}](\zeta_2; \omega) =
\tfrac{\del}{\del \omega}\mathscr{P}_{\stext{EM}}[\bv{\chi}^{\stext{in}\ast}](\zeta_2; \omega) \ge 0.
\ee
It then follows that
\be\label{eqn:power_balance_2}
\begin{split}
-\smallhalf\tfrac{\del}{\del \omega}\mathscr{P}_{ \text{\tiny{mech}}}[ \bv{\varepsilon}_{\bv{\chi}\,\perp}; \bv{g}](\omega) &=
-\smallhalf\realpart \int d^3 \bv{\zeta} \, \left[\bv{g}\cc\!\cdot\!\bv{\varepsilon}_{\bv{\chi}\, \perp}\right]
= \smallhalf\!\int d^3 \bv{\zeta} \,\, \left[\bv{g}\cc\!\cdot\!\bv{\varepsilon}_{\bv{\chi}\, \perp}\right] \\
&=  \smallhalf\!\int \!d^3 \bv{\zeta}\!\int \!d^3 \bv{\zeta}' \,\,  i\omega D(\bv{\zeta}; \bv{\zeta}'; \omega)\,\bv{g}_{\perp}(\bv{\zeta}; \omega)\cc \!\cdot\! \bv{g}_{\perp}(\bv{\zeta}'; \omega)\\
&= \int \!d^3 \bv{\zeta}\!\int \!d^3 \bv{\zeta}' \,\,  K(\bv{\zeta}; \bv{\zeta}'; \omega)\,\bv{g}_{\perp}(\bv{\zeta}; \omega)\cc \!\cdot\! \bv{g}_{\perp}(\bv{\zeta}'; \omega)\\
&= \qamp{\bv{g}_{\perp}(\omega}{K(\omega)}{\bv{g}_{\perp}(\omega)} = -\tfrac{\del}{\del \omega}\mathscr{P}_{ \text{\tiny{mech}}}[ \bv{\varepsilon}^{\stext{out}}_{\bv{\chi}\,\perp}; \bv{g}](\omega) \ge 0,
\end{split}
\ee
which is the generalization of the paraxial result (\ref{eqn:paraxial_energy_balance_2}).

If we take $\bv{g}(\bv{\zeta}; \omega) = \bv{j}(\bv{\zeta}; \omega)$ to be the physical current density of interest, then the homogeneous approximation 
\be
\bv{\chi}_{\bv{j}}(\bv{\zeta}; \omega) = \int d^3 \bv{\zeta}' \, D(\bv{\zeta}; \bv{\zeta}'; \omega) \bv{j}_{\perp}(\bv{\zeta}'; \omega)
\ee
agrees exactly in its outgoing components with the actual radiation:
\be
\bv{\chi}_{\bv{j}}^{\stext{out}}(\bv{\zeta}; \omega) = \bv{a}(\bv{\zeta}; \omega),
\ee
but in order to actually satisfy the homogeneous Helmholtz equation everywhere, it must possess an equal magnitude of net power in the corresponding (\ie, conjugate) incoming modes making up $\bv{\chi}^{\stext{in}}(\bv{\zeta}; \omega),$ power which will flow in the opposite direction through any surface enclosing the sources.  This extra power appears as extra work which would be done by the actual source charges on these incoming fields, if they were present.  (These advanced fields do not lead to absorption by the sources making up $\bv{j}$ because, again, they are the time-reversal of the retarded fields which would be produced by $-\bv{j}\cc,$ not $\bv{j}$ itself.)  Using Green's identities and the governing Helmholtz equations, one can show that this homogeneous solution may also be formally expressed in the Kirchoff-integral form, only with the Green function $G^{\stext{ret}}(\bv{\zeta}; \bv{\zeta}'; \omega)$ replaced with the fundamental homogeneous solution $D(\bv{\zeta}; \bv{\zeta}'; \omega):$
\be\label{eqn:kirchoff_hom}
\bv{\chi}_{\bv{j}}(\bv{\zeta}; \omega) = \!\!\int\limits_{\del V(\zeta_3)} \!\!\!d\sigma' \, \bigl[ \bv{a}(\bv{\zeta}'; \omega)(\unitvec{n}'\!\cdot\!\bv{\del}') D(\bv{\zeta}; \bv{\zeta}'; \omega) - D(\bv{\zeta}; \bv{\zeta}'; \omega)(\unitvec{n}'\!\cdot\!\bv{\del}')\bv{a}(\bv{\zeta}'; \omega) \bigr].
\ee
Unlike the inhomogeneous case, this remains valid for any boundary radius $\zeta_3 > 0$ and any observation position $\bv{\zeta}.$

Not surprisingly, the homogeneous approximation $\bv{\chi}_{\bv{j}}(\bv{\zeta}; \omega)$ to  $\bv{a}(\bv{\zeta}; \omega)$ is just the result of the above-mentioned one-to-one mapping between the isomorphic spaces $\mathcal{H}_{\stext{out}}$ and  $\mathcal{H}_{\stext{hom}},$ involving the replacement of each ``outgoing'' spherical Hankel function $h^{(1)}_{\ell}(k\zeta)$ with the corresponding ``standing-wave'' spherical Bessel function $j_{\ell}(k\zeta)$ in the spherical wave expansion for $\bv{a}(\bv{\zeta}; \omega),$ without altering any of the multipole expansion coefficients or basis vector fields for the angular dependence.  Schematically, we have:
\begin{subequations}
\begin{align}
\mathcal{H}_{\stext{out}} &\to \mathcal{H}_{\stext{hom}}\\
\bv{a}(\bv{\zeta}; \omega) &\to \bv{\chi}_{\bv{j}}(\bv{\zeta}; \omega)\\
h^{(1)}_{\ell}(k\zeta) &\to j_{\ell}(k\zeta)\\
\bv{W}_{\ell\, m}(\unitvec{\zeta}) &\to \bv{W}_{\ell\, m}(\unitvec{\zeta})\\
a^{\stext{E}}_{\ell\, m}(\omega) &\to \chi^{\stext{E}}_{\ell\, m}(\omega) =  a^{\stext{E}}_{\ell\, m}(\omega)\\
a^{\stext{M}}_{\ell\, m}(\omega) &\to \chi^{\stext{M}}_{\ell\, m}(\omega) =  a^{\stext{M}}_{\ell\, m}(\omega).
\end{align}
\end{subequations}
We expect that this $\bv{\chi}_{\bv{j}}(\bv{\zeta}; \omega)$ is in some sense the closest possible homogeneous approximation to $\bv{a}(\bv{\zeta}; \omega),$ which will be confirmed below by the non-paraxial form of the Maximum-Power Variational Principle.

We can also express the various radiated power spectral densitities directly in terms of the multipole expansion coefficients or associated inner product, by using (\ref{eqn:radiated_power_alt}) in the limit $\zeta_2 \to \infty.$  Asymptotically, \ie, for $kr \gg \ell,$ the limiting forms of the spherical Bessel, Neumann, and Hankel functions are:
\begin{subequations}
\begin{align}
j_{\ell}(kr) &\to \phantom{-}\frac{\sin(kr - \tfrac{\ell \pi}{2} )}{kr},\\
n_{\ell}(kr) &\to -\frac{\cos(kr - \tfrac{\ell \pi}{2} )}{kr},\\
h_{\ell}^{(1)}(kr) &\to (-i)^{\ell+1}\,\frac{e^{+ikr}}{kr},\\
h_{\ell}^{(2)}(kr) &\to (+i)^{\ell+1}\,\frac{e^{-ikr}}{kr}.
\end{align}
\end{subequations} 
Examining the form of the Sommerfeld radiation conditions, we see that, asymptotically as $k\zeta \to \infty,$ any vector potential $\bv{\psi}(\bv{\zeta}; \omega)$ (not necessarily a source-free solution) may be locally decomposed into its incoming and outgoing components on the spherical surface $\del B(\zeta)$ by the projections:
\begin{subequations}\label{eqn:in_out_project}
\begin{align}
\bv{\psi}^{\stext{out}}(\bv{\zeta}; \omega) &\to \smallhalf \left(1 - \tfrac{i}{k}\tfrac{\del}{\del \zeta} \right)\bv{\psi}(\bv{\zeta}; \omega),\\
\bv{\psi}^{\stext{in}}(\bv{\zeta}; \omega) &\to \smallhalf \left(1 + \tfrac{i}{k}\tfrac{\del}{\del \zeta}\right)\bv{\psi}(\bv{\zeta}; \omega).
\end{align}
\end{subequations}
Equipped with these results, it is straightforward to show that for any solution $\bv{\psi} \in \mathcal{H}_{\stext{out}} \oplus \mathcal{H}_{\stext{in}},$ the outgoing power spectral density is given by
\be\label{eqn:radiated_power_alt2}
\begin{split}
\tfrac{\del}{\del \omega}\mathscr{P}_{\stext{EM}}[\bv{\psi}^{\stext{out}}](\infty; \omega)  &= 
\lim_{\zeta_2 \to \infty} \frac{i\omega}{2}\!\!\!\int\limits_{\del B(\zeta_2)}\!\!d^{2}\sigma\, \bigl[ \bv{\psi}^{\stext{out}}\!\cdot\!(\unitvec{n}\!\cdot\!\bv{\del})\bv{\psi}^{\stext{out}\ast} -      \bv{\psi}^{\stext{out}\ast}\!\cdot\!(\unitvec{n}\!\cdot\!\bv{\del})\bv{\psi}^{\stext{out}} \bigr]\\
&= \lim_{\zeta_2 \to \infty} \biggl[\frac{i\omega}{8}\!\!\!\int\limits_{\del B(\zeta_2)}\!\!\!d^{2}\sigma\, \bigl[(1 - \tfrac{i}{k}\tfrac{\del}{\del \zeta})\bv{\psi}\!\cdot\!(\tfrac{\del}{\del \zeta} - \tfrac{i}{k}\tfrac{\del^2}{\del \zeta^2})\bv{\psi}\cc  \bigr] \biggr] + c.c\\
&= \lim_{\zeta_2 \to \infty} \left[\frac{i\omega}{8} \,\zeta_2^2  \int d\varphi \int d\theta\, \sin\theta \bigl[(1 - \tfrac{i}{k}\tfrac{\del}{\del \zeta})\bv{\psi}\!\cdot\!(\tfrac{\del}{\del \zeta} - \tfrac{i}{k}\tfrac{\del^2}{\del \zeta^2})\bv{\psi}\cc  \bigr] \right] + c.c,
\end{split}
\ee
which, after a little algebra, reduces to the inner product over the outgoing multipole coefficients:
\be
\label{eqn:power_out_alt1}
\tfrac{\del}{\del \omega}\mathscr{P}_{\stext{EM}}[\bv{\psi}^{\stext{out}}](\infty; \omega) = 
\sum_{\ell\, m} \left[ \abs{\psi^{\stext{E (out)}}_{\ell\, m}(\omega)}^2 +    \abs{\psi^{\stext{M (out)}}_{\ell\, m}(\omega)}^2 \right] = \innerpa{\bv{\psi}^{\stext{out}}(\omega)}{\bv{\psi}^{\stext{out}}(\omega)}_{\stext{out}} \ge 0,
\ee
with equality (vanishing outgoing power) in the last step if and only if $\bv{\psi}^{\stext{out}}(\bv{\zeta}; \omega)$ vanishes identically.
Similarly, we find
\be\label{eqn:power_in_alt}
-\tfrac{\del}{\del \omega}\mathscr{P}_{\stext{EM}}[\bv{\psi}^{\stext{in}}](\infty; \omega) = \innerpa{\bv{\psi}^{\stext{in}}(\omega)}{\bv{\psi}^{\stext{in}}(\omega)}_{\stext{in}}  \ge 0,
\ee
so we may write
\be
\begin{split}
\innerpa{\bv{\psi}(\omega)}{\bv{\psi}(\omega)} &\equiv 
\innerpa{\bv{\psi}^{\stext{out}}(\omega)}{\bv{\psi}^{\stext{out}}(\omega)}_{\stext{out}} +\innerpa{\bv{\psi}^{\stext{in}}(\omega)}{\bv{\psi}^{\stext{in}}(\omega)}_{\stext{in}}\\ &= 
\tfrac{\del}{\del \omega}\mathscr{P}_{\stext{EM}}[\bv{\psi}^{\stext{out}}](\infty; \omega) - \tfrac{\del}{\del \omega}\mathscr{P}_{\stext{EM}}[\bv{\psi}^{\stext{in}}](\infty; \omega)\\ &=  \abs{\tfrac{\del}{\del \omega}\mathscr{P}_{\stext{EM}}[\bv{\psi}^{\stext{out}}](\infty; \omega)} + \abs{\tfrac{\del}{\del \omega}\mathscr{P}_{\stext{EM}}[\bv{\psi}^{\stext{in}}](\infty; \omega)}.
\end{split}
\ee

The inner products previously introduced for $\mathcal{H}_{\stext{out}}$ and $\mathcal{H}_{\stext{in}},$ and $\mathcal{H}_{\stext{out}} \oplus \mathcal{H}_{\stext{in}}$ are just equal, respectively, to the (scaled) outgoing power spectral density, the absolute value of the incoming spectral density, and the magnitude of total (not net) in-flowing and out-flowing power spectral density, passing through an arbitrarily remote boundary somewhere  outside the support of the (real or effective) sources.  Choosing $\bv{\psi} = \bv{\chi}_{\bv{j}},$  the relations (\ref{eqn:power_out_alt1}) and (\ref{eqn:power_in_alt}) provide an explicit verfication of (\ref{eqn:hom_power_2}).  As an aside, we note that these separate incoming and outgoing power spectral densities and their (incoherent) sum, are often of more physical interest than the net power $\tfrac{\del}{\del \omega}\mathscr{P}_{\stext{EM}}[\bv{\psi}](\infty; \omega).$  For UV and lower frequencies, detectors  (such as photomultipliers or solid state photoelectric devices) are typically directionally sensitive, in the sense of an acceptance solid angle  $\Omega \le 2\pi,$ so measurements of the power in the radiation $\bv{\psi}$ will typically reveal something closer to either $\abs{\tfrac{\del}{\del \omega}\mathscr{P}_{\stext{EM}}[\bv{\psi}^{\stext{out}}](\infty; \omega)}$ or $\abs{\tfrac{\del}{\del \omega}\mathscr{P}_{\stext{EM}}[\bv{\psi}^{\stext{in}}](\infty; \omega)},$ depending on the orientation, (or some fraction thereof, with less than full $4\pi$ solid-angular coverage of the localized source), rather than $\tfrac{\del}{\del \omega}\mathscr{P}_{\stext{EM}}[\bv{\psi}](\infty; \omega).$  In the X-ray regime, shielding is difficult and detectors are often not directionally sensitive, and the response will likely depend on the total fluence into the detector volume, \ie, on $\abs{\tfrac{\del}{\del \omega}\mathscr{P}_{\stext{EM}}[\bv{\psi}^{\stext{out}}](\infty; \omega)} + \abs{\tfrac{\del}{\del \omega}\mathscr{P}_{\stext{EM}}[\bv{\psi}^{\stext{in}}](\infty; \omega)}$ rather than $\tfrac{\del}{\del \omega}\mathscr{P}_{\stext{EM}}[\bv{\psi}](\infty; \omega).$

In a similar fashion, we may formally express  the spectral density of mechanical work in terms of the multipole coefficients for the exact radiation $\bv{a}$ and the homogeneous approximant $\bv{\chi},$ by using (\ref{eqn:work_alt2}), choosing a spherical boundary $\del B(\zeta_2) $ and letting $\zeta_2 \to \infty.$  After the algebraic dust settles, we find
\be\label{eqn:work_inner_alt}
\begin{split}
-\smallhalf\tfrac{\del}{\del \omega}\mathscr{P}_{ \text{\tiny{mech}}}[ \bv{\varepsilon}_{\bv{\chi}\,\perp}; \bv{j}](\omega) &=
 \realpart \sum_{\ell\, m} \left[ \chi^{\stext{E}}_{\ell\, m}(\omega)\cc a^{\stext{E}}_{\ell\, m}(\omega) + \chi^{\stext{M}}_{\ell\, m}(\omega)\cc a^{\stext{M}}_{\ell\, m}(\omega) \right] \\
&=  \realpart \innerpa{\bv{\chi}^{\stext{out}}(\omega)}{\bv{a}(\omega)}_{\stext{out}} 
\end{split}
\ee
In effect, we have managed to re-express the three-dimensional, ``volume''  inner product $\innerp{\bv{j}(\omega)}{\bv{a}(\omega)}$ in terms of the two-dimensional, ``boundary'' inner product $\innerpa{\bv{\chi}^{\stext{out}}(\omega)}{\bv{a}(\omega)}_{\stext{out}}.$   If we choose $\bv{\chi} = \bv{\chi}_{\bv{j}},$ then by applying (\ref{eqn:power_out_alt1}), the power spectrial density in the form (\ref{eqn:work_inner_alt}) is seen to satisfy (\ref{eqn:power_balance_2}) explicitly.
 
\subsection{Variational Principle}

Our general variational principle follows directly from the nonnegative-definiteness of the operator $K(\omega),$ \ie, equation (\ref{eqn:pos_def1}), and the above energy conservation considerations.  Re-tracing the standard proof of the Cauchy-Schwarz inequality, we find that, with a little extra care, it goes through almost unchanged if we replace the positive definite inner product with a positive-semidefinite pseudo-product.  In particular, with the bilinear form associated with 
$K(\omega),$ one can show
\be
\abs{ \qamp{\bv{g}(\omega)}{K(\omega)}{\bv{g}'(\omega)}   }^2  \le   \qamp{\bv{g}(\omega)}{K(\omega)}{\bv{g}(\omega)}  \qamp{\bv{g}'(\omega)}{K(\omega)}{\bv{g}'(\omega)},
\ee
with strict equality if and only if there exist constants $\delta,\, \delta',\,\delta_0 \in \complexsymbol,$ not all zero, and a non-radiating source $\bv{g}_0(\bv{\zeta}; \omega)$ which is not identically zero but is otherwise arbitrary,  such that 
\be
\delta\, \bv{g}(\bv{\zeta}; \omega) + \delta'\,\bv{g}'(\bv{\zeta}; \omega) + \delta_0\,\bv{g}_{0}(\bv{\zeta}; \omega) = \bv{0}. 
\ee
We also have, for any complex number $c \in \complexsymbol,$ the inequalities
\be
\realpart [ c ] \le \abs{ \realpart [ c ] } \le \abs{c},
\ee
with overall equality if and only if $c$ is real and nonnegative.
Choosing $\bv{g}'= \bv{j}_{\perp}(\bv{\zeta}; \omega)$ to be the solenoidal source responsible for the actual radiation  $\bv{a}(\bv{\zeta}; \omega)$ and $\bv{g} = \bv{g}_{\perp}(\bv{\zeta}; \omega; \bv{\alpha})$ to be an effective source corresponding to a solenoidal, source-free trial vector field $\bv{\chi}(\bv{\zeta}; \omega;  \bv{\alpha})$ depending on certain variational parameters collected in the vector $\bv{\alpha},$ these inequalities yield:
\be
\begin{split}
-\smallhalf \tfrac{\del}{\del \omega}\mathscr{P}_{\stext{mech}}[ \bv{\varepsilon}_{\bv{\chi}\,\perp}; \bv{j}](\omega; \bv{\alpha}) &\le
\smallhalf \abs{\tfrac{\del}{\del \omega}\mathscr{P}_{\stext{mech}}[ \bv{\varepsilon}_{\bv{\chi}\,\perp}; \bv{j}](\omega; \bv{\alpha}) }\\
&\le 
\bigl[  \tfrac{\del}{\del \omega}\mathscr{P}_{\stext{EM}}[\bv{\chi}^{\stext{out}}](\zeta_2; \omega; \bv{\alpha}) \bigr]^{1/2}
\bigl[  \tfrac{\del}{\del \omega}\mathscr{P}_{\stext{EM}}[\bv{a}](\zeta_2; \omega) \bigr]^{1/2},
\end{split}
\ee
with overall equality assuming $\tfrac{\del}{\del \omega}\mathscr{P}_{\stext{EM}}[\bv{a}](\zeta_2; \omega) \neq 0,$ if and only if 
\be
\bv{\chi}^{\stext{out}}(\bv{\zeta}; \omega; \bv{\alpha}) \propto \bv{a}(\bv{\zeta}; \omega).
\ee
(In the trivial case where $\tfrac{\del}{\del \omega}\mathscr{P}_{\stext{EM}}[\bv{a}](\zeta_2; \omega) = 0,$ equality is achieved trivially for $\bv{\chi}(\bv{\zeta}; \omega) = \bv{0}$ or indeed for any $\bv{\chi}(\bv{\zeta}; \omega)$ which is Hilbert space orthogonal to $\bv{j}_{\perp}(\bv{\zeta}; \omega).$)

In practice, we can obtain an actual variational approximation with one of two different but ultimately equivalent procedures, each arriving at the same final result when optimizing over equivalent families of trial solutions.  In the first approach, we may partition the variational parameters as $\bv{\alpha} = \left(\eta_0, \theta_0, \bv{\alpha}' \right)\trans,$ where $\bv{\alpha}' = \left(\alpha'_1, \alpha'_2, \ldots, \right)\trans,$ and consider a restricted  family of solenoidal, homogeneous trial solutions $\bv{v}(\bv{\zeta}; \omega; \bv{\alpha}')$ normalized to unit power:
\be
\tfrac{\del}{\del \omega}\mathscr{P}_{\stext{EM}}[\bv{v}^{\stext{out}}](\zeta_2; \omega; \bv{\alpha}') = 1,
\ee
or really to any conveniently-chosen but \textit{constant} value.  (Recall that this power is independent of the precise choice of surface $V(\zeta_2)$ as long as $\zeta_2 > \zeta_1.$)  We then maximize $\abs{\tfrac{\del}{\del \omega}\mathscr{P}_{\stext{mech}}[ \bv{\varepsilon}_{\bv{v}\,\perp}; \bv{j}](\omega; \bv{\alpha}') }$ for fixed output power, in order to determine the optimal parameter values $\tilde{\bv{\alpha}}' = \tilde{\bv{\alpha}}'[\bv{j}](\omega)$ and therefore the  \textit{relative} shape (spatial profile and polarization) of the radiation. Once the relative profile is determined, we then renormalize this trial solution,  taking 
\be
\bv{\chi}(\bv{\zeta}; \omega; \tilde{\bv{\alpha}}) = \tilde{\eta}_0 e^{i\tilde{\theta}_0}\,\bv{v}(\bv{\zeta}; \omega; \tilde{\bv{\alpha}}'),
\ee
for some real phase $\tilde{\theta}_0,\, 0 \le \theta_0 < 2\pi,$ and nonnegative amplitude $\tilde{\eta}_0  \ge 0.$ 
To maximize the mechanical work term, the overlap integral (3D inner product) between (scaled) solenoidal electric field and current density must necessarily be real and negative, so $\tilde{\theta}_0$ is chosen so as to ensure
\begin{subequations}
\begin{align}
-\smallhalf \realpart \Bigl[e^{-i\tilde{\theta}_0}\innerp{\bv{\varepsilon}_{\bv{\chi}\,\perp}(\omega; \tilde{\bv{\alpha}}')}{ \bv{j}(\omega)}\Bigr] &\ge 0\\
\phantom{-\smallhalf} \impart \Bigl[e^{-i\tilde{\theta}_0}\innerp{\bv{\varepsilon}_{\bv{\chi}\,\perp}(\omega); \tilde{\bv{\alpha}}'}{ \bv{j}(\omega)}\Bigr]  &=  0.
\end{align}
\end{subequations}
Finally , the positive amplitude $\tilde{\eta}_0 > 0$  is chosen to ensure energy balance between the power (spectral density) in the approximate outgoing fields and the work which would be done on these fields, if present in the region of the sources, by those actual sources:
\be 
-\smallhalf  \eta_0 e^{-i\tilde{\theta}_0} \innerp{ \bv{\varepsilon}_{\bv{\chi}\,\perp}(\omega)}{ \bv{j}(\omega)}  = 
\eta_0^2 \tfrac{\del}{\del \omega}\mathscr{P}_{\stext{EM}}[\bv{v}^{\stext{out}}](\zeta_2; \omega; \tilde{\bv{\alpha}}),
\ee
or
\be
\tilde{\eta}_0 = \frac{ -\smallhalf  \tfrac{\del}{\del \omega}\mathscr{P}_{\stext{mech}}[ \bv{\varepsilon}_{\bv{\chi}\,\perp}; \bv{j}](\omega; \tilde{\bv{\alpha}}') }{ \tfrac{\del}{\del \omega}\mathscr{P}_{\stext{EM}}[\bv{\chi}^{\stext{out}}](\zeta_2; \omega; \tilde{\bv{\alpha}}) }.
\ee

Note that under the assumptions of energy conservation given the sources, the overall amplitude, set by $\tilde{\eta}_0,$  is uniquely determined, and the overall phase $\tilde{\theta}_0$ is determined modulo $2\pi$, because the mechanical work done on/by the sources is linear in the fields, but the radiated power is quadratic in the fields.  As in the paraxial case, we can also interpret this as a maximization of the small-signal, no-recoil gain which would be experienced by the trial field if it were to be incident on the sources.  Upon substituting back into the Cauchy-Schwarz inequality, canceling common terms, and squaring, we deduce
\be
-\smallhalf \tfrac{\del}{\del \omega}\mathscr{P}_{\stext{mech}}[ \bv{\varepsilon}_{\bv{\chi}\,\perp}; \bv{j}](\omega; \tilde{\bv{\alpha}}) \le
\tfrac{\del}{\del \omega}\mathscr{P}_{\stext{EM}}[\bv{\chi}^{\stext{out}}](\zeta_2; \omega; \tilde{\bv{\alpha}}),
\ee
with equality if and only if 
\be
\bv{\chi}^{\stext{out}}(\bv{\zeta}; \omega; \tilde{\bv{\alpha}}) =  \bv{a}(\bv{\zeta}; \omega),
\ee
or equivalently,
\be
\bv{\chi}(\bv{\zeta}; \omega; \tilde{\bv{\alpha}}) = \bv{\chi}_{\bv{j}}(\bv{\zeta}; \omega) = \int d^3 \bv{\zeta}'\, D(\bv{\zeta}; \bv{\zeta}';\omega)\,\bv{j}_{\perp}(\bv{\zeta}; \omega).
\ee
Thus, we see that the variational approximation $\bv{\chi}^{\stext{out}}(\zeta_2; \omega; \tilde{\bv{\alpha}})$ may also be obtained more directly from a simultaneous optimization, with respect to all parameters in $\bv{\alpha},$ of the outgoing power, subject to the constraint of energy conservation:
\begin{subequations}\label{eqn:non-paraxial_max_4}
\begin{align}
\tilde{\bv{\alpha}} &= \argmax_{\bv{\alpha}} \Bigl[ \tfrac{\del}{\del \omega}\mathscr{P}_{\stext{EM}}[\bv{\chi}^{\stext{out}}](\zeta_2; \omega; \bv{\alpha}) \Bigr] \\
&\mbox{ s.t. }  -\smallhalf \tfrac{\del}{\del \omega}\mathscr{P}_{\stext{mech}}[ \bv{\varepsilon}_{\bv{\chi}\,\perp}; \bv{j}](\omega; \bv{\alpha}) =
\tfrac{\del}{\del \omega}\mathscr{P}_{\stext{EM}}[\bv{\chi}^{\stext{out}}](\zeta_2; \omega; \bv{\alpha}).
\end{align}
\end{subequations}
This is our desired result.  As anticipated, we have arrived at a variational principle which holds in the general, non-paraxial case, which is  exactly analogous to that found previously in the paraxial limit.  Unfortunately, there seems to be no simple way to express the constrained optimization problem (\ref{eqn:non-paraxial_max_4}) as an unconstrained maximization of a ratio between the mechanical work and radiated power (or some functions thereof). 

It is also instructive to also see how this all works out explicitly in the spherical wave basis.  Applying the usual Cauchy-Schwarz inequality directly in $\mathcal{H}_{\stext{out}},$ and using equations (\ref{eqn:power_out_alt1}) and (\ref{eqn:work_inner_alt}), we have
\be
\begin{split}
-\smallhalf\tfrac{\del}{\del \omega}\mathscr{P}_{ \text{\tiny{mech}}}[ \bv{\varepsilon}_{\bv{\chi}\,\perp}; \bv{j}](\omega; \bv{\alpha}) 
&=  \realpart \innerpa{\bv{\chi}^{\stext{out}}(\omega;  \bv{\alpha})}{\bv{a}(\omega)}_{\stext{out}} 
\le \abs{ \innerpa{\bv{\chi}^{\stext{out}}(\omega;  \bv{\alpha})}{\bv{a}(\omega)}_{\stext{out}}  }\\
&\le \innerpa{\bv{\chi}^{\stext{out}}(\omega;  \bv{\alpha})}{\bv{\chi}^{\stext{out}}(\omega;  \bv{\alpha})}_{\stext{out}}^{1/2} \innerpa{\bv{a}(\omega)}{\bv{a}(\omega)}_{\stext{out}}^{1/2}\\
&= \tfrac{\del}{\del \omega}\mathscr{P}_{\stext{EM}}[\bv{\chi}^{\stext{out}}](\infty; \omega)^{1/2}\tfrac{\del}{\del \omega}\mathscr{P}_{\stext{EM}}[\bv{a}](\infty; \omega)^{1/2},
\end{split}
\ee
with overall equality if and only if $\bv{\chi}^{\stext{out}}(\omega;  \bv{\alpha}) = \kappa(\omega)\, \bv{a}(\omega)$ for some positive real scalar $\kappa(\omega) > 0,$ or equivalently  when
\be
\frac{a^{\stext{E}}_{\ell\, m}(\omega)}{\chi^{\stext{E}}_{\ell\, m}(\omega; \bv{\alpha})} = \frac{a^{\stext{M}}_{\ell\, m}(\omega)}{\chi^{\stext{M}}_{\ell\, m}(\omega; \bv{\alpha})} = \frac{1}{\kappa(\omega)}  > 0.
\ee
for all mutlipole coefficients of $\bv{a}(\bv{\zeta}; \omega).$ Imposing the additional energy conservation relation on $\bv{\chi}(\bv{\zeta}; \omega; \bv{\alpha}),$ this becomes
\be
-\smallhalf\tfrac{\del}{\del \omega}\mathscr{P}_{ \text{\tiny{mech}}}[ \bv{\varepsilon}_{\bv{\chi}\,\perp}; \bv{j}](\omega; \bv{\alpha})
\le \mathscr{P}_{\stext{EM}}[\bv{a}](\infty; \omega)
\ee
with equality if and only if $\bv{\chi}^{\stext{out}}(\bv{\zeta}; \omega; \bv{\alpha}) = \bv{a}(\bv{\zeta}; \omega).$  Because in practice we will not have knowledge of the actual multipole expansion coefficients for $\bv{a}$ in situations where we are resorting to a variational approximation, this ``angular-representation'' or ``boundary-value'' formulation is not directly applicable for calculation, but it does lead directly to the relevant power inequality without any worries about a semidefinite or ``pseudo'' inner product.  We also see that the MPVP  follows immediately as a consequence of the minimum distance property of orthogonal projectors in these multipolar Hilbert spaces: the variational approximation $\bv{\chi}(\bv{\zeta}; \omega; \tilde{\bv{\alpha}})$ is simply that which minimizes the distance to the actual solution:
\be
\tilde{\bv{\alpha}} =  \argmin_{\bv{\alpha}} \Bigl[ \innerpa{ \bv{a}(\omega) - \bv{\chi}(\omega; \bv{\alpha})   }{ \bv{a}(\omega) - \bv{\chi}(\omega; \bv{\alpha})  }^{1/2}\Bigr]
\ee
within the considered family of solenoidal, source-free, trial wavefields.

\section{Discussion}\label{sec:discussion}

Here we assess advantages and limitations of the MPVP, briefly consider possible variations and extensions, and compare the method to other variational techniques commonly encountered in electromagnetic theory, in particular  those which have been used for paraxial radiation problems, laser-plasma interactions, and FEL-type devices.

\subsection{Summary}

The derived MPVP (\ref{eqn:non-paraxial_max_4}), valid in the paraxial or non-paraxial regimes, says, in effect, that classical (accelerating) charges spontaneously radiate ``as much as possible,'' consistent with energy conservation.  By spontaneous we mean that charges are assumed to follow {\it prescribed} classical trajectories determined by the external fields and possibly (averaged) quasi-static self-fields, but are not affected by recoil/radiation-reaction, multiple scattering, absorption, or other feedback from the emitted radiation (although arbitrary pre-bunching due to previously-imposed or radiated fields may be included, if it can be characterized).  The sources are assumed to be localized in space, so that the far-field (\ie, radiation zone , or wave zone) is defined, and at least weakly localized in time, so that Fourier transforms (\ie, frequency-domain representations) of the current density exist.  In turn, the radiation, once emitted by any part of the source, propagates as in free space, without further scattering or absorption.  The optimized trial mode-shape (or more accurately, the outgoing component thereof) is then the best guess, within the manifold of possibilities allowed by the adjustable parameters ($\bv{\alpha}$), of the exact radiation field profile radiated by those prescribed sources, and the optimized outgoing Poynting flux yields a {\it lower bound} on the actual spontaneously-radiated power spectral density at the frequency under consideration (apart from any numerical/roundoff errors in the computation).  The approximation will improve monotonically as additional independent parameters are included to allow for more general envelope shapes.  As with other variational principles, the relative error in the value of the variational functional (here, the power) at the stationary point is generally smaller than the error in the parameterized trial function (here, the radiation profile).  The adjustable parameters may appear linearly or non-linearly in the trial-functions; if they are taken to be the linear expansion coefficients in a sum over orthogonal modes, then the MPVP is equivalent to a truncated basis-set approach.
(In the paraxial case, ``outgoing'' implies propagation nearly along the optic axis away from the localized sources, defining the ``downstream'' or ``post-source'' direction, while ``incoming'' means propagating towards the sources from the upstream direction.  In this case, say, a right-going pulse has both ingoing and outgoing components.  In the general non-paraxial case, ``outgoing'' implies wavefronts which, at least asymptotically, diverge in time and propagate away from the localized sources, while ``incoming'' denotes the reverse, \ie, asymptotically converging wavefronts propagating towards the sources.)

This variational principle can be variously interpreted according to one's tastes or application.  As we have seen, the best variational approximation maximizes the radiated power consistent with the constraint that this energy could have arisen from work extracted from the actual sources.  It also minimizes a Hilbert-space distance between the actual fields and the parameterized family of solenoidal, free-space radiation fields, and in many cases may actually be regarded as an orthogonal projection into this manifold of trial solutions.  It also maximizes, for each frequency component, the spatial overlap, or physical resemblance, between the actual current density and the trial fields, when the latter are extrapolated back into the region of the sources assuming source-free propagation.  Equivalently, one can say the optimal free-space field profile is that which, if it actually were to be incident on the sources, would maximally couple to the given sources and would experience maximal small-signal gain, and furthermore the ÒvirtualÓ gain delivered would be equal to the estimated power spontaneously radiated.

Whatever the shape of the trial function, and whether the variational parameters appear linearly or non-linearly, after optimization this profile may be regarded as one of a complete set of solenoidal basis functions solving the source-free wave equation.  After introducing an  inner product associated with the radiated power (not energy),  these basis functions can be orthonormalized via a Gram-Schmidt procedure, leaving the original trial function unchanged.  Bessel's inequality then confirms that the outgoing power radiated in (the outgoing component of) this one mode can be no greater than the total outgoing power radiated in all the modes.  All this is obvious.  What is perhaps more surprising (if we do not think too hard about the Fundamental Theorem of Calculus or its multidimensional generalizations) is that we can exactly relate the outgoing power radiated in the far-field to the ``virtual'' work which would be performed by the localized sources on the ``source-free extrapolant'' of this radiation field, as if it and it alone were present in the region of the source, \ie, to the overlap integral between source and extrapolated electric field, even though the extrapolated field may not closely resemble the actual emitted field there, as it  must necessarily solve the inhomogeneous wave equation.  The factor of $\tfrac{1}{2}$ in the resulting conservation constraint emerges in order to avoid double-counting in the energetics.  The radiated power in the outgoing component of the variational approximation is being related to the mechanical power which would be delivered by the sources to the full source-free fields, if they were actually present in the region of the sources, rather than the actual, inhomogeneous fields; and, any wave-field satisfying the source-free wave equation everywhere in space must have the power in each outgoing mode exactly balanced by that  in a corresponding incoming mode, or else singularities would arise somewhere in space.
(Without any sources, in the paraxial case, radiation observed to emerge from, say, the right side of any longitudinal interval must have previously traveled into that interval from the left.  In the free-space non-paraxial case, analogous conclusions can be drawn for plane waves traveling along any chosen direction, or, thinking in terms of multipole radiation, it is necessary that any diverging spherical wave must be matched to a converging spherical wave to avoid a singularity at their common center of wavefront curvature.)

\subsection{Assessment}

Again, the present approach involves approximation of the actual radiation fields by trial modes which are solenoidal and satisfy the homogeneous (source-free) wave equation.  As we are primarily interested in the net radiation from the given sources observed in the free-space region beyond the sources, it is appropriate to use solutions of the source-free equation there, but then it is natural, efficient, and even almost inevitable, in the context of an approximation scheme, to effectively extrapolate the fields back into the support of the sources by free-space propagation.  (Otherwise, if we could somehow propagate according to the actual dynamics, we would not need to resort to any approximation -- but we would also be carrying around unneeded information about the near-fields.)  In the paraxial limit, these free-space solutions are uniquely specified by the carrier frequency and the (complex) profile in any transverse plane, which can be decomposed into a convenient, countable set of modes consisting of the eigenmodes of some quantum-like operator(s).  In the general case the solutions are more complicated, but from the spherical wave expansion, we know at least that they also form a separable Hilbert space, parameterized by the multipole expansion coefficients.  Because there is no known closed-form, analytic solution for a general focused radiation beam outside the paraxial approximation, the formalism and variational principle are probably of most use in the paraxial case, although may also be applied, with little added complication, to cases where the radiation consists of a superposition of multiple beams, each paraxial individually, but with sufficiently divergent directions to violate overall paraxiality, or to cases where higher-order terms in the generalized paraxial (\ie, diffraction-angle) expansion are included to describe a moderately-collimated beam.

The output $\bv{\chi}(\bv{\zeta}; \omega; \tilde{\bv{\alpha}})$ of the MPVP is therefore actually better considered a variational approximation to the source-free ``embedding'' or ``extrapolation''  $\bv{\chi}_{\bv{j}}(\bv{\zeta}; \omega)$ rather than to the actual outgoing vector potential $\bv{a}(\bv{\zeta}; \omega).$  If the radiation is sufficiently directional, then the outgoing component $\bv{\chi}^{\stext{out}}(\bv{\zeta}; \omega; \tilde{\bv{\alpha}})$ may be determined simply by restricting attention to some finite solid angle.  In the most general case, where the ``spurious'' incoming radiation may overlap with the ``physical'' outgoing radiation, it is more difficult to disentangle $\bv{\chi}^{\stext{out}}$ and $\bv{\chi}^{\stext{in}}$ at an arbitrary position.  If the spherical wave expansion is known, then the outgoing projection may be easily determined everywhere in space, but direct use of the the spherical-waves as the variational basis is not generically expected to prove convenient.  (This would correspond to just using a truncated multipole expansion for the radiation.)  In the far-field limit $k \zeta \to \infty,$ at least, the outgoing component may be determined easily by using the Sommerfeld projections (\ref{eqn:in_out_project}).

Because of these difficulties, and remaining mindful of the proportionality between the calculated power delivered from the source $\bv{j}_{\perp}(\bv{\zeta}; \omega)$ to the radiation fields corresponding to the vector potential  $\bv{a}(\bv{\zeta}; \omega)$ or to the fields of its source-free approximant $\bv{\chi}(\bv{\zeta}, \omega),$ one might be tempted to apply the variational method directly (just without the extra factor of $\smallhalf$) to outgoing, solenoidal solutions of some convenient inhomogeneous or homogeneous Helmholtz equation.  However, purely outgoing-wave, free-space solutions cannot exist everywhere in space, and will inevitably possess one or more singularities in the vicinity of the sources, preventing calculation of the required mechanical work integrals.   Furthermore, any solenoidal vector field whatsoever, say  $\bv{\psi}(\bv{\zeta}; \omega),$ is trivially the solution to some inhomogeneous Helmholtz equation for some boundary conditions and some solenoidal source, namely, for those boundary conditions satisfied by $\bv{\psi}$ itself, and for a solenoidal source given by $-(\del^2 + \omega^2) \bv{\psi}.$  Without further constraints, this class of trial solutions is simply too general to be useful.  If we naively apply the MPVP using homogeneous trial functions, the method will converge not to $\bv{a}$ but to something proportional to $\bv{j},$ since, for given norm, no other vector field can have greater overlap with $\bv{j}$ than a multiple of $\bv{j}$ itself.  By using the source-free solutions, we effectively regularize the problem, so that we can converge to something which maximizes the overlap with the source under the assumption that the trial  look like a radiation field, and not a current density.
 
In any case, the method requires not only the parameterization of trial modes but the calculation of inner products between these trial modes and the current density, or else some other, problem-specific, means to determine the absolute level of the power, either analytically, in terms of the free parameters, or numerically and indeed repeatedly, as the parameters are varied in search of the maximum.   One can easily imagine counter-examples where it is simply easier to work directly with the Li\'{e}nard-Wichart potentials or corresponding expressions for the fields, than use the variational principle.

We also have some options (or rather, trade-offs) in handling stochastic sources.  So far, we have assumed a specific, prescribed, deterministic current density $\bv{j}(\bv{\zeta}; \tau),$ but in practice we must contend with statistical uncertainty in the exact electron trajectories in any one realization, meaning $\bv{j}(\bv{\zeta}; \omega)$ is in principle stochastic, with statistical properties determined by the phase space distribution of the particle beam (in the reduced six-dimensional phase space, if we may neglecting direct two-body or higher-order interactions, but in an even higher-dimensional phase space if particle-particle correlations are to be included).  In certain cases, we can get by with using an averaged or course-grained current directly as the source, without need for further probabilistic considerations.  In the Vlasov equation, for example, the fields that appear are self-consistent mean fields, \ie, solutions to Maxwell's equations for the \textit{averaged} sources.  The exact fields are linear in the sources, so it makes no difference in the end result whether we average the currents first and then calculate the resulting fields, or calculate the fields for a general instance of the currents and then average the result -- the averaging procedure will commute with the convolution over the Green function.  (Pragmatically, the former approach will almost always be more efficient if the average is all we need, since the random variables determining the stochastic shape of the current need not carried through the convolution, but the latter approach allows calculation of higher-order correlations or moments.)  When using the  MPVP to approximate the radiation, the averaged trial fields will equal the trial fields from the averaged sources if the variational parameters appear as linear expansion coefficients in a basis-set expansion, but not, in general, if they occur non-linearly in the trial wavefields.  Convolution is a linear process, but optimization is in general a nonlinear one.

Unfortunately, it is often the case that we desire to estimate the fields from given sources, then calculate moments, rather than to determine the fields from averaged sources.  This is particularly true of spontaneous wiggler radiation, where the radiation is almost entirely ``fluctuational,'' both in the sense that the squared-magnitude of the average field is small compared to the average of the squared-magnitude, and in that the relative variance in the spectral intensity approaches $100\%$ in each sufficiently narrow frequency-band for radiation from sufficiently long, unbunched electron beams.  At least second-order moments are needed to estimate the power spectrum and transverse coherence properties, and fourth-order  moments are necessary if we wish to to estimate the uncertainty in the estimates for these spectra or coherence functions.  Statistical estimates for the power spectral density, based on the moments of the MPVP trial fields, cannot in general be guaranteed to be lower bounds, as in the deterministic case.

Before even calculating moments, we may either approximate the radiation fields from the total current density using the MPVP, or else apply the MPVP separately to each particle trajectory or family of trajectories, then superimpose the resulting variational fields.  If identical, linear basis-sets are used in the  expansions for each particle's radiation, identical results will be achieved (although the amount of calculation or computational effort may differ), but if nonlinear variational parameters are used or even if different truncated linear basis sets are used for different particles, the two approaches will be inequivalent.

A simple example should clarify the issues.  Suppose we seek to approximate the wiggler radiation from some electron beam with negligible normalized energy spread $\tfrac{\delta}{\gamma} \ll 1$ but some non-zero transverse emittance $\epsilon > \lambda_0,$ where $\lambda_0 \sim \tfrac{1}{2\gamma^2} \lambda_u(1 + \smallhalf a_u^2)$ is the central wavelength of the wiggler radiation in the lab frame, $\lambda_u$ is the wiggler wavelength, and $a_u$ is the normalized wiggler strength parameter.   For radiation from a single on-axis electron, confined within the relative bandwidth $\tfrac{\delta \lambda}{\lambda_0} \sim \tfrac{1}{N_u},$ where $N_u$ is the number of wiggler periods, we know the radiation is approximately diffraction-limited, with characteristic angle $\delta \theta_{\stext{D}} \sim \tfrac{1}{\sqrt{N_u}\gamma}$ and a waist located at the midpoint of the wiggler.  However, the radiation for the beam as a whole will not be diffraction-limited, but will have a transverse degree of coherence $\Gamma_{\stext{coh}\, \perp} \sim \tfrac{\lambda_0}{4\pi\epsilon} < 1,$ due to averaging, or convolution, over the transverse particle distribution.

Suppose first we use the total current density as source, and use paraxial Gauss-Hermite or Gauss-Laguerre modes as trial solutions, with optic axis coinciding with the wiggler axis.  If we attempt to include, for $\lambda \approx \lambda_0,$ only the fundamental Gaussian mode with undetermined waist size $w_0$ and location $\xi_{\stext{w}},$  then, regardless of the exact values of the optimized parameters, the approximation cannot simultaneously predict both the transverse size and angular divergence with accuracy, or equivalently, the partial transverse incoherence cannot be adequately captured, since we are putting all the power in a single, diffraction-limited transverse mode.  In fact, we expect that we must include at least $n_{\perp} \sim \tfrac{4\pi\epsilon}{\lambda_0}$ diffraction-limited modes to resolve the transverse coherence, even if the average transverse \textit{intensity} profile continues to look Gaussian.  If the particle phase space distribution function is also taken to be Gaussian, then the needed overlap integrals in the MPVP may be calculated analytically.

Alternatively, we could decompose the current density into contributions from each electron, and employ a single Gaussian mode for each, only with the optic axis aligned along the average single-particle trajectory determined by its spatial and angular displacement from the reference orbit.  The needed overlap integrals in this case are easily perfomed, at least  for typical values of $\lambda_0,$ $\gamma,$ and $a_u,$ where the transverse spatial extent of an electron orbit is small compared to $w_0,$ but after determining the single-particle MPVP fields, each with a different propagation axis, they all must be superimposed to determine the full field.  
This can be done easily in the far-field so as to determine the angular spectrum, or at arbitrary positions, to a good approximation, provided the angular deviations for the electrons remain moderately small, \ie, $\Delta \theta \lesssim \tfrac{1}{\gamma}.$  The result will agree approximately, but not exactly, with the previous approach.  The details of a simplified version of this calculation will be presented in elsewhere.

\subsection{Extensions}

The results derived here are restricted to radiation from localized sources otherwise propagating in vacuum, which includes the cases of charged particle beams in bending magnets or undulators, but it would be desirable to extend applicability to more general radiation problems.  Using the macroscopic Maxwell's equations, it should be straightforward to include a  linear, dielectric background $\epsilon_0(\omega) \ge 0$ which is spatially homogeneous, but possibly frequency-dependent.  This would be sufficient to handle many \v{C}erenkov radiation problems.  To treat some forms of transition radiation and waveguide problems, piecewise-slowly-varying spatial variations in the dielectric tensor and the possibility of (perfectly) conducting boundaries should both be included, but in general the ``source-free'' solutions become difficult to calculate and parameterize -- a single planar jump, or other simple geometries, might still be tractable.

One important class of geometries for which this formalism should be almost immediately applicable is that of conducting-wall and/or dielectric waveguides with translational symmetry along the optic axis.  In such circumstances, it is well known \cite{collin:60, waldron:70, jackson:75,vanbladel:91} that such systems share much of the mathematical Hilbert-space structure as that elaborated above for paraxial radiation -- in particular, solenoidal modes which are uniquely specified by their cross-section in any transverse plane, and which are orthonormalizable with respect to the natural $\mathscr{L}_2$/Euclidean inner product in that plane.  (However, these modes are not complete for the near-fields, as discussed, for example, in \cite{rahmat-samii:75}.)  In addition, it is easily shown that, in any such waveguides, the power in any source-free, propagating mode, measured sufficiently far downstream from a \textit{localized} source inserted inside the waveguide, can be related to the overlap integral between the actual current density of the source and the electric field profile of this free-space mode extrapolated back into the support region of the sources as if no sources were present.  Modifications due to finite conductivity effects might be then treated perturbatively.  Again, this is all familiar, and is exactly analogous  to our basis-set expansion technique.  What is perhaps not widely appreciated is that the Cauchy-Schwarz inequality then allows one to transform this basis-set approach into a maximum principle, where the exact form of the normal modes are not needed.

Attempts to use a more general  complex-valued dielectric tensor or index of refraction in order to incorporate linear loss or gain will encounter fundamental difficulties.  Our derivations have exploited the fact that the free-space retarded and advanced Green functions are just related by complex conjugation in the frequency domain, but with any dissipation or gain, this time-reversal symmetry will be lost.  Ideally, one would like to fully generalize the treatment to allow for dielectric tensors $\epsilon_{\mu\nu}(\bv{x}, t; \omega)$ with any frequency dispersion consistent with the Kramers-Kroning relations and/or piecewise-slowly-varying spacetime dependence (inhomogeneity), and possibly anisotropy, in order to model the effects of essentially arbitrary conducting or dielectric boundaries, obstacles, or passive optical elements, such as lenses, prisms, polarizers, waveplates, apertures, waveguides, cavities,  transmission lines, fibers, gratings, etc., either ideal or lossy, but it not even clear whether this is possible, or if so, practical.  Many variational techniques have been used for waveguides, cavities, and other structures, but these are usually of a stationary, rather than extremal, character, an important distinction which is examined below in Section~\ref{sec:lagrangian}.   All of these directions are left as open problems.

\subsection{Connections to Stimulated Emission}

One is also naturally led to speculations as to whether similar ideas may be applied to stimulated emission.  Naively, one might imagine some iterative procedure, where the approximated radiation, assuming given sources, is somehow used to deduce corresponding energy changes that must have taken place in those sources, which leads to revised estimates for the radiation, and so on.  It is not clear whether this is applicable to any real problems, or even if it can converge to a self-consistent solution describing an active medium.

However, results for the case of spontaneous emission from a classical particle beam may be of direct relevance to the problem of stimulated emission, namely in the so-called small-signal regime where saturation and depletion effects are ignored.  In his celebrated 1917 quantum analysis of radiation \cite{einstein:17}, Einstein first classified the radiative processes involving photons (or really any bosons) into spontaneous emission of, stimulated emission by, and stimulated absorption of the photons by atoms or other material sources.  By exploiting the fact that the matter and radiation can be in thermodynamic equilibrium, Einstein used the properties of blackbody radiation and the principle of detailed balance to establish the proportionalities between the intrinsic rates, within each mode, for each of these processes, as summarized in the famous $A$ and $B$ coefficients, results which must hold for arbitrary initial conditions, not just those consistent with thermal equilibrium.  Less widely appreciated is that these relations persist in a purely classical limit of the matter and fields where Planck's constant $h$ never appears, clearly discussed, for example, by Beckefi in Chapter 2 of \cite{beckefi:66}, and derived using somewhat different arguments in \cite{litvinenko_vinokurov:93}.

As a simple example, consider the case of radiation transport in an electron plasma, where direct many-body effects are neglected, and with an isotropic, single-particle momentum distribution function $f(p),$  where $p = \vnorm{\bv{p}}$ is the magnitude of kinetic momentum, and uniform density $n_0$ over some region. (This might also describe a long electron beam in its average rest frame.)  Then, for each type of radiative process (or the aggregate of all processes), the spontaneous emission rate $\ell_{\omega}$ defined as the power emitted in polarization $\unitvec{e}$ per unit volume of medium per unit bandwidth per solid angle in the direction of $\unitvec{s},$ may be written as
\be
\ell_{\omega}[\unitvec{e}; \unitvec{s}] = n_0 \int d^3 \bv{p}\, \eta_{\omega}(\bv{p}; \unitvec{e}; \unitvec{s}) f(p),
\ee
where $\eta_{\omega}(\bv{p}; \unitvec{e}; \unitvec{s})$ is the single-particle,  intrinsic emission coefficient, which is determined by the details of the microscopic physics, but is {\it independent} of the particle distribution function or the incident radiation state.  The {\it net} absorption rate, \ie, bare stimulated absorption less stimulated emission, is given by $\alpha_\omega I_{\omega},$ where the radiant brightness $I_{\omega}$ is the incident power per unit area per unit solid angle per unit bandwidth, and the absorption coefficient is given by
\be
\alpha_{\omega}[\unitvec{e}; \unitvec{s}] = - \frac{8\pi c^2}{\mathsf{n}^2_{\stext{ir}}\omega^2} n_0 \int d^3 \bv{p}\, \eta_{\omega}(\bv{p}; \unitvec{e}; \unitvec{s})\tfrac{\del }{\del \varepsilon} f(p),
\ee
where $\varepsilon = \varepsilon(p) = \sqrt{c^2 p^2 + m^2 c^4}$ is the particle energy, $\mathsf{n}_{\stext{ir}}$ is the background index of refraction,
and $\eta_{\omega}(\bv{p}; \unitvec{e}; \unitvec{s})$ is the same quantity that appears in the spontaneous emission.  These results are completely classical, as no factors of $\hbar$ anywhere appear.  The dependence on the derivative of the distribution function is the only reminder that this expression represents the net difference between stimulated absorption and emission.  These connections between spontaneous and stimulated emission/absorption are essentially a generalized manifestation of the well-known fluctuation-dissipation theorem, which relates the response of a system when perturbed to the spontaneous fluctuations which occur in the absence of external perturbation \cite{beckefi:66, krinsky_et_al:82, litvinenko_vinokurov:93}.

In FEL physics, first described quantum-mechanically but now understood classically, such generalizations of Einstein's arguments are the basis for Madey's theorem in one-dimensional theory \cite{madey:79} and its generalizations to higher dimensions (see, for example, \cite{krinsky_et_al:82, kim:92, litvinenko_vinokurov:93}, or else \cite{nikonov_et_al:98} and references therein), where the gain curve (specifically, the relative change in intensity versus de-tuning) in the small-signal regime is proportional to the \textit{derivative} of  the spontaneous emission spectrum.

Given these connections, certain properties of the spontaneous wiggler radiation, or approximations thereof, can yield information about the stimulated emission, or gain, in the small-signal regime.  Perhaps more satisfying is the obverse relation, whereby our intuitions about stimulated emission provide the clearest justification and interpretation for the appearance of a maximum-power variational principle for classical spontaneous emission.  In situations with gain, we might naturally approximate the radiation profile by that which maximizes the extraction of energy from the electrons to the laser.   For example, this is precisely the heuristic justification for the variational principle suggested in \cite{xie_deacon:86} for optically-guided modes in FELs, where the fundamental mode is estimated by maximizing the imaginary part of the effective wavenumber.  If the radiation is to be estimated by one mode, the mode should be chosen to be whatever shape is expected to have the fastest growth (or slowest loss).

In fact, we saw above that the MPVP can be interpreted precisely in this manner -- as finding the mode shape which, if actually incident, would maximally couple to the given sources, and furthermore the ``virtual'' gain delivered would be equal to the estimated power spontaneously radiated.  The only essential difference between our case, and say, Madey's theorem is  that by taking completely prescribed sources, we implicitly assume that any radiation, once emitted by some part of the source,  can then propagate to the far-field where it may interfere constructively or destructively with other radiation, but cannot cause any recoil in that source nor subsequently be re-scattered or absorbed by any other part of the source.  So in fact we find a relationship between the spontaneous emission spectrum and that of the \textit{bare} stimulated emission, not the \textit{net} response given by the difference between stimulated emission and absorption as in Madey's theorem.

\subsection{Relation to Lagrangian Formulation and Other Variational Approaches}\label{sec:lagrangian}

A survey of variational techniques in electrostatics, electrodynamics, and optics reveals that, at some fundamental level, they seem to derive from only a very few general concepts:  Hamilton's principle for systems derivable from a Lagrangian/action, energy conservation/optimization, electromagnetic reciprocity, and entropy maximization (for problems of a thermodynamic nature.)  Of course these principles are interrelated -- both energy conservation and reciprocity follow from the structure of the governing Lagrangian, while entropy maximization usually relies on energy conservation or other dynamical invariants as constraints and automatically satisfies certain reciprocity relations --  but nevertheless they remain useful and conceptually distinct organizational categories.  Each variational principle may be further classified by whether it demands (generally constrained) optimality, or mere stationarity, of the relevant function or functional.

We examine in turn the relevance or relation of each of these principles to the MPVP.  Obviously the MPVP is most closely related to energy principles, although in the case of classical spontaneous radiation, we maximize the rate of energy transfer (power) in each spectral band, rather than, say, minimize potential energy as in electrostatics, or free energy in thermodynamics.

Entropy maximization might be a useful technique to determine or approximate certain source distributions given only macroscopic thermodynamic constraints or other incomplete information, but is not directly relevant for the problem of finding the radiation from given deterministic sources, which is more a problem of dynamics than thermodynamics.  However, as we saw above, new connections and rationales are revealed by generalization of the arguments leading to Einstein's $A$ and $B$ coefficients to classical beams or plasmas, and these arguments rely on the equilibrium properties of black-body radiation, even when the actual  system of interest may be far from thermodynamic equillibrium.

Reciprocity-based variational principles are commonly used to approximate resonant frequencies in cavities or waveguides, impedances of cavities, transmission lines, apertures, or other structures, and scattering cross sections by conductors or dielectrics \cite{rumsey:54, harrington:61, kong:86}.  They can be suitably generalized to many situations where the background medium is not necessarily homogeneous or isotropic.  Despite superificial similarities between the MPVP and reciprocity-based techniques, a closer look reveals them to be quite distinct.  Consider the solenoidal electric field $\bv{\varepsilon}_{\bv{\psi}\, \perp}(\bv{\zeta}; \omega)$ derived from some Coulomb-gauge vector potential $\bv{\psi}(\bv{\zeta}; \omega),$ and any source $\bv{j}'(\bv{\zeta}; \omega),$ not necessarily that associated with $\bv{\psi}.$
For our purposes, we may define the complex \textit{reaction} of the field $\bv{\varepsilon}_{\bv{a}\, \perp}$ on the source $\bv{j}'$ as 
\be
\mathcal{R}[\bv{\varepsilon}_{\bv{\psi}\, \perp}, \bv{j}'] \equiv \int d^3\bv{\zeta}\, \bv{\varepsilon}_{\bv{\psi}\, \perp}(\bv{\zeta}; \omega)\!\cdot\! \bv{j}'(\bv{\zeta}; \omega) = \innerp{\bv{\varepsilon}_{\bv{\psi}\, \perp}(\omega)\cc}{\bv{j}'(\omega)},
\ee
Which is similar to the original definition in \cite{rumsey:54}, but involves only the solenoidal electric fields, and drops the analogous terms involving the magnetic field and magnetization density, because we are working with the microscopic fields from classical sources.   Given any source $\bv{j}(\bv{\zeta}; \omega),$ let  $\bv{a}(\bv{\zeta}; \omega)$ denote the corresponding Coulomb-gauge, retarded vector potential, and let $\bv{\chi}(\bv{\zeta}; \omega) = \bv{\chi}_{\bv{j}}(\bv{\zeta}; \omega)$ denote the closest homogeneous approximant (\ie, that vector potential obtained by using $\bv{j}_{\perp}$ as an effective source in convolution with the fundamental solution, rather than an actual source in convolution with the retarded Green function); and define $\bv{a}'$ and $\bv{\chi}'$ in an analogous fashion for the source $\bv{j}'.$ 

By using the symmetry properties of the Green functions, it is then straightforward to establish the Raleigh-Carson reciprocity relation for the inhomogeneous case:
\be\label{eqn:reciprocity1}
\mathcal{R}[\bv{\varepsilon}_{\bv{a}\, \perp}, \bv{j}'](\omega) = \innerp{\bv{\varepsilon}_{\bv{a}\, \perp}(\omega)\cc}{\bv{j}'(\omega)}
= \innerp{\bv{\varepsilon}_{\bv{a}'\, \perp}(\omega)\cc}{\bv{j}(\omega)} = \mathcal{R}[\bv{\varepsilon}_{\bv{a}'\, \perp}, \bv{j}](\omega), 
\ee
and similarly for the homogeneous case:
\be\label{eqn:reciprocity2}
\mathcal{R}[\bv{\varepsilon}_{\bv{\chi}\, \perp}, \bv{j}'](\omega) = \innerp{\bv{\varepsilon}_{\bv{\chi}\, \perp}(\omega)\cc}{\bv{j}'(\omega)}
= \innerp{\bv{\varepsilon}_{\bv{\chi}'\, \perp}(\omega)\cc}{\bv{j}(\omega)} = \mathcal{R}[\bv{\varepsilon}_{\bv{\chi}'\, \perp}, \bv{j}](\omega).
\ee

Such a quantity was introduced in \cite{rumsey:54} as a measure of  ``reaction,'' or coupling, between the sources $\bv{j}$ and $\bv{j}',$ and in many contexts a variety of physical observables such as fields or forces, impedances, etc., may be proportional to or otherwise related to reactions.  Despite resemblances, the complex self-reaction $\innerp{\bv{\varepsilon}_{\bv{\psi}\, \perp}(\omega)\cc}{\bv{j}(\omega)}$ is distinct from the complex power $\innerp{\bv{\varepsilon}_{\bv{\psi}\, \perp}(\omega)}{\bv{j}(\omega)},$ because no complex conjugates appear in the overlap integral in the former case.  (Adding to the potential for confusion, the imaginary part of the complex power is referred to as the ``reactive power,'' following the conventions of circuit theory.)  Said another way, the complex power is associated with a standard complex inner product, which is conjugate-symmetric and positive-definite, while the reaction is associated with a symmetric, bilinear form which is not a true inner product.  Physically, the concepts are also distinct: energy conservation arises from the invariance properties of the full electromagnetic Lagrangian (including self-consistent particle dynamics) under time-translations, while reciprocity arises from invariance under time reversals.  The real part of the complex power is of course proportional to the time-average (over an optical period) of the power exchanged between source and field.  The real part of the reaction is actually proportional to the fluctuation about this average.
 
The concept of reaction is of interest precisely because of reciprocity properties like (\ref{eqn:reciprocity1}) or (\ref{eqn:reciprocity2}), and the possibilities for variational approximation that emerge, by demanding that all trial sources ``look'' the same to themselves as to the correct sources.
For example, suppose we seek a trial-function approximation $\tilde{\bv{\psi}}$ to the unknown vector potential $\bv{\psi},$ given the actual source $\bv{j}.$  Let $\tilde{\bv{j}}$ correspond to a source associated with $\tilde{\bv{a}}$ (the existence of non-radiating sources implies non-uniqueness, which is of no consequence here.)  From a reaction point-of-view, we would like an approximation such that 
\be
\mathcal{R}[\bv{\varepsilon}_{\tilde{\bv{\psi}}\, \perp}, \tilde{\bv{j}}](\omega) \approx \mathcal{R}[\bv{\varepsilon}_{\bv{\psi}\, \perp}, \bv{j}](\omega),
\ee
but by assumption we have no tractable way of calculating the right-hand side, so
instead that approximation is then ``best'' if we can impose the constraint  
\be
\mathcal{R}[\bv{\varepsilon}_{\tilde{\bv{\psi}}\, \perp}, \tilde{\bv{j}}](\omega) \approx \mathcal{R}[\bv{\varepsilon}_{\tilde{\bv{\psi}}\, \perp}, \bv{j}](\omega).
\ee
In fact, one can show that under this constraint, the reaction $\mathcal{R}[\bv{\varepsilon}_{\tilde{\bv{\psi}}\, \perp}, \bv{j}](\omega)$ is then stationary for first-order variations of $\tilde{\bv{\psi}}$ about the actual solution $\bv{\psi}.$ 

Note that this leads to a stationarity condition, not an optimization condition.  Because reaction and power are not equivalent, in general they lead to different variational principles.  In the Method of Moments, which encompasses both the Finite-Element and the Ritz-Galerkin basis-expansion techniques, one can obtain algebraic (typically linear) stationarity conditions using either a symmetric form between trial functions and weight functions, and thus automatically satisfy reciprocity relations, or using a conjugate-symmetric inner product,  and satisfy energy constraints, but typically not both simultaneously.  Perhaps the reaction approach may also lead to some useful variational principle for our problem, but we do not pursue this question further here.  One immediate difficulty is that, if we use inhomogeneous solutions $\tilde{\psi}$ as trial vector fields, the corresponding sources are easily determined by substitution into the Helmholtz equation, but the fields themselves are difficult to parameterize in any economical manner, while if we use source-free solutions for $\tilde{\bv{\psi}},$ variational families may be more easily characterized, but the effective sources appearing in the reactions are then difficult to determine.

Now we turn to action principles.  Because the fundamental dynamics of charge-carrying matter and electromagnetic  fields are ultimately derivable from a Lagrangian, we had anticipated that our maximum-power variational principle would be traced back to Hamilton's Principle, but this appears not to be the case.  (Perhaps this should not have been so surprising; after all, the Lagrangian involves differences between kinetic and potential energies, while an an energy-based principle should involve the sum.)  To understand the issues encountered, let us proceed by re-framing the governing dynamical equations in terms of a standard Lagrangian formulation.  For prescribed current and charge densities, the dynamics of the fields are derivable from the action functional, which in terms of the Coulomb-gauge potentials and scaled coordinates may be written as a space-time integral over the Lagrangian densities for the free electromagnetic fields and the source-field interaction:
\be
\mathcal{S} = \int \!\!d\tau \!\!\int \!\!d^{3}\bv{\zeta}\, \left[\mathcal{L}_0 +  \mathcal{L}_{\text{\tiny{int}}}\right],
\ee
where
\be
\mathcal{L}_0 = \vnorm{\bv{\varepsilon}(\bv{\zeta}; \tau)}^2 -\vnorm{\bv{b}(\bv{\zeta}; \tau)}^2
=\vnorm{-\tfrac{\del}{\del \tau}\bv{a}(\bv{\zeta}; \tau) - \bv{\del}\phi(\bv{\zeta}; \tau)}^2
- \vnorm{\bv{\del}\times\bv{a}(\bv{\zeta}; \tau)}^2,
\ee
and
\be
\mathcal{L}_{\text{\tiny{int}}} = \bv{j}(\bv{\zeta}; \tau)\cdot \bv{a}(\bv{\zeta}; \tau) - \mu(\bv{\zeta}; \tau)\phi(\bv{\zeta};\tau),
\ee
subject as always to the solenoidal gauge constraint
\be
\bv{\del}\cdot \bv{a}(\bv{\zeta}; \tau) = 0.
\ee
Using the reality of the physical potentials and sources and the (Hilbert-space) orthogonality between the transverse vector field $\bv{a}(\bv{\zeta}; \tau)$ and longitudinal electric field $-\bv{\del}\phi(\bv{\zeta}; \tau)$ when integrated over all space, we can re-write this as
\be
\mathcal{S} =\int \!\!d\tau \!\!\int \!\!d^{3}\bv{\zeta}\left[(\tfrac{\del}{\del \tau}\bv{a}\cc \!\cdot\! \tfrac{\del}{\del \tau}\bv{a}) + (\bv{\del}\phi\cc\!\cdot\!\bv{\del}\phi) - (\bv{\del}\!\times\!\bv{a}\cc)\!\cdot\!(\bv{\del}\!\times\!\bv{a}) +  (\bv{a}\cc\!\cdot\!\bv{j})
- (\phi\cc\mu)\right],
\ee
where we may use $\bv{j}$ or $\bv{j}_{\perp}$ interchangeably within the action functional because the current density only appears in an inner product with the solenoidal vector potential $\bv{a}.$  The equations of motion (Euler-Lagrange equations) for the vector and scalar potentials (and their complex conjugates) are obtained as the conditions for this action remaining stationary under independent, infinitesimal variations $\delta \bv{a}(\bv{\zeta}; \tau),$ $\delta \bv{a}(\bv{\zeta}; \tau)\cc,$ $\delta \phi(\bv{\zeta}; \tau),$ and $\delta \phi(\bv{\zeta}; \tau)\cc,$ which are fixed at the space-time boundaries (here taken to be at infinity), are consistent with the gauge constraint, and which do not destroy integrability of the Lagrangian density, but are otherwise arbitrary.

By interchanging the order of temporal and spatial integrations and using the Parseval-Plancherel Identity, we may express this action in the (scaled) frequency domain:
\be
\mathcal{S} = \int \!\!d\omega \!\!\int \!\!d^{3}\bv{\zeta}\left[(i\omega \bv{a})\cc \!\cdot\! (i\omega\bv{a}) + (\bv{\del}\phi\cc\!\cdot\!\bv{\del}\phi) - (\bv{\del}\!\times\!\bv{a}\cc)\!\cdot\!(\bv{\del}\!\times\!\bv{a}) + (\bv{a}\cc\!\cdot\!\bv{j}) - (\phi\cc\mu)\right],
\ee
with additional gauge constraint
\be
\bv{\del}\cdot \bv{a}(\bv{\zeta}; \omega) = 0,
\ee
where we have used integration by parts on the time-derivatives, and where each potential and source term appearing in the action integral is now interpreted as the Fourier transform in scaled time (assumed to exist) of the function of the same name appearing in the time-domain version above, and where coordinate dependence has been suppressed for the sake of brevity and readability in the equations.  Using standard vector identities, this may be written in the more transparent form:
\be
\mathcal{S} = \int \!\!d\omega \!\!\int \!\!d^{3}\bv{\zeta}\left[
\bv{a}\cc\!\cdot\!( \del^2 + \omega^2)\bv{a} + \bv{a}\cc\!\cdot\!\bv{j}
- \phi\cc(\del^2 \phi) -  \phi\cc\mu + \bv{\del}\!\cdot\!( \phi\cc\bv{\del}\phi -\bv{a}\cc\!\times\!\bv{\del}\!\times\!\bv{a})
\right]
\ee
Demanding stationarity under independent, infinitesimal variations $\delta \bv{a}(\bv{\zeta}; \omega),$ $\delta \bv{a}(\bv{\zeta}; \omega)\cc,$ $\delta \phi(\bv{\zeta}; \omega),$ and $\delta \phi(\bv{\zeta}' \omega)$ satisfying the gauge constraint and fixed boundary conditions (as $\vnorm{\bv{\zeta}} \to \infty$ and $\abs{\omega} \to \infty$) leads to the frequency-domain version of Poisson's equation and the wave equation in the Coulomb gauge.  Now, by Gauss's theorem, the final term involving the pure divergence can be expressed as a surface integral over the spatial boundary (at infinity), which will not affect the stationarity conditions (Euler-Lagrange equations) under the assumption of  infinitesimal variations with fixed boundary conditions as dictated by Hamilton's principle, so it may be dropped without any change in the resulting dynamical equations.  Likewise, we may then introduce the divergence $\bv{\del}\cdot\bv{q}$ of some other, arbitrary differentiable vector field $\bv{q}$ which may be a function of $\bv{\zeta}$ and $\omega$ and a functional of $\bv{a}$ and $\bv{a}\cc$ and their spatial derivatives.  The significance of this auxiliary field will emerge shortly.  Since we are ultimately only interested in the radiative component of the fields, the terms in the action involving the scalar potential $\phi$ may also be suppressed, as they are constant with respect to variations in the vector potential.  Recalling that the time-domain potentials and sources are all real, we can then use as the relevant action the modified functional
\be
\tilde{\mathcal{S}} = \int\limits_{0}^{\infty} \!d\omega \!\int \!d^{3}\bv{\zeta}\,\left[
\bv{a}\cc\!\cdot\!( \del^2 + \omega^2)\bv{a} + \bv{a}\cc\!\cdot \!\bv{j} + \bv{\del} \!\cdot \! \bv{q} \right] + c.c.,
\ee
and, noting the additivity with respect to contributions from the various frequencies $\omega,$ we may use as the variational, or ``objective'' or ``cost''  function, for each distinct (positive) frequency component the spectral density associated with this expression:
\be\label{eqn:action2}
\tfrac{\del}{\del \omega}\tilde{\mathcal{S}}[\bv{a}, \bv{a}\cc, \bv{j}, \bv{j}\cc](\omega) \equiv \int \!d^{3}\bv{\zeta}\,\left[
\bv{a}\cc\!\cdot\!( \del^2 + \omega^2)\bv{a} + \bv{a}\cc\!\cdot \!\bv{j} + \bv{\del} \!\cdot \! \bv{q} \right] + c.c.,
\ee
whose variations, subject to the solenoidal gauge constraint and given spatial boundary conditions,  lead to Euler-Lagrange equations consisting of the inhomogeneous Helmholtz equation for the vector potential and its complex conjugate at the frequency $\omega > 0.$  Choosing
\be
\bv{q}  =
 \smallhalf \lambda \left[ \tfrac{1}{4}\left(1 - \tfrac{i}{k} \tfrac{\del }{\del \zeta}\right)\bv{a} \!\times\! \left(1 + \tfrac{i}{k} \tfrac{\del }{\del \zeta}\right)\bv{a}\cc  + \tfrac{\mathcal{P}'_{0}}{4\pi} \, \bv{\del} \, \zeta^{-1}   \right] ,
\ee
or equivalently (at least asymptotically, as $\zeta \to \infty$),
\be
\bv{q}  = \smallhalf \lambda \left[ \bv{s}^{\stext{out}}_{\bv{a}}   -  \tfrac{\mathcal{P}'_{0}}{4\pi}\tfrac{\bv{\zeta}\phantom{^3}}{\zeta^3}  \right], 
\ee 
for some undetermined real parameters $\lambda= \lambda(\omega)$ and $\mathcal{P}'_0 = \mathcal{P}'_0(\omega),$
equation  (\ref{eqn:action2}) becomes:
\be\label{eqn:action3}
\tfrac{\del}{\del \omega}\tilde{\mathcal{S}}[\bv{a}, \bv{a}\cc, \bv{j}, \bv{j}\cc](\omega) \equiv 2\! \int \!d^{3}\bv{\zeta}\,\realpart \bigl[
\bv{a}\cc\!\cdot\!( \del^2 + \omega^2)\bv{a} + \bv{a}\cc\!\cdot \!\bv{j}\bigr] + \lambda(\omega) \bigl[
\tfrac{\del}{\del \omega}\mathscr{P}_{\stext{EM}}[\bv{a}^{\stext{out}}](\infty; \omega)  - \mathcal{P}'_0 (\omega)\bigr].
\ee
The chosen boundary term amounts to the addition of a Lagrange multiplier enforcing a specified outgoing power spectral density in the radiation fields at the stationary points of $\tfrac{\del}{\del \omega}\tilde{\mathcal{S}}.$

Now, a natural approach to approximation, associated variously with the names of Raleigh, Ritz, and Galerkin in slightly different contexts, consists in simply replacing this infinite-dimensional variational problem with a finite-dimensional restriction or projection of it onto some more easily characterized space of possibilities.

That is, we restrict the space of possible vector potentials to some parameterized family $\bv{a}(\bv{\zeta}; \omega; \bv{\alpha})$ for some finite-dimensional vector $\bv{\alpha}$ of continuous parameters (over some specified domain), and thereby replace a variational problem involving functional derivatives of the action functional with respect to the vector potential with one involving  ordinary derivatives of an action function with respect to the variational shape parameters determining the form of the radiation and the Lagrange multiplier enforcing a power normalization.  That is, assuming the family of variational trial functions is implicitly constrained to satisfy the gauge constraint, the inhomogeneous Helmholtz PDE is replaced for each $\omega > 0$ with a system of algebraic equations,
\begin{subequations}
\begin{align}
\tfrac{\del }{\del \bv{\alpha}}\tfrac{\del}{\del \omega} \tilde{\mathcal{S}}(\bv{\alpha}) &= \bv{0},\\
\tfrac{\del }{\del \lambda}\tfrac{\del}{\del \omega} \tilde{\mathcal{S}}(\bv{\alpha}) &= 0,
\end{align}
\end{subequations}
 whose solution $\tilde{\alpha} = \tilde{\alpha}[\bv{j}](\omega;  \mathcal{P}'_0),$ $\tilde{\lambda} = \tilde{\lambda}[\bv{j}](\omega;  \mathcal{P}'_0)$ determines the variational approximation $\tilde{\bv{a}}(\bv{\zeta};\omega; \tilde{\bv{\alpha}})$ to $\bv{a}(\bv{\zeta}; \omega).$

Now, following our earlier development, suppose we consider a family  $\bv{\chi}(\bv{\zeta}; \omega; \bv{\alpha})$ of trial vector fields for the general inhomogeneous problem which are explicitly constrained to be solenoidal and to satisfy the source-free Helmholtz equation, so the first term in the action density vanishes identically. Then the variational action spectral density becomes
\be
\tfrac{\del}{\del \omega}\tilde{\mathcal{S}}(\omega; \bv{\alpha}; \lambda) = \tfrac{2}{\omega} \impart \innerp{ \bv{\varepsilon}_{\bv{\chi}\, \perp}(\omega; \bv{\alpha}) }{\bv{j}(\omega)} 
+ \lambda(\omega) \bigl[
\tfrac{\del}{\del \omega}\mathscr{P}_{\stext{EM}}[\bv{\chi}^{\stext{out}}](\infty; \omega)  - \mathcal{P}'_0 (\omega)\bigr].
\ee
Despite certain similarities, the Ritz-Galerkin-type variational principle associated with this action is distinct form the MPVP.
The action principle involves finding constrained stationary points of $\impart \innerp{ \bv{\varepsilon}_{\bv{\chi}\, \perp}(\omega) }{\bv{j}(\omega)},$ while the MPVP involves finding the constrained maxima of  $\mathscr{P}_{\stext{mech}}[\bv{\varepsilon}_{\bv{\chi}\, \perp}, \bv{j}](\omega; \bv{\alpha}) = 
\realpart \innerp{ \bv{\varepsilon}_{\bv{\chi}\, \perp}(\omega; \bv{\alpha}) }{\bv{j}(\omega)}.$ 

Although generically the ``true'' Euler-Lagrange solutions correspond to saddle-points, when the trial functions are restricted to a parameterized source-free family and when the power-normalization constraint is imposed via a Lagrange multiplier, the overall problem inherits convexity from the constraint alone, and the critical points turn out to be power-constrained local maxima or minima of  $\impart \innerp{ \bv{\varepsilon}_{\bv{\chi}\, \perp}(\omega; \bv{\alpha}) }{\bv{j}(\omega)},$ which implies that the overall phase should be chosen such that $\realpart \innerp{ \bv{\varepsilon}_{\bv{\chi}\, \perp}(\omega; \bv{\alpha}) }{\bv{j}(\omega)} = 0.$  In the MPVP case, the variational solution corresponds to constrained local maxima of $-\realpart \innerp{ \bv{\varepsilon}_{\bv{\chi}\, \perp}(\omega; \bv{\alpha}) }{\bv{j}(\omega)},$ so that $\impart \innerp{ \bv{\varepsilon}_{\bv{\chi}\, \perp}(\omega; \bv{\alpha}) }{\bv{j}(\omega)} = 0.$  Any attempt to employ directly the action-based variational principle with the source-free trial solutions thus encounters two related problems: the ostensible method converges to a solution with a global phase error of $\pi/2;$ and as an immediate result, the calculated work performed on/by the sources vanishes, so the problem becomes degenerate, and we are left with no means to consistently choose the \textit{absolute} power level. 

It is not difficult to trace the origin of this phase error.  Naive application of Fourier transforms to the Green function yields the reciprocal-space, frequency-domain kernel
\be\label{eqn:green_4}
G^{\stext{out}}(\bv{k}, \bv{k}; \omega) = \delta(\bv{k} - \bv{k}') \frac{1}{\omega^2 -\vnorm{\bv{k}}^2},
\ee
but this neglects the requirement of causality demanded of the retarded Green function, wherein the response can appear only after the source is applied, implying that the Green function must be analytic for $\impart \omega \ge 0.$  We must be slightly more careful, and here treat the (temporal) Fourier transform as the limiting case of a Laplace transform, where the initial conditions are pushed back into the arbitrarily remote past rather than future, before being forgotten altogether.  The Green function (\ref{eqn:green_4}) must be understood in the sense
\be
G^{\stext{out}}(\bv{k}, \bv{k}; \omega) = \lim\limits_{\epsilon \to 0^{+}} \delta(\bv{k} - \bv{k}') \frac{1}{\vnorm{\bv{k}}^2 - (\omega + i \epsilon)^2}
\ee
where the limit is taken only \textit{after} $G^{\stext{out}}$ is convolved with the Fourier transform of the source.  In the full Fourier representation, the  vector potential and transverse source are then related by
\be
\bv{a}(\bv{k}; \omega) =  \lim\limits_{\epsilon \to 0^{+}} \frac{\bv{j}_{\perp}(\bv{k}; \omega)}{\vnorm{\bv{k}}^2 - (\omega + i \epsilon)^2},
\ee
where
\be
\bv{j}_{\perp}(\bv{k}; \omega) = (1 - \unitvec{k}\unitvec{k}\trans ) \bv{j}(\bv{k}; \omega),
\ee
and
\begin{subequations}
\begin{align}
\bv{a}(\bv{k}; \omega) &= \tfrac{1}{(2\pi)^{3/2}}\int d^{3}\bv{\zeta}\, \bv{a}(\bv{\zeta}; \omega) e^{-i \bv{k}\cdot \bv{\zeta}},\\
\bv{j}(\bv{k}; \omega) &= \tfrac{1}{(2\pi)^{3/2}}\int d^{3}\bv{\zeta}\, \bv{j}(\bv{\zeta}; \omega) e^{-i \bv{k}\cdot \bv{\zeta}}\\
\end{align}
\end{subequations}
are the (scaled) Fourier transforms.
For real $\omega$ and $\vnorm{\bv{k}} > \abs{\omega},$ we see that 
\be
\arg\left[\frac{ \unitvec{e}\!\cdot\! \bv{a}(\bv{k}; \omega)}{\unitvec{e} \!\cdot\! \bv{j}_{\perp}(\bv{k}; \omega)} \right] = 0,
\ee
while for $\vnorm{\bv{k}} < \abs{\omega},$
\be
\arg\left[\frac{ \unitvec{e}\!\cdot\! \bv{a}(\bv{k}; \omega)}{\unitvec{e} 
\!\cdot\! \bv{j}_{\perp}(\bv{k}; \omega)} \right] = \pi.
\ee
In either case, the associated electric field for these spectral components is then in quadrature with the current, and no time-averaged work is performed or energy exchanged.
However, on the singular resonant manifold where $\vnorm{\bv{k}} = \abs{\omega}$ exactly, the magnitude of the Green function diverges in the limit as $\epsilon \to 0^{+},$ but the phase is such that
\be
\arg\left[\frac{ \unitvec{e}\!\cdot\! \bv{a}(\bv{k}; \omega = \vnorm{\bv{k}})}{\unitvec{e} \!\cdot\! \bv{j}_{\perp}(\bv{k}; \omega = \vnorm{\bv{k}})} \right] = \frac{\pi}{2}.
\ee
The relative phase between field and source then jumps discontinuously, to where the currents deliver time-averaged energy to the fields.  Our purported action principle fails to capture this, while the MPVP, although operating entirely within this singular manifold of source-free solutions, relies on energy conservation, which then enforces the correct phase relation.  This is not to imply that action-based variational  principles are not also useful for radiation problems, only that they will not naturally work with source-free solutions, and generically will involve finding saddle-points, not extrema. 

In fact, most of the variational principles previously employed for FEL analysis, paraxial wave propagation, or laser-plasma problems are based on precisely this type of generalized  Ritz-Galerkin approximations to the underlying Lagrangian dynamics, although some of the authors have erroneously claimed extremal instead of mere stationary character for their techniques.  A Lagrangian-based variational principle is developed in \cite{firth:77} for paraxial optical propagation in an inhomogeneous gain medium, which is generalized in \cite{anderson_bonnedal:79} to include nonlinear self-focusing effects, but the authors erroneously claim extremal, not just stationary, properties.  This approach was further extended to include more general laser-plasma interaction terms in \cite{duda_mori:00}, where the authors correctly state that their trial solutions derive from a stationary principle.  In \cite{xie_deacon:86} the authors state without proof an extremal principle for the fundamental FEL mode in the small-signal regime, which in fact appears to be perfectly correct, but higher modes will merely be stationary.  These ideas are further generalized in \cite{amir_greenzweig:86, luchini_motz:88a, luchini_motz:88a, amir:88}, where the variational techniques are erroneously described as extremal principles.  Similar trial-function-based variational principles are also used in \cite{yu_et_al:90, hafizi_roberson:92, xie:00} for which, correctly, only stationarity is claimed.

A number of other authors have made extensive use of (correctly-stated) \textit{stationary} action principles both for general wave-wave and wave-particle interactions in kinetic and fluid treatments of plasmas and particle beams  \cite{kaufman_holm:84, pfirsch_morrison:85, turski_kaufman:87, kaufman_et_al:87, brizard:98} and more specifically for analysis of FELs \cite{semilon_wurtele:91}.  However, rather than involving parameter-laden trial functions, these approaches employ the variational principles in order to provide economical descriptions and derivations of dynamical equations and conservation laws  in the form of conventional ODEs or PDEs, and to effect in an efficient and transparent manner certain eikonal expansions or approximations, or oscillation-center/ponderomotive or other averaging procedures \cite{delaney:71, whitham:74} while preserving the Hamiltonian nature of the dynamics.

Such confusion between stationary and extremal principles seems deeply embedded in physics literature and folklore.  Linguistically, ``optimizing,''  ``maximizing,'' or ``minimizing'' simply sound better than ``criticalizing'' or ``making stationary.''  Philosophically, optimality enjoys a certain teleological appeal which mere stationarity lacks.  Historically, the buzzwords ``least action''  have been invoked so often in the discussion of dynamics that  Hamilton's Principle and the Principle of Least Action have been mistakenly conflated, despite being quite distinct concepts.  The Principle of Least Action was first articulated by Maupertuis and formalized by Euler, not Hamilton; is restricted to a smaller class of Lagrangians (no explicit time dependence); uses a different action (the abbreviated action integral) and a different set of constrained variations (the energy is fixed but temporal endpoints are not); and in fact it is itself misnamed, requiring only that the abbreviated action be stationary, not necessarily minimal, for the physical trajectory.  Practically, both optimization principles and stationary principles share an insensitivity of the variational quantity to the trial functions, but the extremal case additionally provides an upper or lower bound.

Numerically, more efficient and/or reliable computational techniques may be employed for optimization problems.  Finding (local) maxima or minima of functions is much easier than finding roots of systems of equations.  The ``Alpine'' analogy is often made: lost in a foggy terrain, the mountain-climber can reliably find a nearby peak (local maximum) by moving uphill, a nearby valley (local minimum) by wandering downhill, but has no guaranteed strategy for finding a mountain pass (saddle-point.)   If the original action density is convex (at least in some sufficiently large region of function space), so that the true solution corresponds to an extremum of the action, then in the Raleigh-Ritz approximation we have a natural metric to measure both ``closeness'' to the true solution and ``progress'' in iteratively determining the approximate one  -- namely, a metric induced by the action itself.  Without convexity, the true and approximate solutions may correspond to saddle-points, and we must introduce some arbitrary external metric to measure similarity or improvement.  Even in the classic formulations which employ only linear expansions in basis functions,  the extremal case leads to the solution of equations corresponding to a symmetric (or Hermitian), positive definite matrix, while in the merely stationary problem the matrix is in general Hermitian but not positive definite.

In the case of electromagnetic radiation, the stationary points are generically saddles, because of the hyperbolic nature of the wave equation.  Alternatively, one can see this by the well-known equivalence of electromagnetic mode dynamics to a collection of harmonic oscillators, for which the action functional is known to be minimal only for sufficiently short times intervals away from turning points along an orbit.

Similar subtleties occur in the usual Raleigh-Ritz approximation for the stationary states in quantum mechanics.  The energy-expectation value is always a local minimum for the ground state, and is always stationary for any excited state, but is only minimal for excited states if the variations are constrained to be orthogonal to all lower states.

To summarize, the MPVP involves finding constrained \textit{extrema} of quantities of the form:
\be
\mathscr{P}'[\omega; \bv{\alpha}]  \sim \impart \left[\int \!d^{3}\bv{x}\,  \bv{\varepsilon}(\bv{x}; \omega; \bv{\alpha})\cc \!\cdot\! \bv{j}(\bv{x}; \omega)\right];
\ee
while Rumsey reaction-based variational principles would involve finding constrained \textit{stationary} points of quantities of the form:
\be
\mathscr{R}[\omega; \bv{\alpha}]   \sim \impart \left[ \int d^{3}\bv{x}\,  \bv{\varepsilon}(\bv{x}; \omega; \bv{\alpha}) \!\cdot\! \bv{j}(\bv{x}; \omega)\right];
\ee
and Lagrangian action-based principles would involve finding constrained \textit{stationary} points of quantities of the form:
\be
\mathcal{S}[\omega; \bv{\alpha}]   \sim \realpart \left[ \int d^{3}\bv{x}\,  \bv{\varepsilon}(\bv{x}; \omega; \bv{\alpha})\cc \!\cdot\! \bv{j}(\bv{x}; \omega)\right];
\ee
so indeed they all appear to be distinct variational principles.


\section{Conclusions}\label{section:conclusions}

We have reviewed in some detail a Hilbert-space and operator-based approach to electromagnetic radiation, and have used this formalism to derive in some detail a maximum-power variational principle (MPVP) for spontaneous radiation from prescribed classical sources, first in the paraxial limit and then in a more general setting, adding to the large family of variational techniques for electromagnetic problems in general, and undulator/wiggler radiation in particular.  Mathematical details aside, at its most essential, the MPVP is really just a straightforward and rather obvious consequence of two simple and rather obvious constraints, together with another fundamental mathematical fact: the power in any one source-free mode of the electromagnetic field may not exceed the total power in all the modes (\ie, Bessel inequality);  and the power radiated must be attributable to power delivered by the sources, even in the regime where we ignore back-action on the sources (\ie, conservation of energy); while information about the fields on some boundary surface, needed to determine this radiated power, can be converted into information about the derivatives of the fields in the interior (\ie, Gauss's law, one of the multidimensional generalizations of the Fundamental Theorem of Calculus).

This approximation, or ones similar to it,  has been frequently used, almost without comment or perceived need for further justification, in classical or quantum stimulated emission situations, where in the presence of gain we naturally expect to observe that mode which grows the fastest, but it is equally applicable in the spontaneous regime, because arguments along the lines of Einstein's derivation of the $A$ and $B$ coefficients lead to definite connections between spontaneous emission, stimulated emission, and stimulated absorption, even when the radiation is completely classical.  However simple, even trivial, these observations are not without practical content or application to undulator and possibly other radiation problems.

\section*{Acknowledgements}\label{section:acknowledgements}

We acknowledge useful discussions with W. Fawley, G. Penn, R. Lindberg, J. Morehead, A. Zholents, and J. Zimba.  This research was supported by the Division of High Energy Physics, U.S. Department of Energy, and by DARPA, U.S. Department of Defense.


\bibliographystyle{unsrt}
\bibliography{maxpowervp}

\end{document}